\documentclass[runningheads,a4paper]{llncs}

\usepackage{placeins}  
\usepackage{color}
\usepackage[usenames,dvipsnames]{xcolor}
\usepackage{url,graphicx,times}
\usepackage{tabularx}
\usepackage{multirow}
\usepackage{booktabs, array, adjustbox} 
\usepackage[square,comma,numbers,sort&compress,sectionbib]{natbib}
\usepackage{enumerate}
\usepackage{amsmath}
\usepackage{hyperref}
\usepackage{amssymb}
\usepackage{graphicx}
\usepackage{hyperref}
\usepackage[ruled,vlined,linesnumbered]{algorithm2e}
\usepackage{cleveref}
\usepackage{enumitem}
\usepackage{floatrow}

\usepackage{fancyhdr}

\fancypagestyle{specialfooter}{%
  \fancyhf{}
  
  \fancyfoot[C]{\vspace{1cm} \hspace*{-.2\textwidth}\parbox{1.4\textwidth}{\small This is the authors version of the work. It is posted here for your personal use. The definitive version is published in: \\ \emph{Proceedings of the 41th European Conference on Information Retrieval} (ECIR '19), April 14--18, 2019, Cologne, Germany}}
}

\begin{document}

\mainmatter

\title{Impact of Training Dataset Size on Neural Answer Selection Models}

\author{Trond Linjordet and Krisztian Balog}
\institute{University of Stavanger, Stavanger, Norway\\
\email{
\{trond.linjordet,krisztian.balog\}@uis.no}
}
\maketitle

\begin{abstract}
It is held as a truism that deep neural networks require large datasets to train effective models. However, large datasets, especially with high-quality labels, can be expensive to obtain. This study sets out to investigate (i) how large a dataset must be to train well-performing models, and (ii) what impact can be shown from fractional changes to the dataset size. A practical method to investigate these questions is to train a collection of deep neural answer selection models using fractional subsets of varying sizes of an initial dataset. We observe that dataset size has a conspicuous lack of effect on the training of some of these models, bringing the underlying algorithms into question.

\end{abstract}

\thispagestyle{specialfooter}

\section{Introduction}
\label{sec:intro}

The impressive performance improvements brought by deep learning applied to certain domains---computer vision, audio speech-to-text, and natural language processing (NLP)~\cite{SCHMIDHUBER201585, LeCun2015}---has motivated a great deal of interest to apply deep learning to other domains as well, including information retrieval (IR). However, the performance improvements from deep learning relative to conventional machine learning approaches have depended on increased computational power, larger datasets to learn from, and some developments on the algorithm and architecture level. Of these three factors, large datasets may represent the least tractable challenge faced by those who would apply deep learning to new domains. Quality training data, especially for supervised learning, requires intensive effort to prepare for the actual learning process. \par

A category of tasks at the intersection of the fields of IR and NLP, question answering (QA) means returning a correct answer sentence in response to a grammatically well-formed, natural language question. In the present work, a specific variant of the QA task is considered, namely answer selection, 
the task of matching single-sentence questions with single-sentence answers. The answer selection task is simply: given a question and a set of candidate answers, select the correct answer. This task has recently been investigated as a neural IR problem \cite{Mitra2017neural-IR-survey, Onal2017}. 

This paper considers a practical approach to investigating the impact of training dataset size on the performance that can be achieved with various deep neural architectures for the task of answer selection. The approach taken by this paper can be summarized as follows: A pre-existing implementation of neural architectures for answer selection is investigated by truncating the training data to fractions of the original training dataset, to quantify the differences in performance by trained models given different amounts of training data from the same distribution. \par

One of the surprising experimental findings of this paper is that most models do not exhibit the expected behavior in terms of performance improvement in response to increased training dataset size. 
\section{Related Work}
\label{sec:related}

The impact of the size of training datasets has been investigated for convolutional neural networks (CNNs) trained on image data \cite{2015arXiv151106348C, Sun_2017_ICCV}. In the latter work, it was observed that model performance improves roughly logarithmically as a function of increased training data. The idea of a logarithmic relationship between performance and dataset size was further corroborated empirically by Hestness, et al. \cite{2017arXiv171200409H}.  \par 

An investigation of the generalization problem in deep neural networks, i.e., the discrepancy between the performance of a trained model on training data and test data, shows that the deep neural models have a representational capacity that enables ``memorization'' of training data: Zhang et al. \cite{2016arXiv161103530Z} show the order-of-magnitude relationship between training dataset size (sample size), input data dimensionality, and the depth of a network with sufficient parameters to fully memorize the training dataset.  
They report a theorem with proof such that for any finite $n$-sized sample of $d$-dimensional inputs, there exists a two-layer ReLU neural network with $2n+d$ weights that can represent any function on the sample. As a corollary, this finding extends from this hypothetical shallow and wide network to a narrow and deep network where the relationship between sample size and number of parameters is conserved. 
This may not be how deep neural networks learn in practice \cite{pmlr-v70-arpit17a}, but the theorem indicates the challenge that finite datasets may present to generalization in deep learning models. 

\section{Approach}
\label{sec:approach}

The approach presented in this paper is practical in that dataset size was manipulated and the effects were evaluated using a pre-existing implementation of multiple neural IR models with a single original dataset. Specifically, this paper presents work on the MatchZoo project\footnote{\url{https://github.com/faneshion/matchzoo}} \cite{MatchZoo2017arXiv170707270F}, where a number of deep neural architectures for text matching have been implemented. Here, answer selection is considered as a form of question answering, where the question text is matched with the text of the correct answer. The original dataset used for training, validation, and testing, was the canonical 
WikiQA dataset \cite{wikiqa-a-challenge-dataset-for-open-domain-question-answering}. The performance of the implemented models on a given dataset was characterized in terms of Mean Average Precision (MAP) over the candidate answer rankings for each question in that dataset. \par 

\subsection{Data preparation}
The training dataset was filtered to provide the models being trained with meaningfully labelled training data. The filtering rule was simply to omit any question and its associated set of candidate answer sentences if the set of candidate answer sentences did not include both true and false candidates. \par 

\begin{table}[t]
  \caption{Summary of datasets.}
  \vspace*{-0.5\baselineskip}
  \label{tab:datasets}
  \begin{tabular}{l@{~~~~}r@{~~~~}r@{~~~~}r@{~~~~}r@{~~~~}r@{~~~~}r@{~~~~}r}
    \toprule
	& \multicolumn{5}{c}{Training} & & \\
    & $10 \%$ & $25\%$ & $50\%$ & $75\%$ & $100\%$ & Valid. & Test\\
    \midrule
    \#Questions & 78 & 209 & 414 & 639 & 857 & 122 & 237 \\
    \#QA pairs & 823 & 2256 & 4321 & 6537 & 8651 &  1126 & 2341 \\
  \bottomrule
\end{tabular}
\end{table}

Table~\ref{tab:datasets} summarizes the datasets used in the training of the various models. Note that the same validation and test sets were used throughout, while the training dataset used was systematically varied between the original (filtered) training set ($100\%$), and various partial training sets truncated to $10 \%$, $25\%$, $50\%$, and $75\%$ of the original (filtered) training set. These partial training sets were made by randomly sampling (without replacement) on the questions in the original (filtered) training set. Each selected question was then included in the respective partial training set along with all corresponding candidate answers and their labels. The percentages thus represent the probability for each question to be included in each partial dataset. However, once the random sub-sampling was accomplished, these partial training sets were fixed. Each of the models was then trained five times independently on each dataset size.

\subsection{Models}
A number of models were able to train and perform nominally with the code provided by the MatchZoo project \cite{MatchZoo2017arXiv170707270F}. The models investigated in the present paper comprised:

\begin{itemize}[leftmargin=*]
\item{\textbf{Deep Structured Semantic Model (DSSM)} \cite{Huang:2013:LDS:2505515.2505665}, which extends latent semantic analysis with deep architectures; a seminal work on neural IR.}
\item{\textbf{Convolutional Deep Structured Semantic Model (CDSSM)} \cite{Shen:2014:LSR}, which uses a convolutional neural network (CNN) to extend DSSM with contextual information at the word n-gram level. }
\item{\textbf{Architecture-I (ARC-I)} \cite{Hu:2014:CNN:2969033.2969055}, an extension of CDSSM whereby siamese CNNs learn to represent two sentences, deferring matching of sentence pairs to a final multi-layer perceptron (MLP).} 
\item{\textbf{Architecture-II (ARC-II)} \cite{Hu:2014:CNN:2969033.2969055}, an alternative to ARC-I where sentences interact by 1D convolution before proceeding through a 2D CNN component which is purported to learn both the representation of the indvidual sentences, as well as the structure of their relationship. Again, matching of the representations is determined by a final MLP.}
\item{\textbf{Multiple positional sentence representations (MV-LSTM)} \cite{Wan:2016:DAS:3016100.3016298}, follows the aforementioned models by capturing local information on multiple levels of granularity within a sentence, using bidirectional long short-term memory networks (bi-LSTMs) to represent input sentences, modeling interactions with a similarity function (tensor layer), and aggregating interactions with $k$-Max Pooling before a final MLP to match the obtained representations.}
\item{\textbf{Deep relevance matching model (DRMM)} \cite{Guo:2016:DRM:2983323.2983769}, distinguishes relevance matching from semantic matching, using pre-trained neural embeddings of terms and building up fixed-length matching histograms from variable-length local interactions between each query term and document. Each query term matching histogram is passed through a matching MLP, and the overall score is aggregated with a query term gate---a softmax function over all terms in that query.}
\item{\textbf{Attention-based neural matching model (aNMM)} \cite{Yang:2016:ARS:2983323.2983818}, which follows a similar structure as ARC-II, except instead of position-shared weighting, aNMM has adopted a value-shared weighting scheme ``to learn the importance of different levels of matching signals,'' and incorporated a query term gate similar to that used in DRMM.}
\item{\textbf{Combined local and distributed representations (DUET)} \cite{Mitra:2017:LMU:3038912.3052579}, which aims to combine local exact matching with embeddings of query-document pairs in semantic space. This relevance matching is enabled by both the local and distributed models, hence a ``duet'' of two parallel neural models. The final matching score is simply the sum of the two outputs.}
\item{\textbf{MatchPyramid} \cite{Pang:2016:TMI:3016100.3016292}, which uses a matching matrix layer to evaluate pairwise term similarity between two texts, followed by 2D convolutional and pooling layers, with a final matching MLP. } 
\item{\textbf{DRMM{\_}TKS}  \cite{MatchZoo2017arXiv170707270F}, which is a variant of DRMM provided by the MatchZoo project, for matching short texts. The architecture is simply described by ``Specifically, the matching histogram is replaced by a top-k max pooling layer and the remaining parts are fixed.''}
\end{itemize} 

\noindent
Some of these models are motivated more by ad hoc search and document retrieval, whereas others were developed specifically for answer selection and the similar task of sentence completion. However, the commonality is that all the models are designed for text matching. 
\section{Experiment/Results} 
\label{sec:experiment}

The following experimental results show the effect of varying training set size. \par 

\subsection{Final performance of trained models}

\begin{figure}[t]
   \centering
   \begin{tabular}{c@{~}c@{~}c@{~}c@{~}}   
   DSSM
   & 
   CDSSM 
   &
   ARC-I
   & 
   ARC-II
   \\
   \includegraphics[width=.25\textwidth]{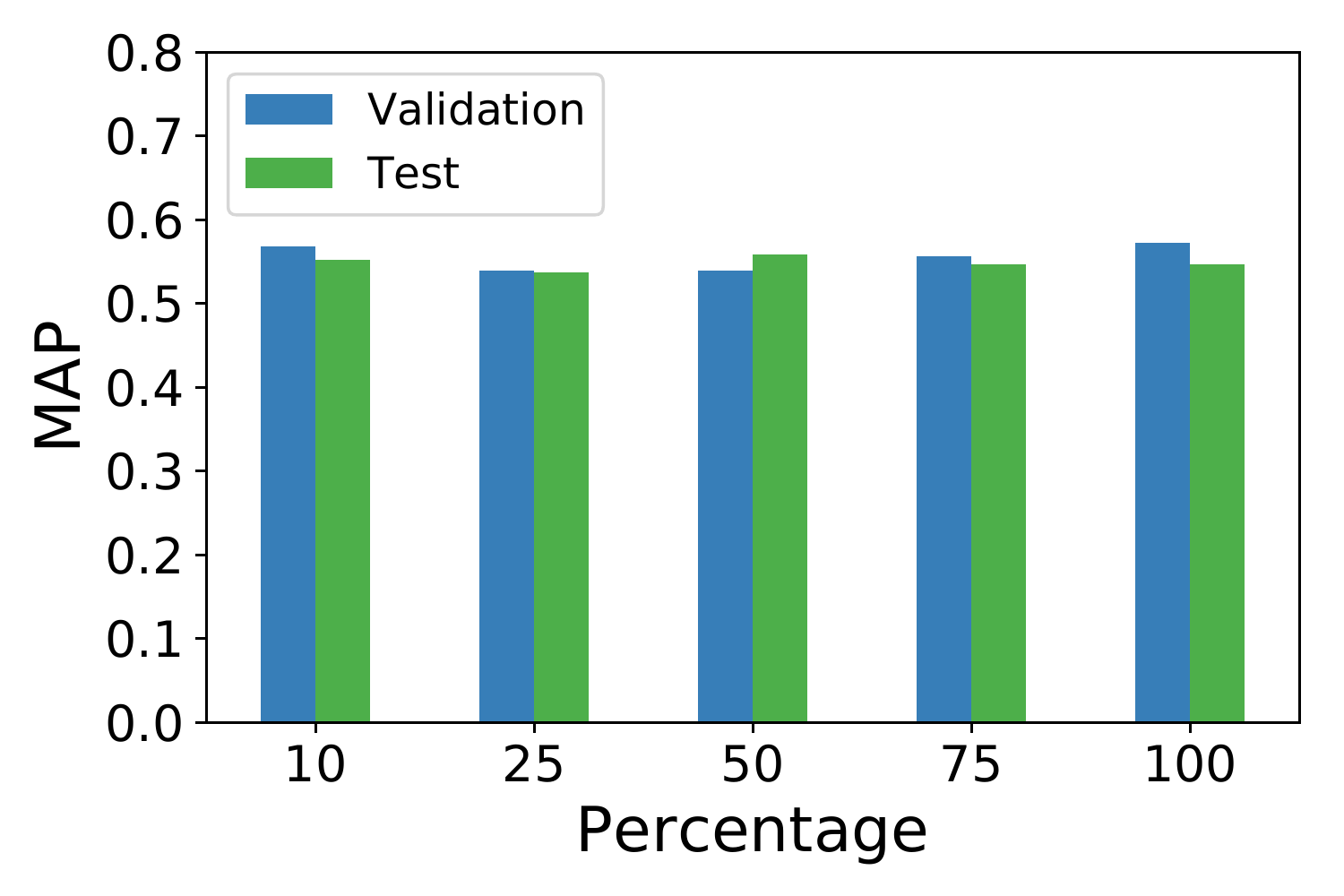}
   & 
   \includegraphics[width=.25\textwidth]{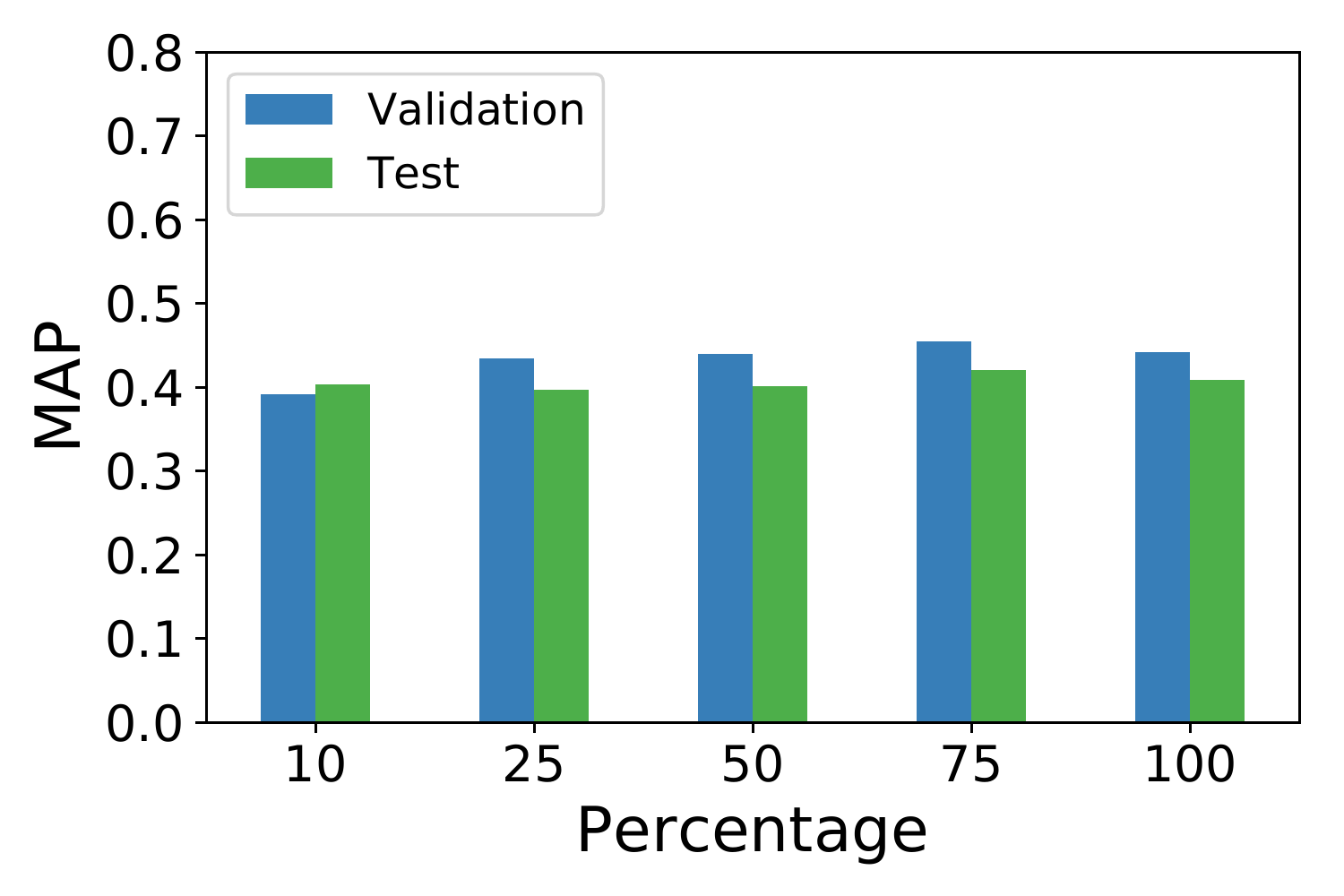}
   &
   \includegraphics[width=.25\textwidth]{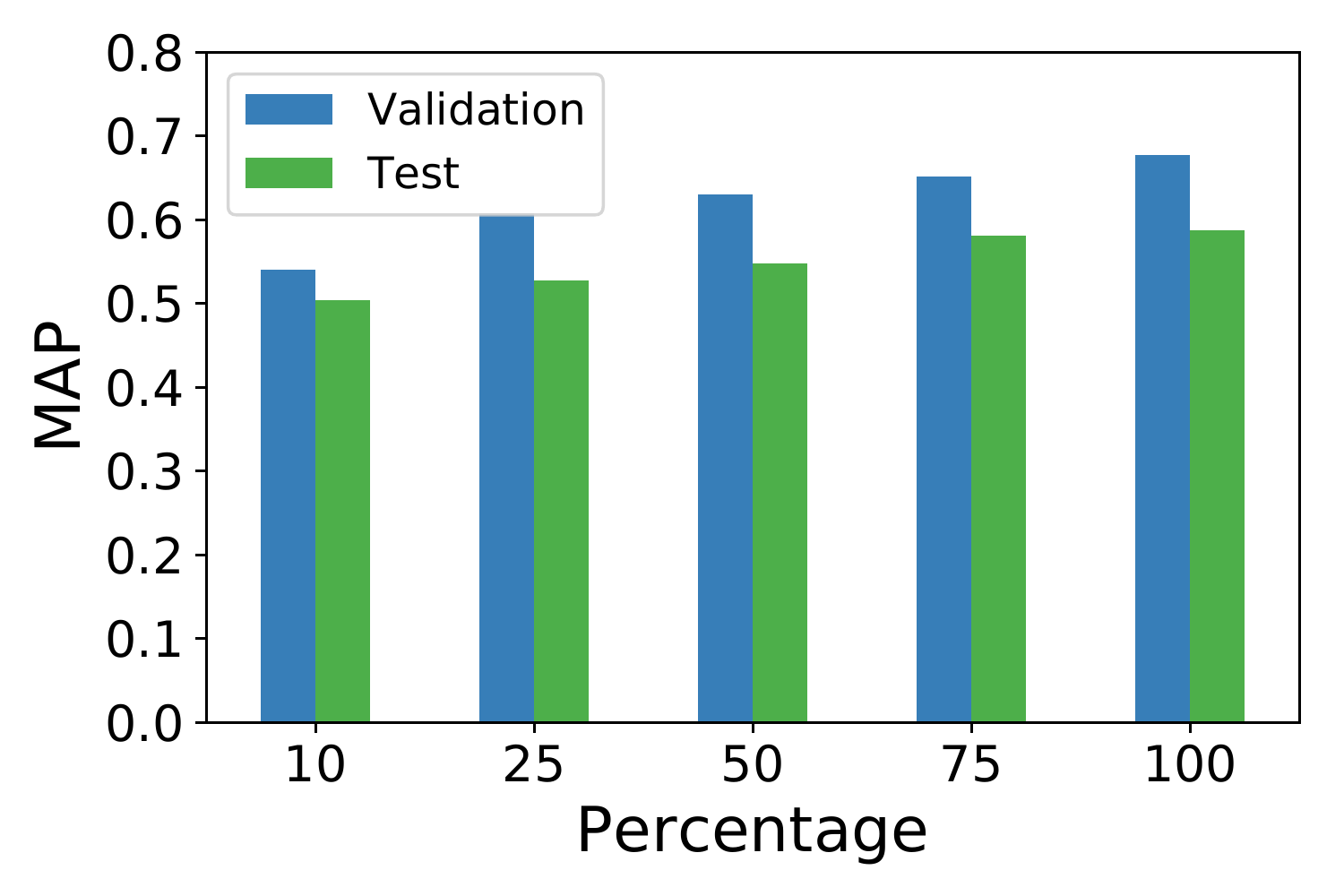}
   &
   \includegraphics[width=.25\textwidth]{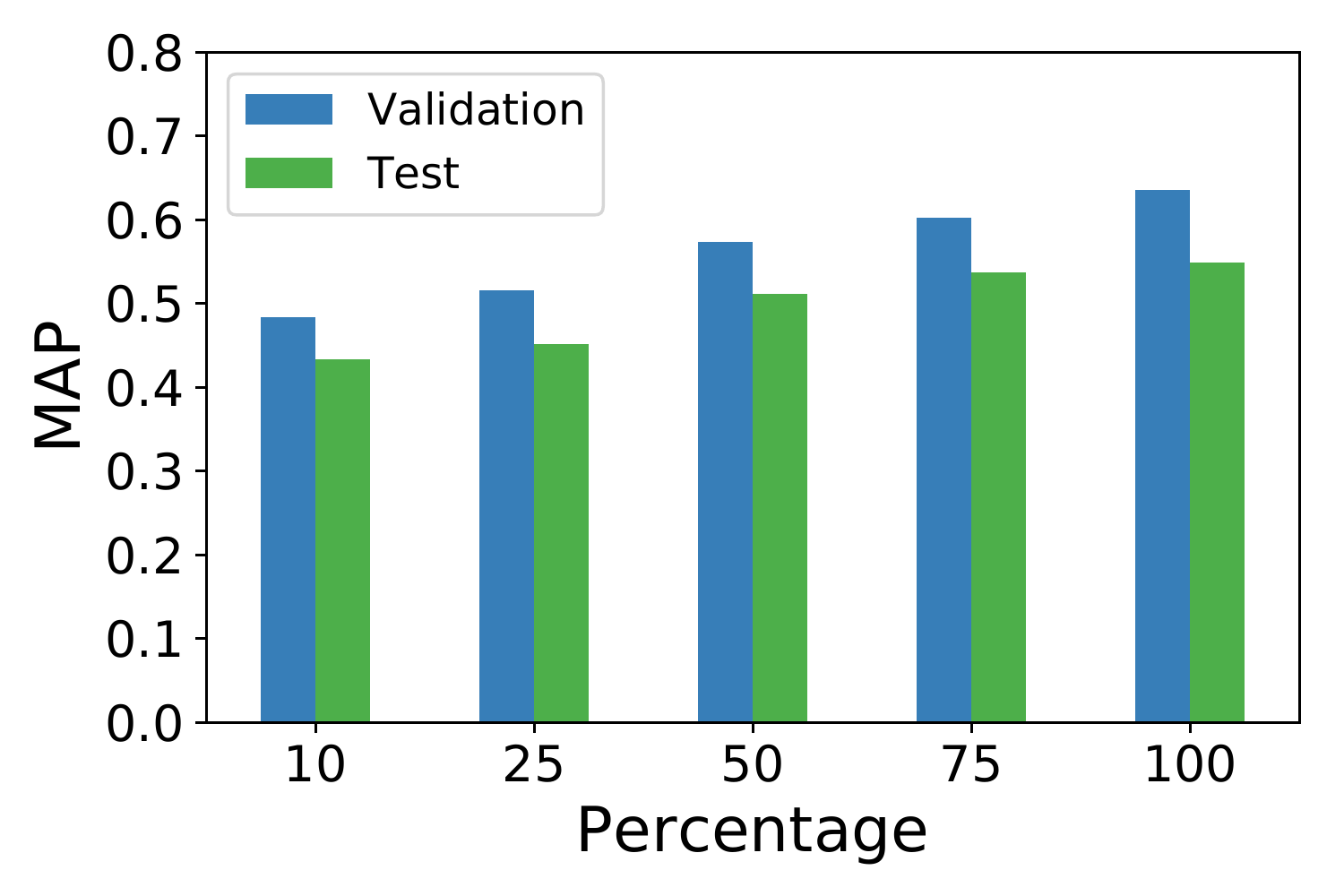}
   \\
   MV-LSTM
   &
   DRMM
   &
   aNMM
   &
   DUET
   \\
   \includegraphics[width=.25\textwidth]{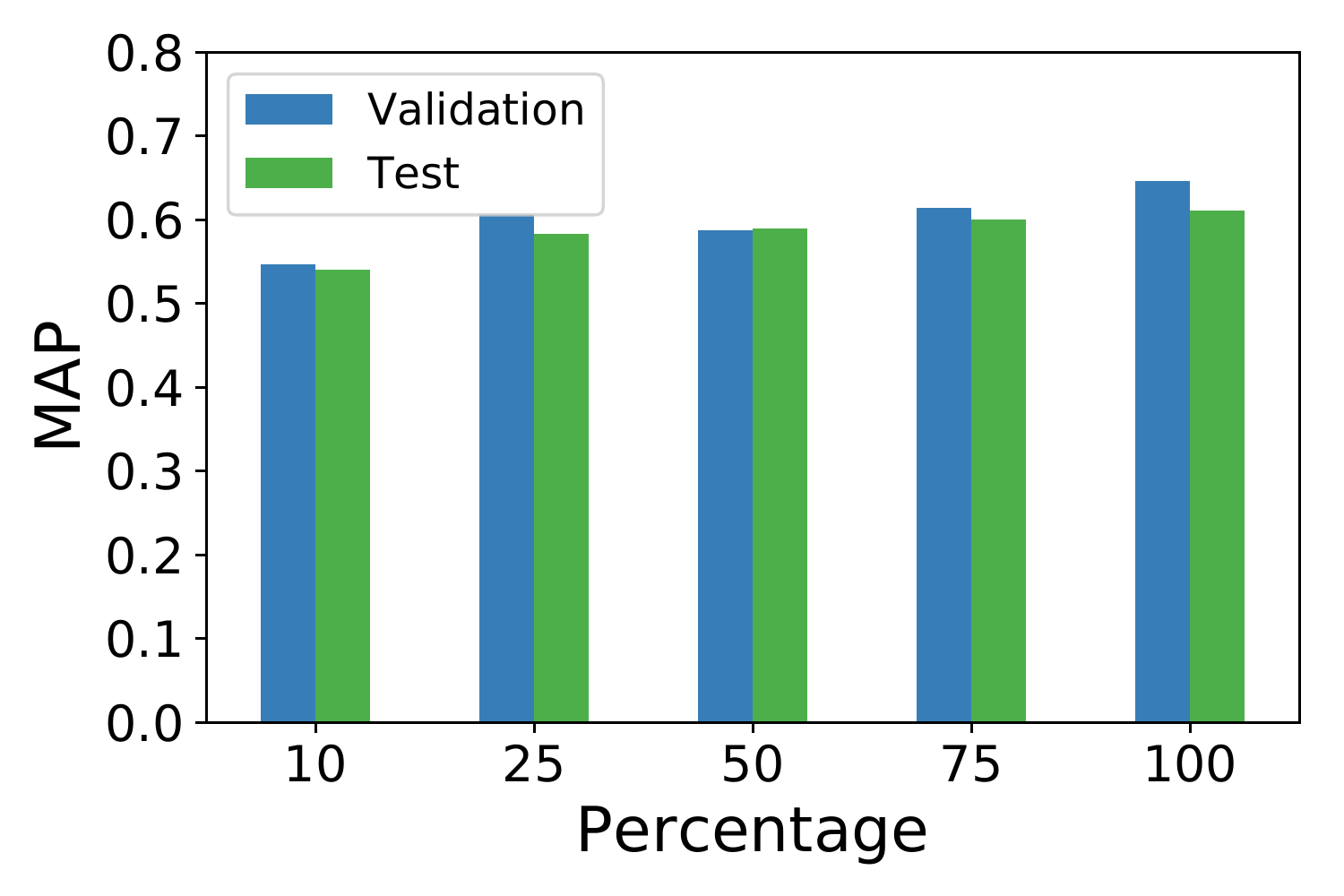}
   &
   \includegraphics[width=.25\textwidth]{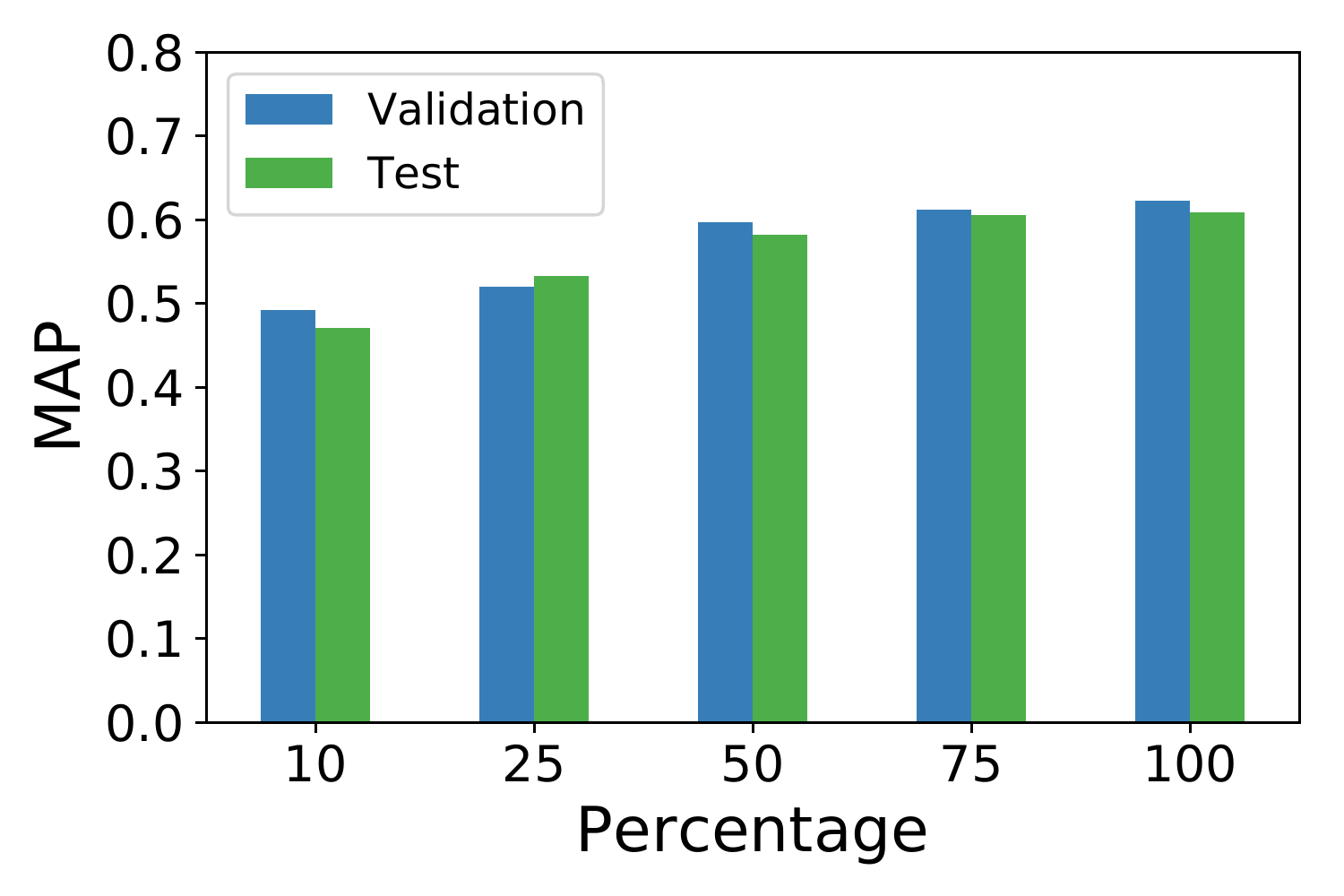}
   &
   \includegraphics[width=.25\textwidth]{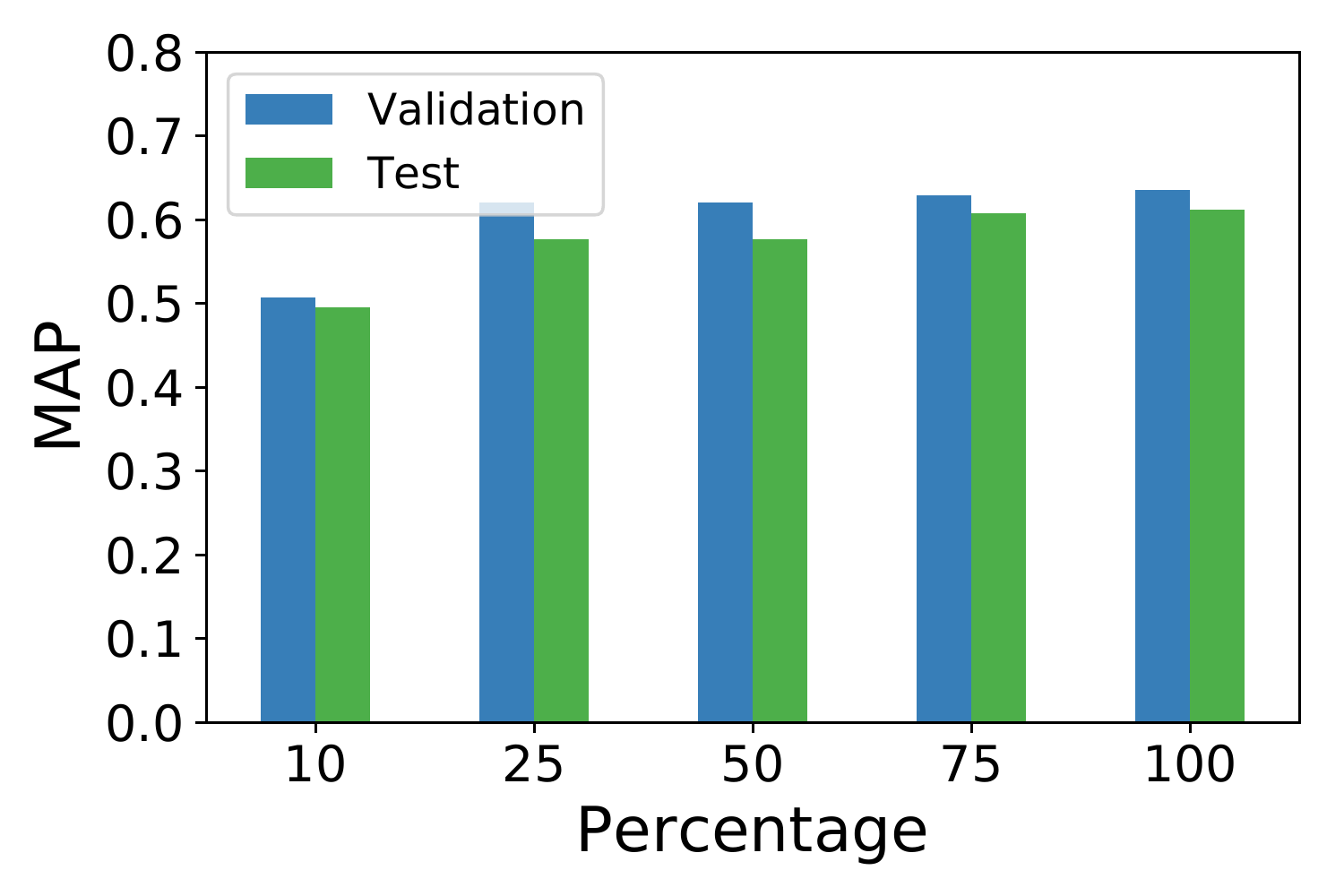}
   &
   \includegraphics[width=.25\textwidth]{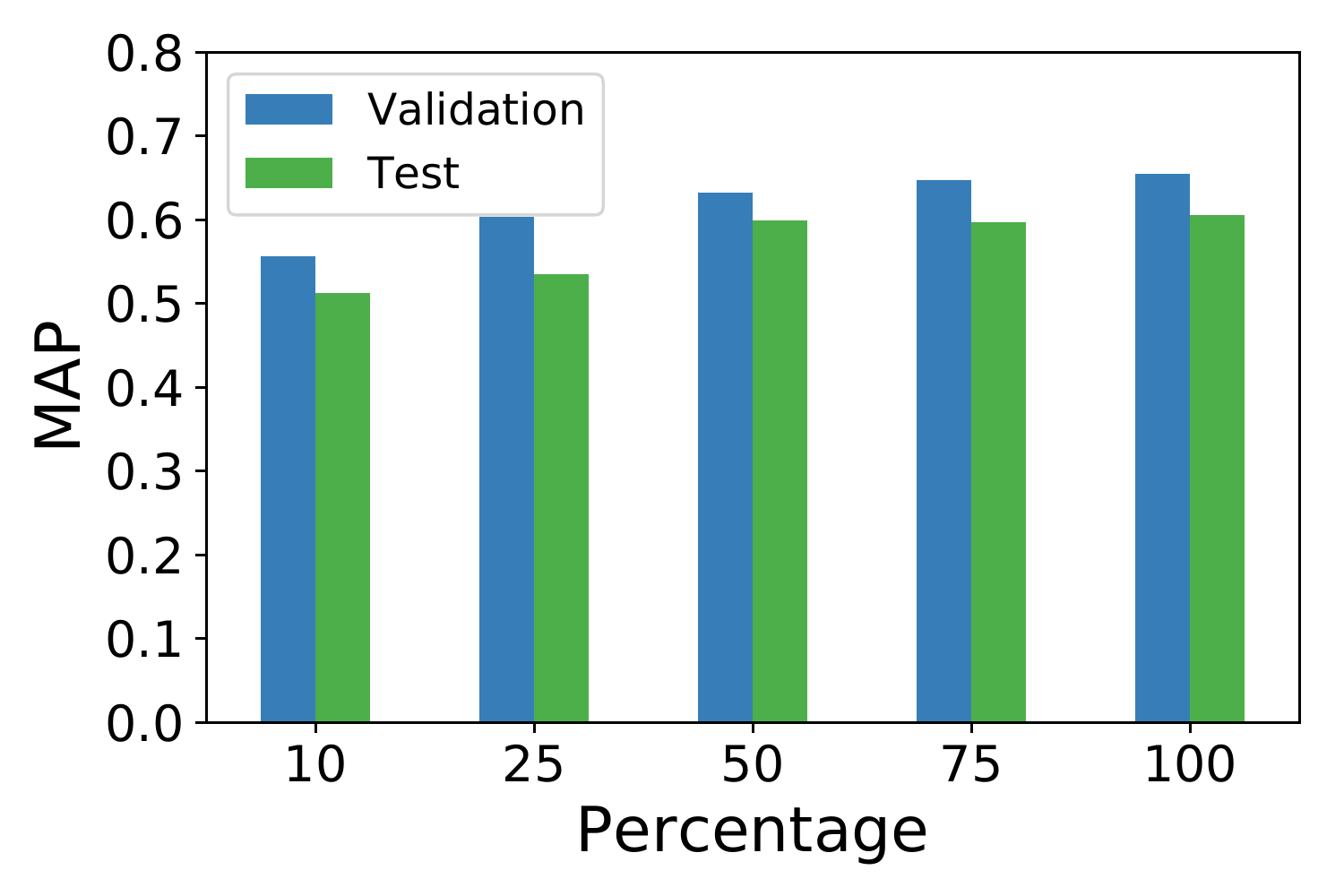}
   \\
   &
   MatchPyramid
   &
   DRMM TKS   
   &
   \\
   &
   \includegraphics[width=.25\textwidth]{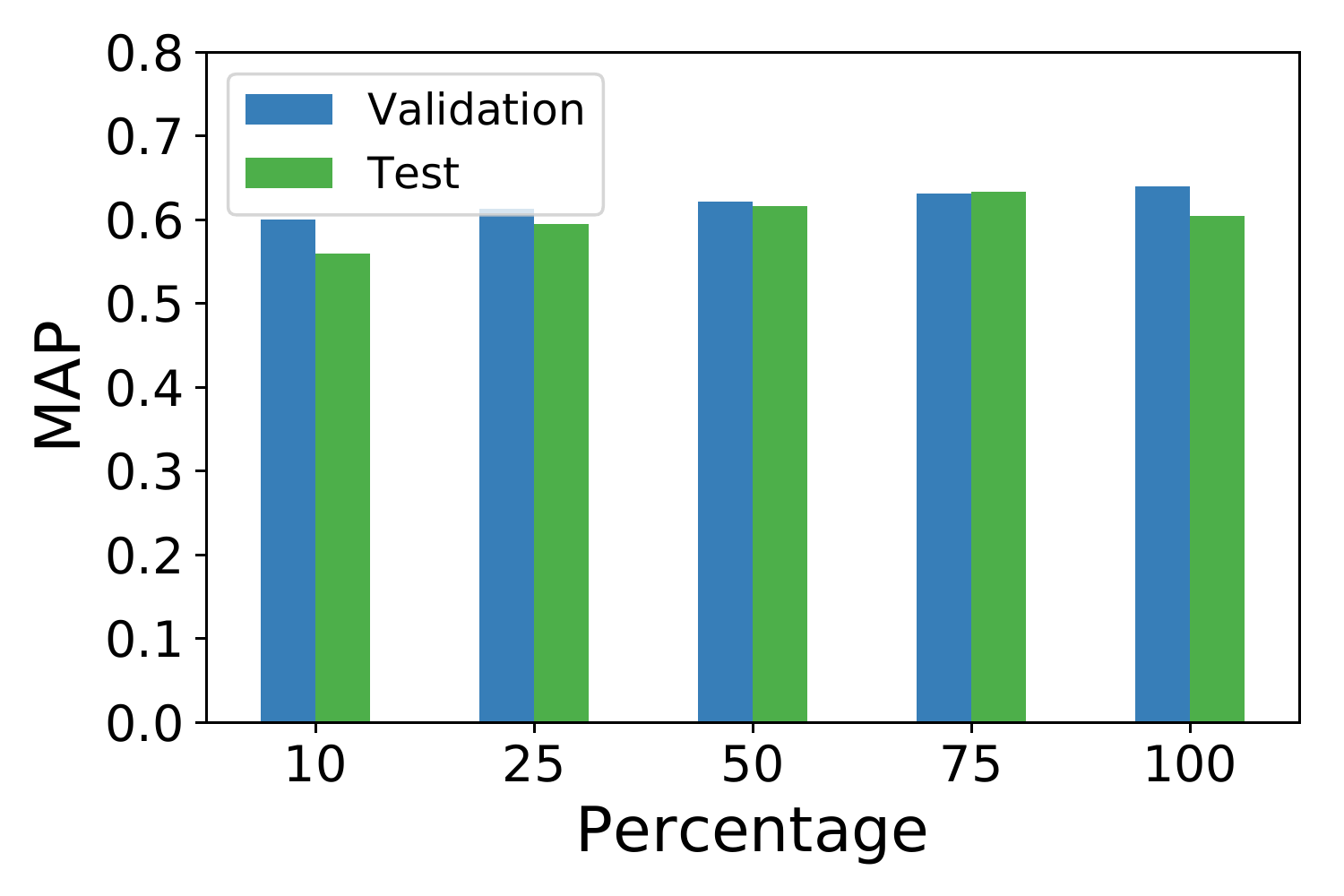}
   &
   \includegraphics[width=.25\textwidth]{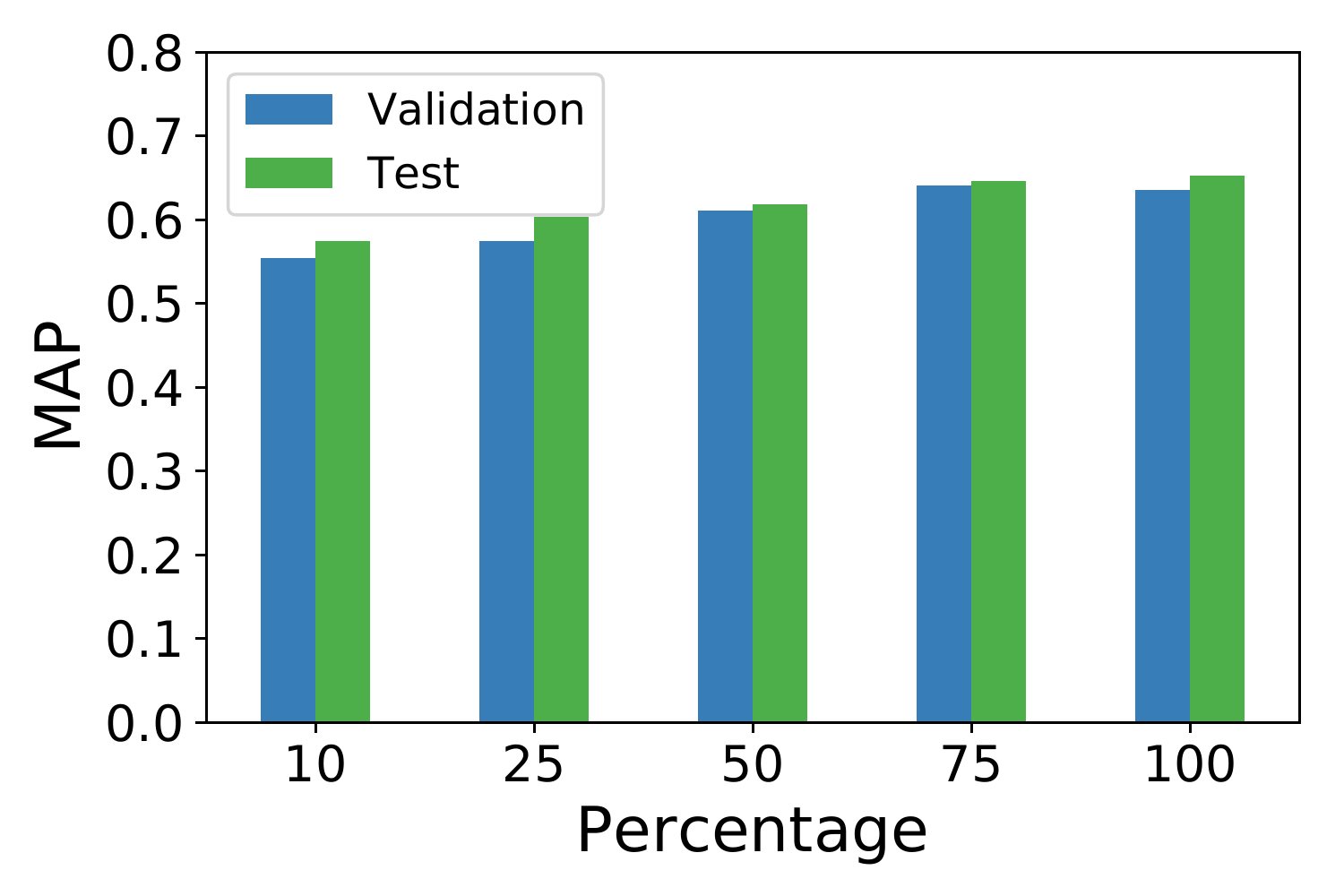}
   &   
   \\
   \end{tabular}    
   \vspace*{-0.5\baselineskip}   
   \caption{Performance (as measured by mean average precision) on the validation (blue) and test (green) datasets with different training dataset sizes.}
   \label{fig:finalperformance}
\end{figure}

Figure~\ref{fig:finalperformance} presents the performance on the test dataset of the different models after training for $400$ iterations on datasets of various sizes.
These figures show that aside from the DSSM, CDSSM, and possibly MatchPyramid models, some improvement does appear to happen with greater training dataset sizes. 
However, by having an order of magnitude more training data (10\% to 100\%), only three models, CDSSM, ARC-II, and DRMM{\_}TKS, achieve a relative improvement above 20\%.  Four more models, DSSM, MV-LSTM, aNMM, and DUET manage to achieve a relative improvement above 10\%.  For DRMM, performance even slightly decreases (by 1\%).
The relative improvements after having doubled (25\% to 50\%), tripled (25\% to 75\%), or quadrupled (25\% to 100\%) the training data size are similarly moderate for most models. Specifically, after doubling, only CDSSM and aNMM showed relative improvement above 10\%, and with tripling and quadrupling, only DSSM, CDSSM, ARC-II, and aNMM showed relative improvement above 10\%.

\subsection{Model training histories}

\begin{figure}[h!]
   \centering
   \begin{tabular}{l@{~}c@{~}c@{~}c@{~}c@{}c@{~}c@{~}c@{~}c@{~}c@{~}c@{}}
    & {\tiny DSSM} & {\tiny CDSSM} & {\tiny ARC-I} & {\tiny ARC-II} & {\tiny MV-LSTM} & {\tiny DRMM} & {\tiny aNMM} & {\tiny DUET} & {\tiny MatchPyramid}  & {\tiny DRMM TKS}\\
   \raisebox{.3cm}{\tiny $10\%$} & 
   \includegraphics[width=.090\textwidth]{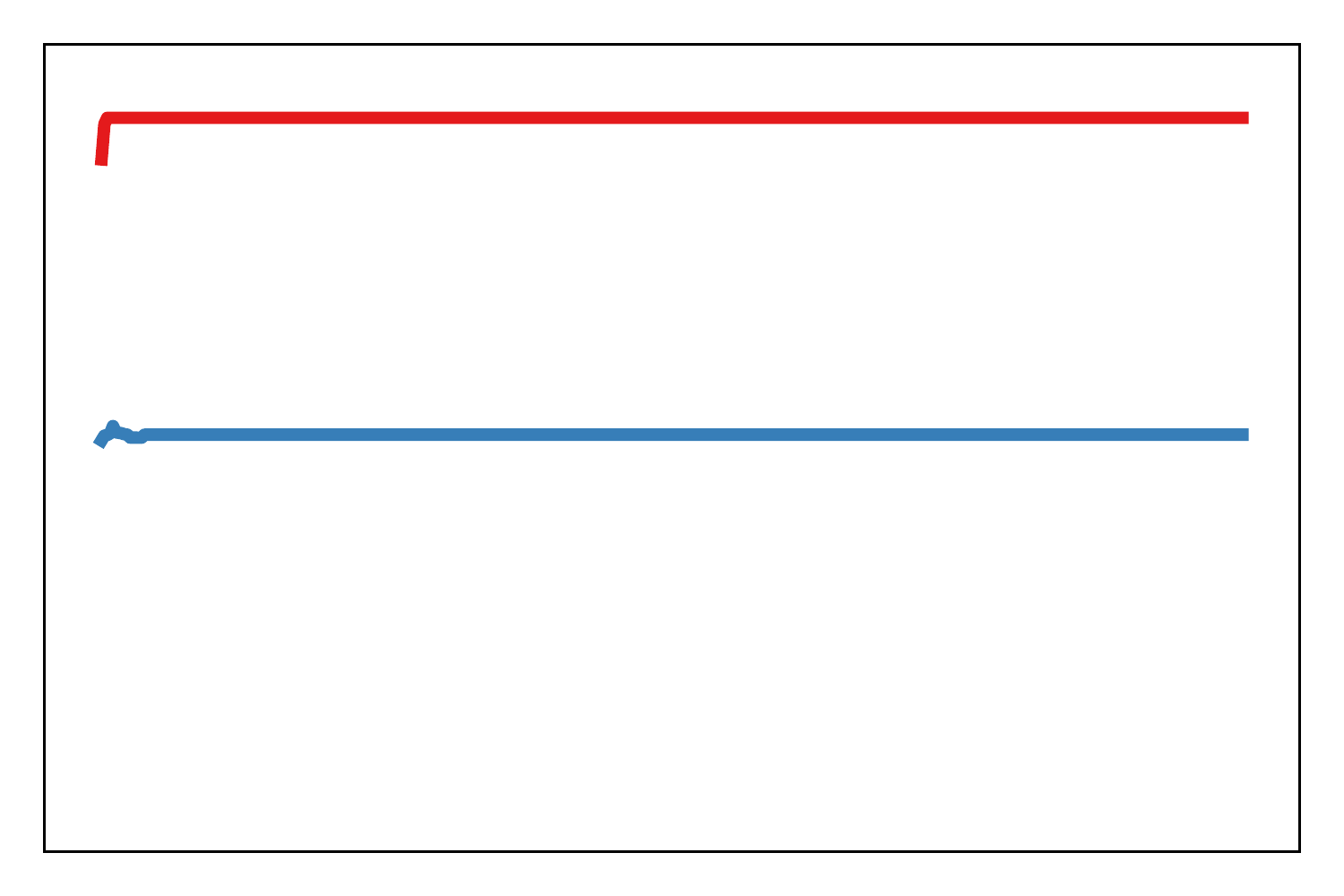}
   &
   \includegraphics[width=.090\textwidth]{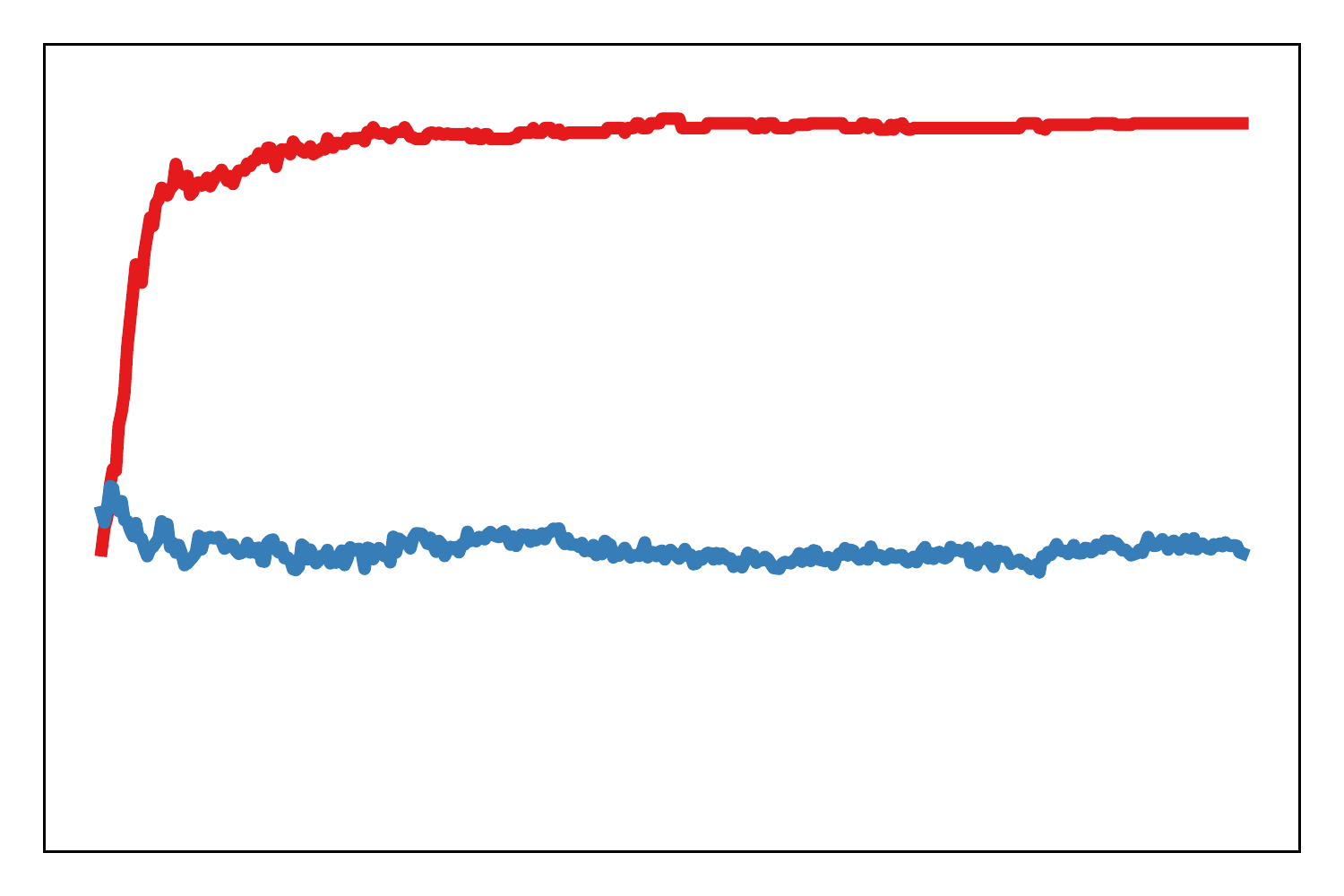}
   &
   \includegraphics[width=.090\textwidth]{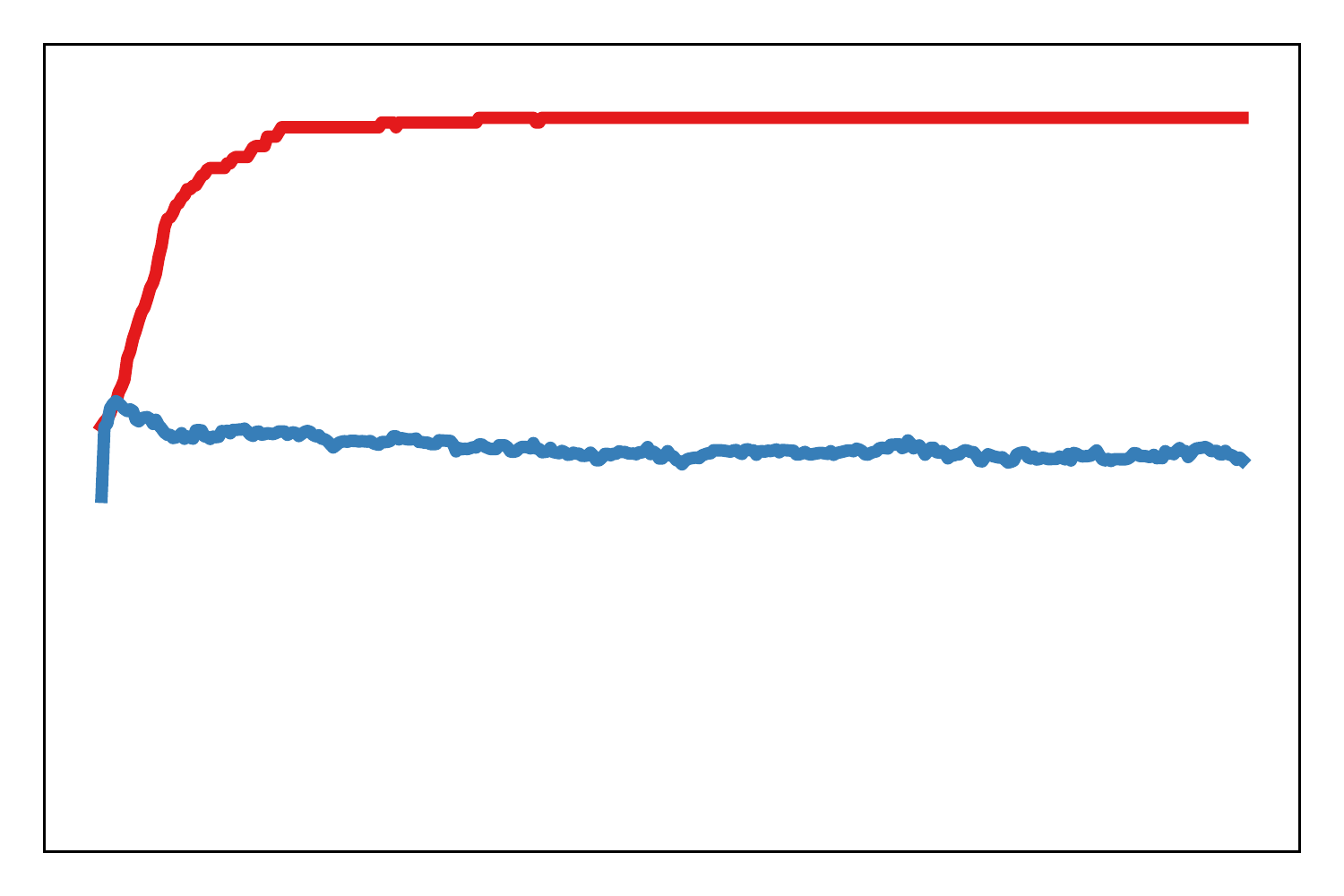}
   &
   \includegraphics[width=.090\textwidth]{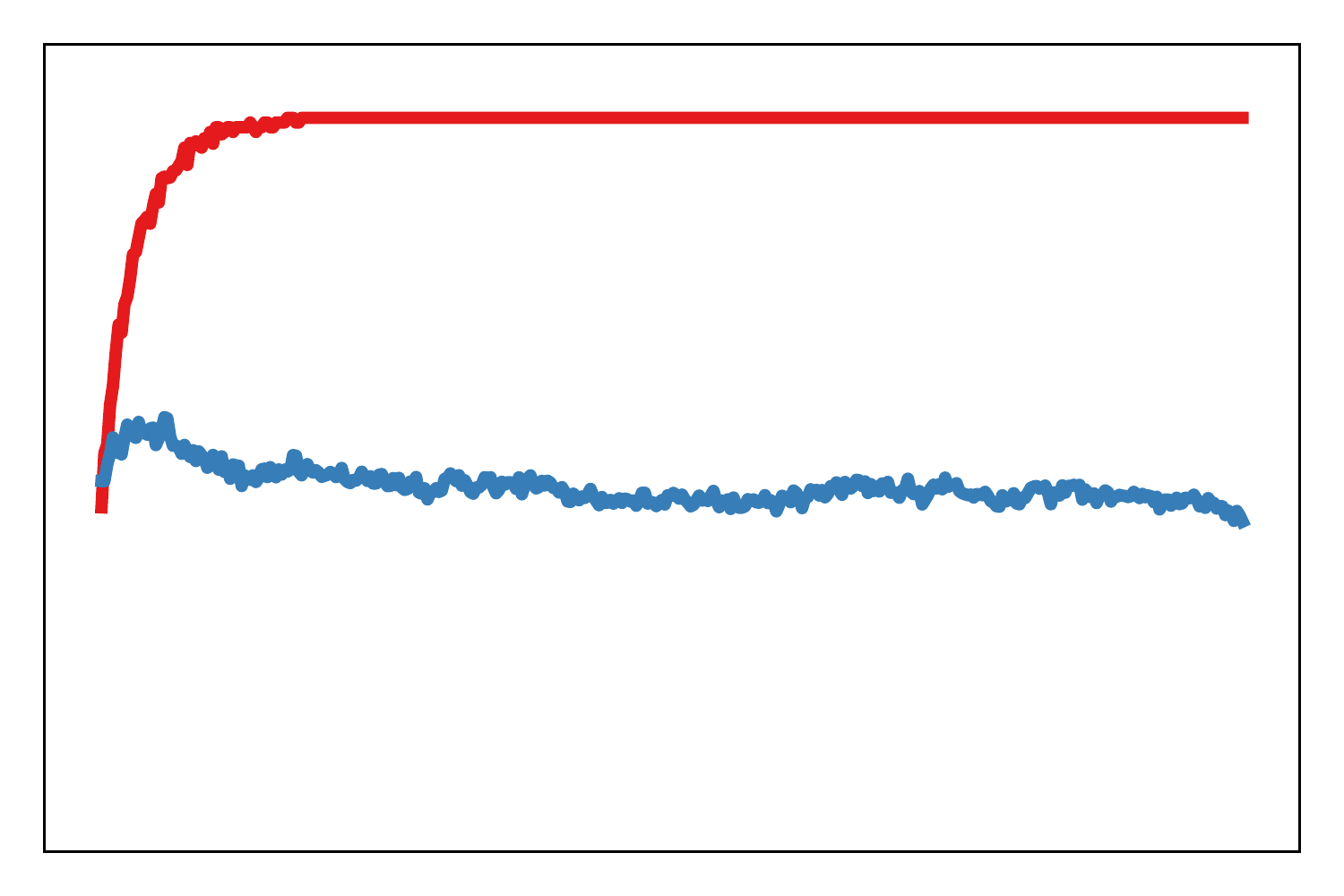}
   &
   \includegraphics[width=.090\textwidth]{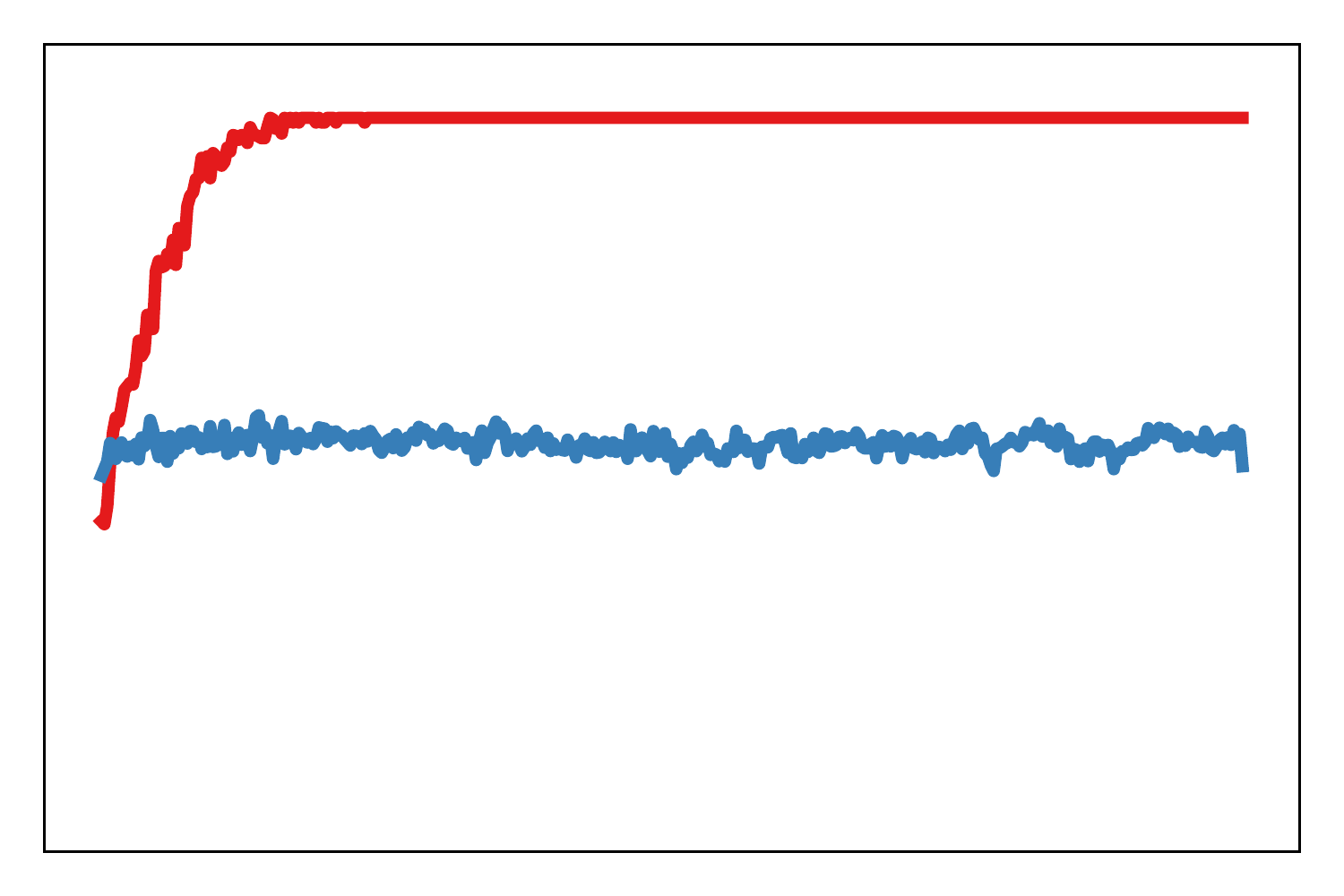}
   &
   \includegraphics[width=.090\textwidth]{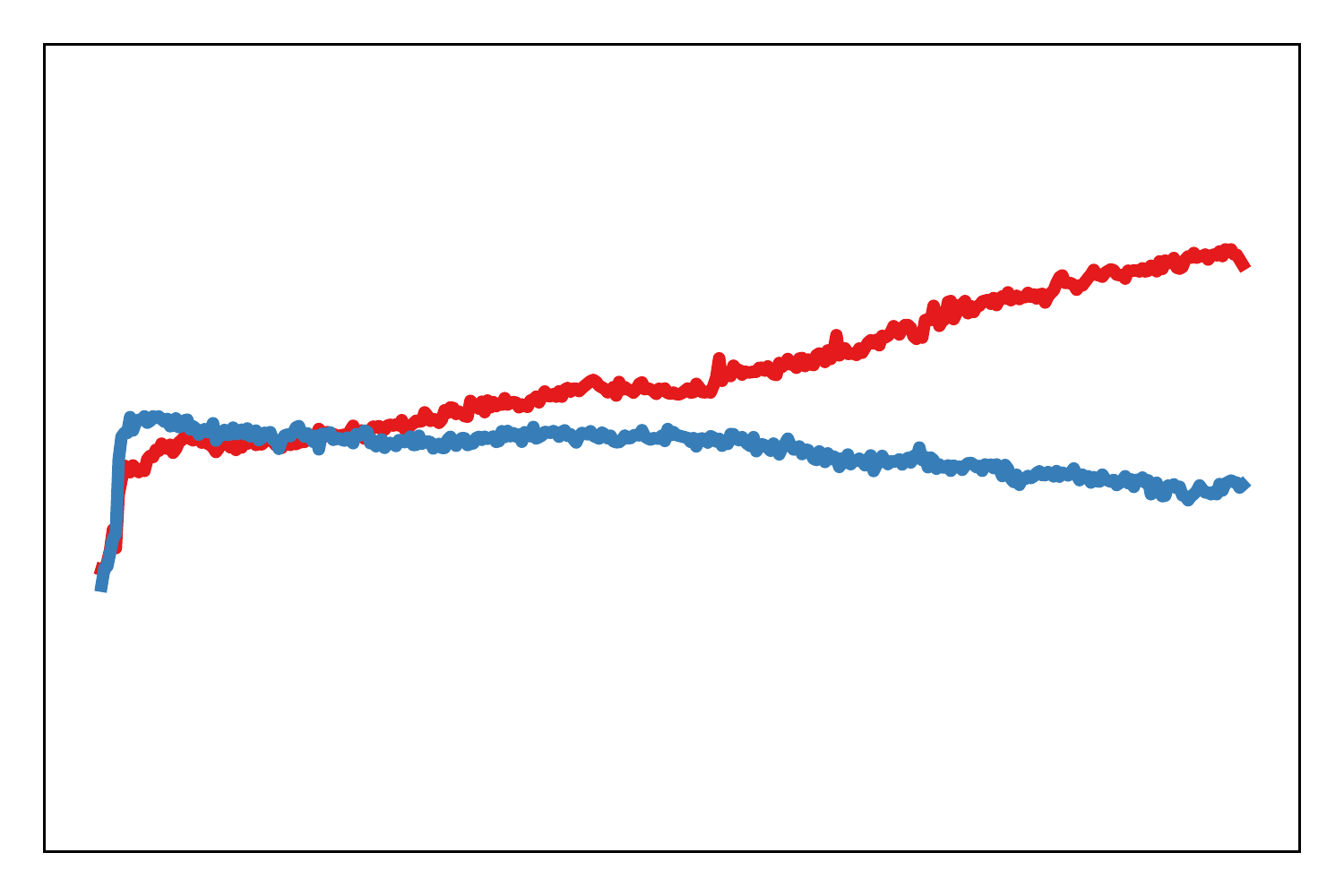}
   &
   \includegraphics[width=.090\textwidth]{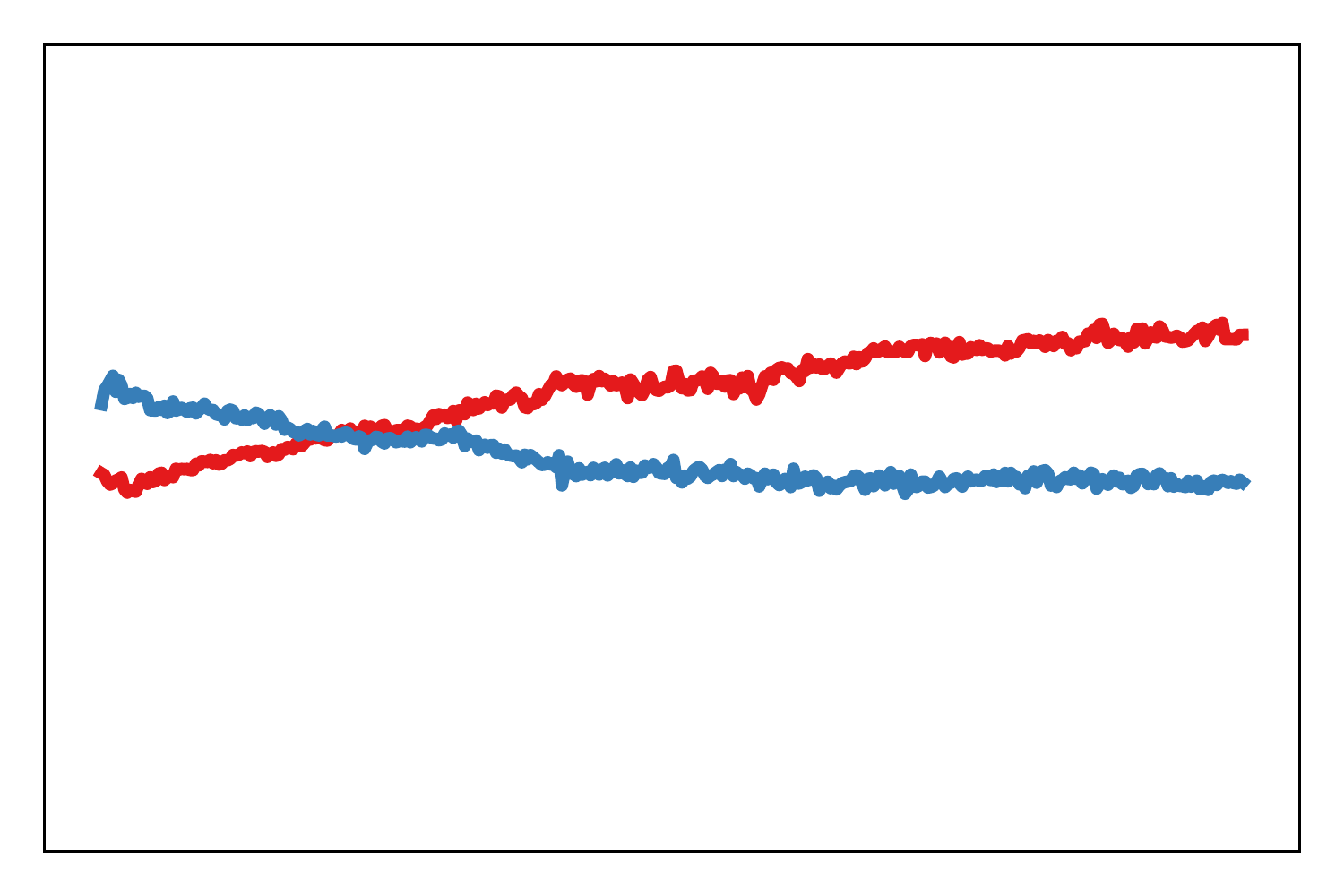}
   &
   \includegraphics[width=.090\textwidth]{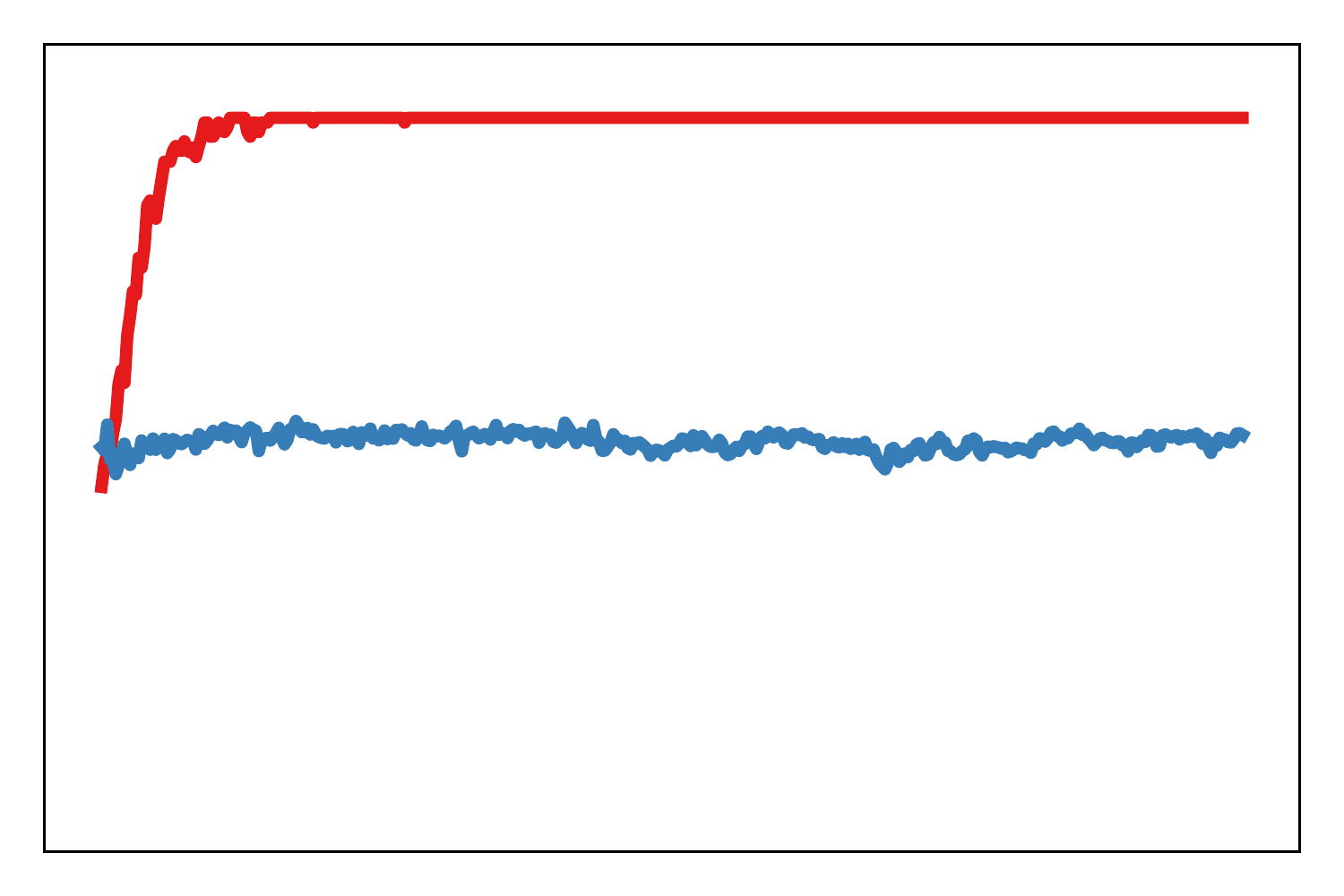}
   &
   \includegraphics[width=.090\textwidth]{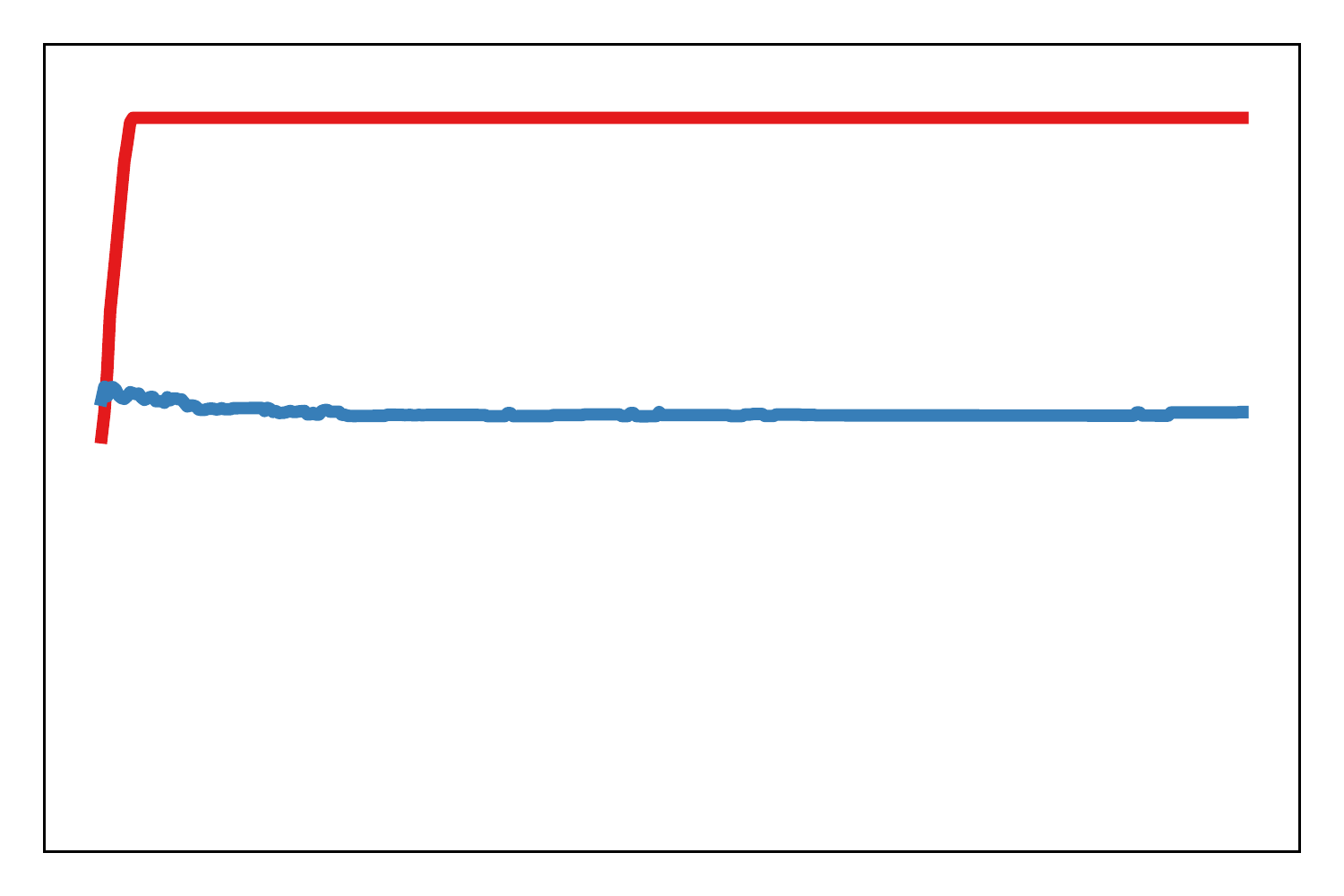}
   &
   \includegraphics[width=.090\textwidth]{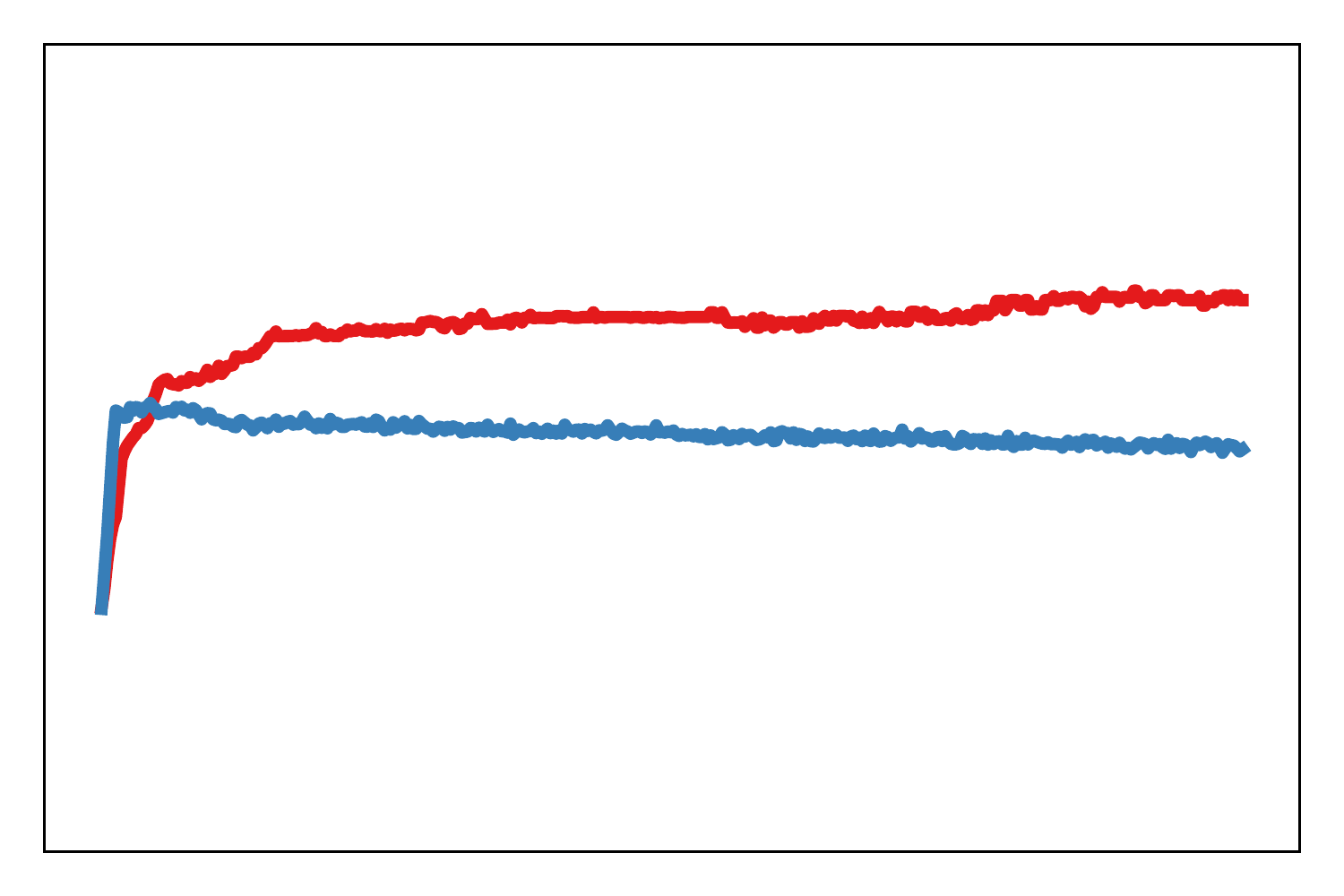}
   \\
   \raisebox{.3cm}{\tiny $25\%$} & 
   \includegraphics[width=.090\textwidth]{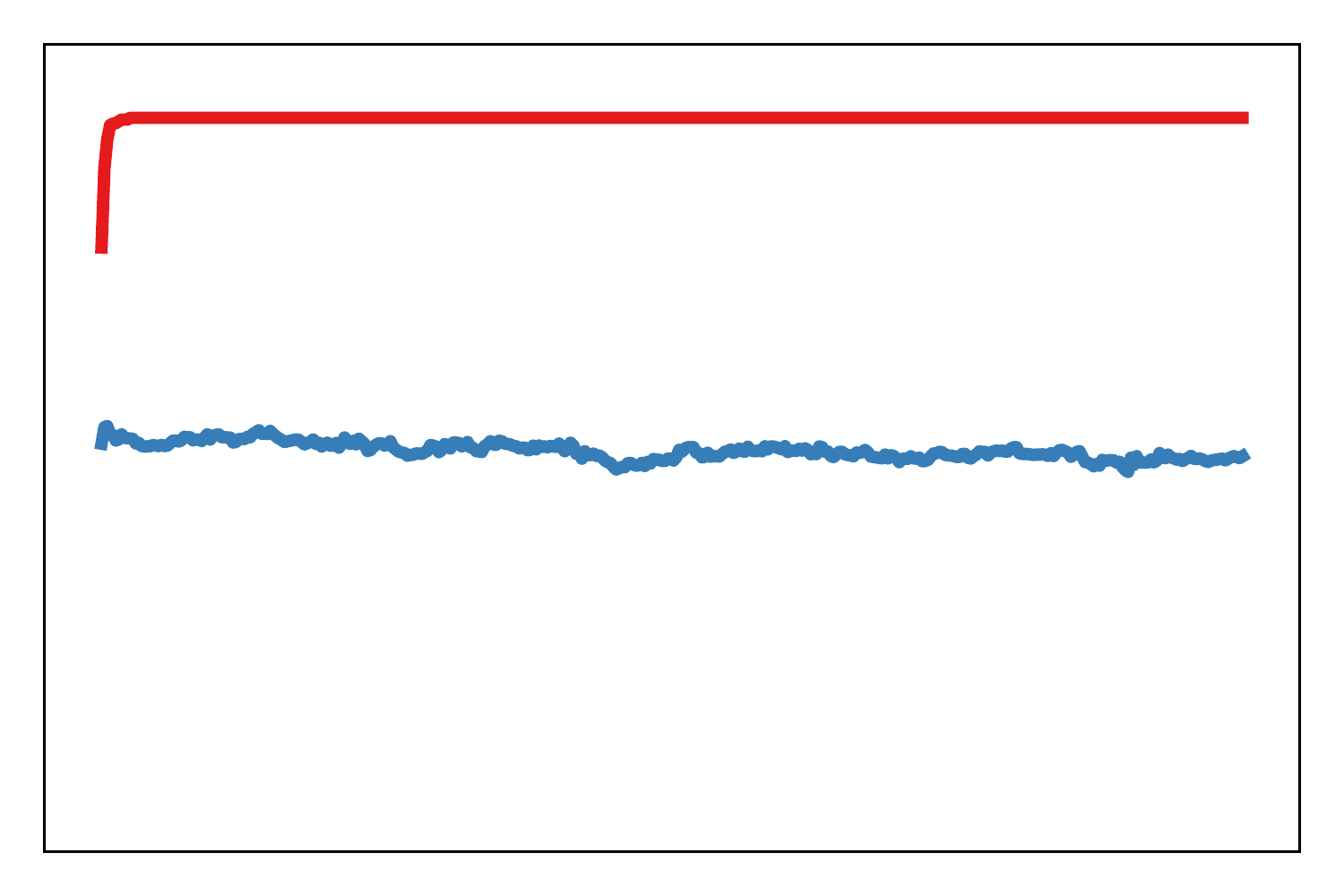}
   &
   \includegraphics[width=.090\textwidth]{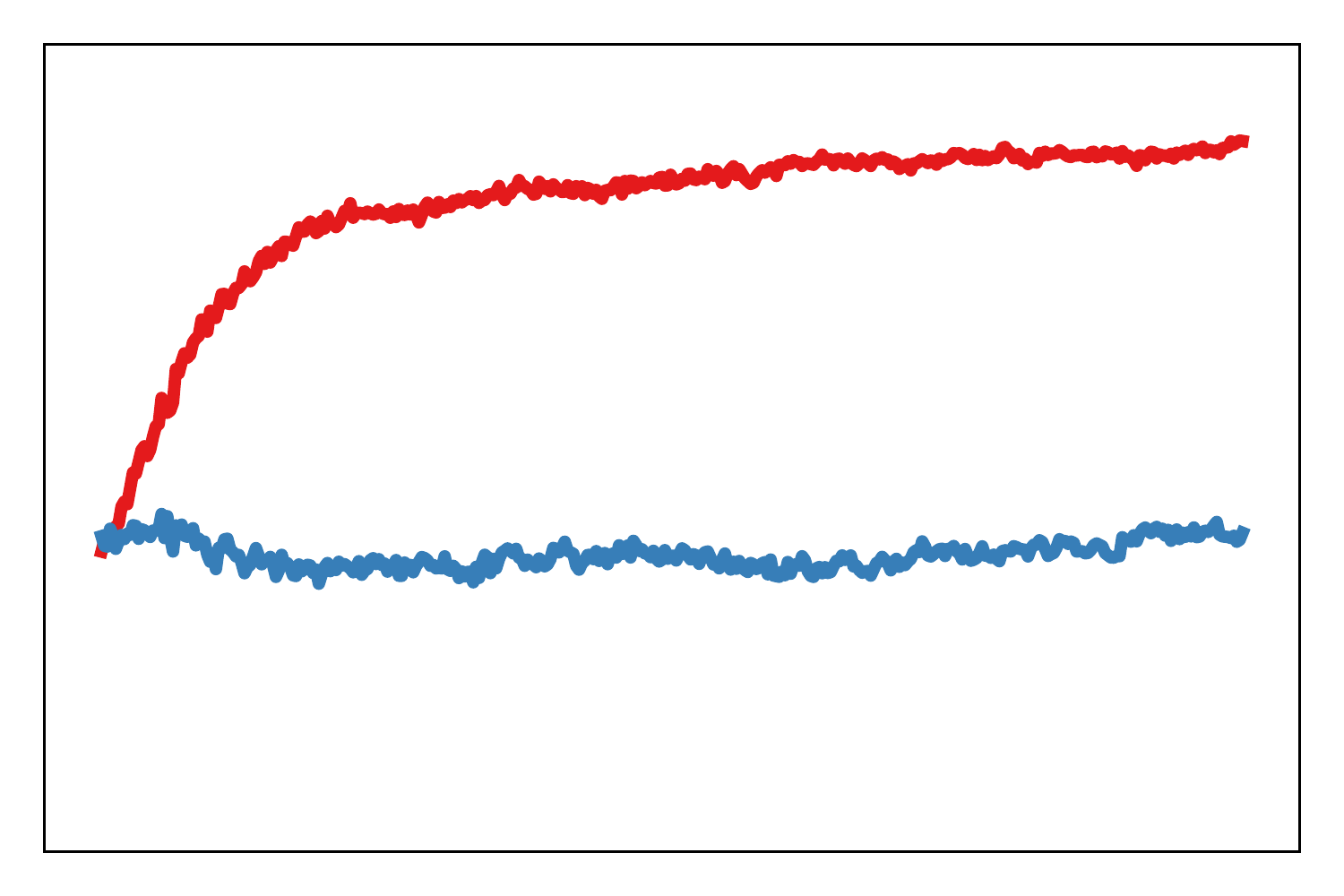}
   &
   \includegraphics[width=.090\textwidth]{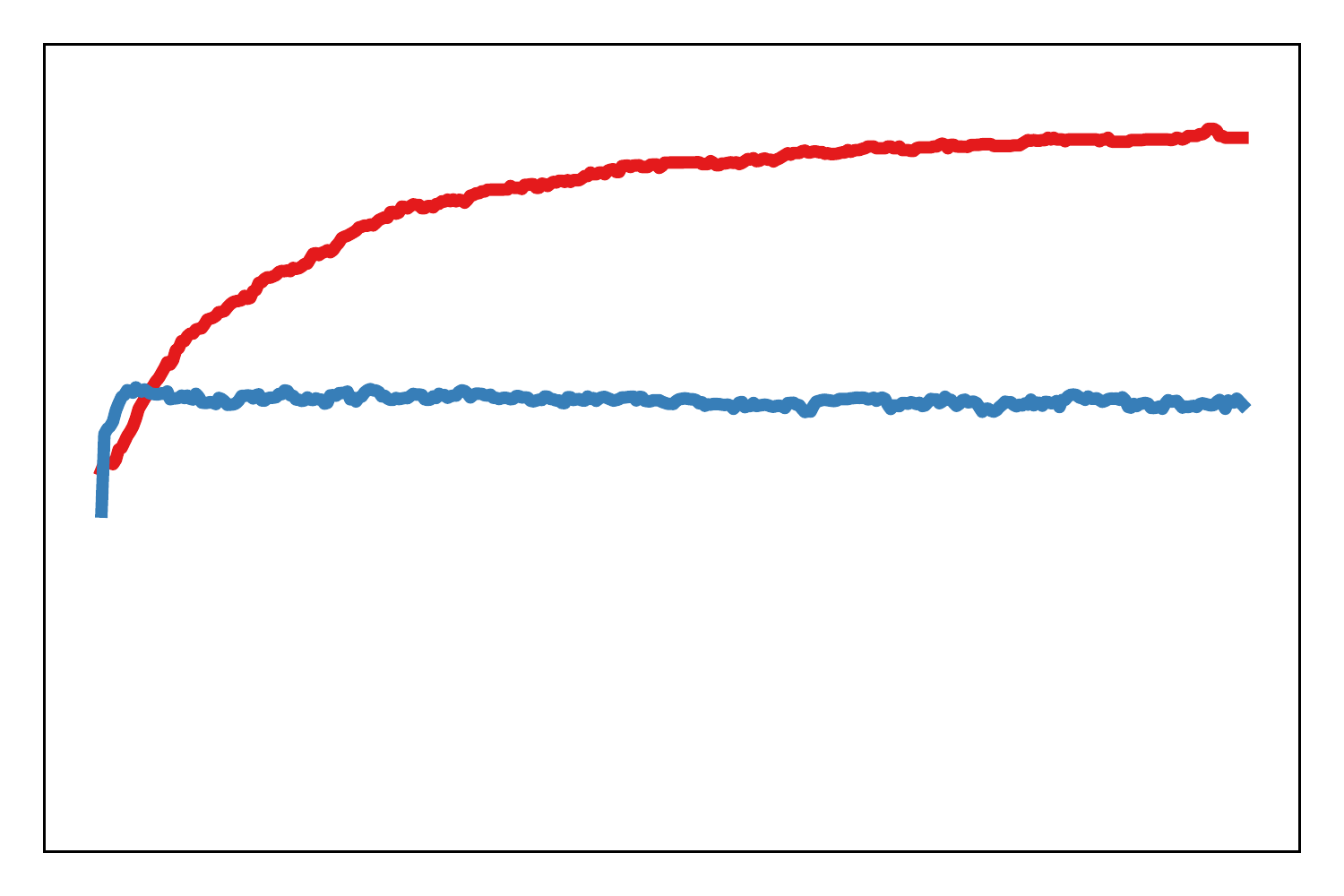}
   &
   \includegraphics[width=.090\textwidth]{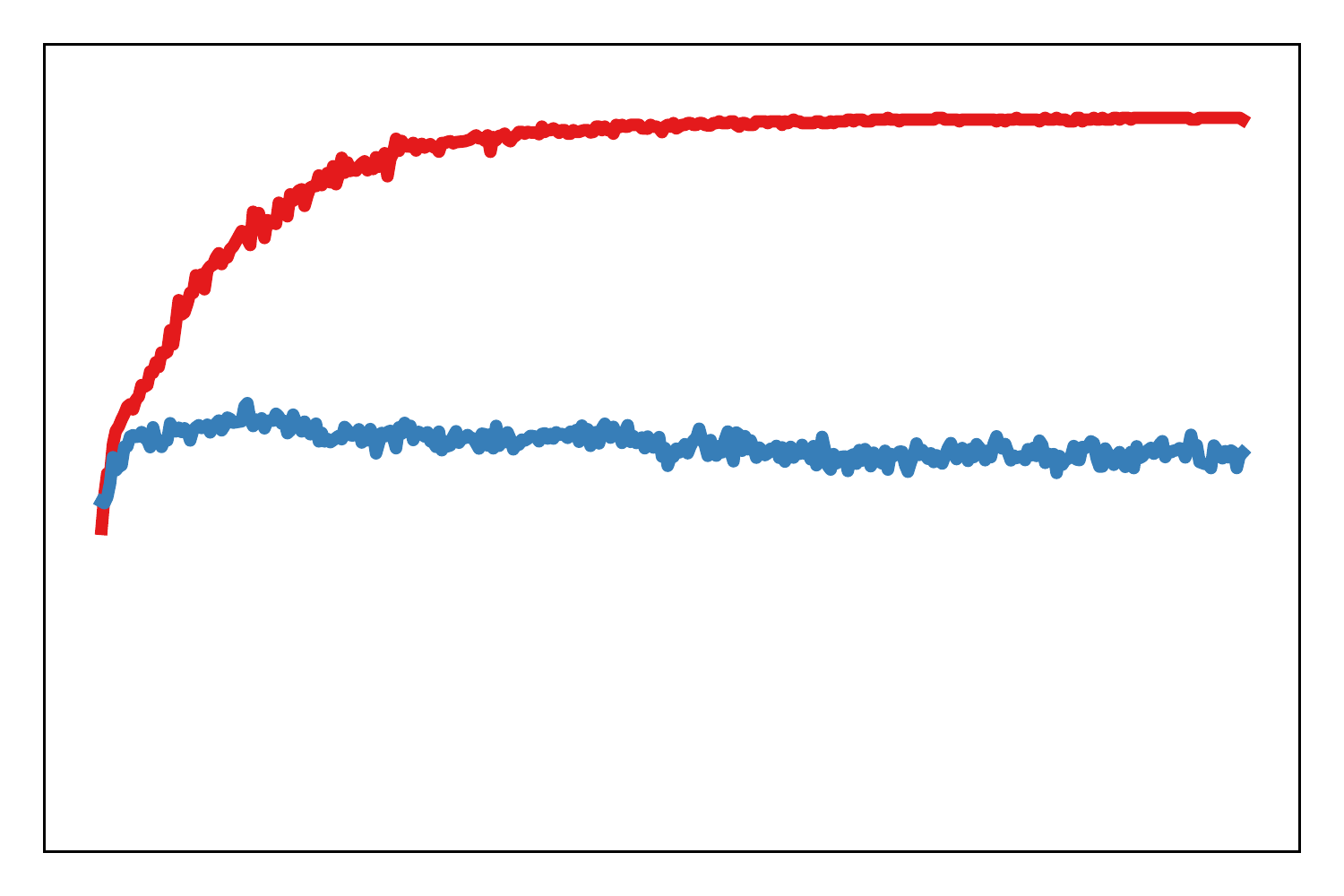}
   &
   \includegraphics[width=.090\textwidth]{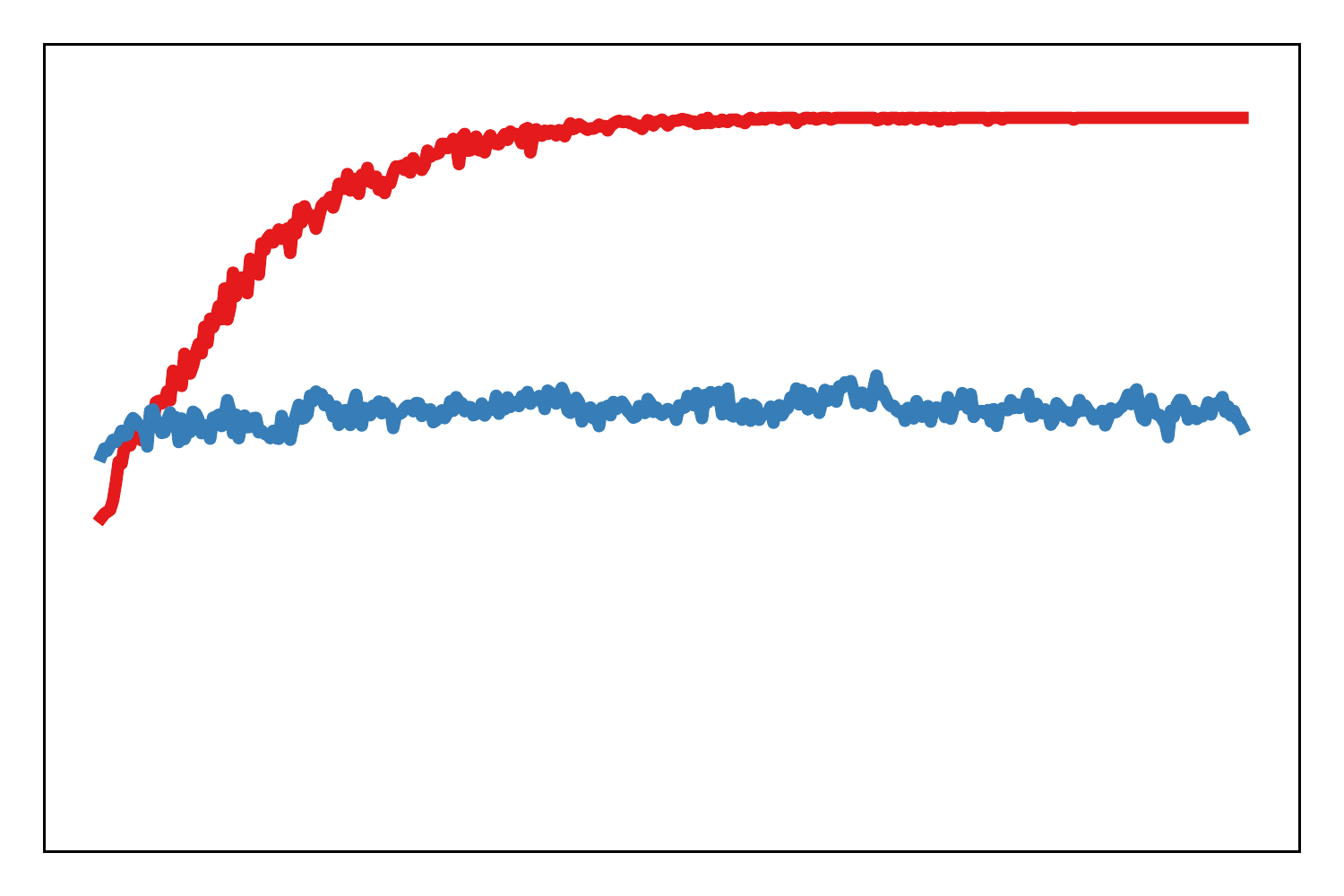}
   &
   \includegraphics[width=.090\textwidth]{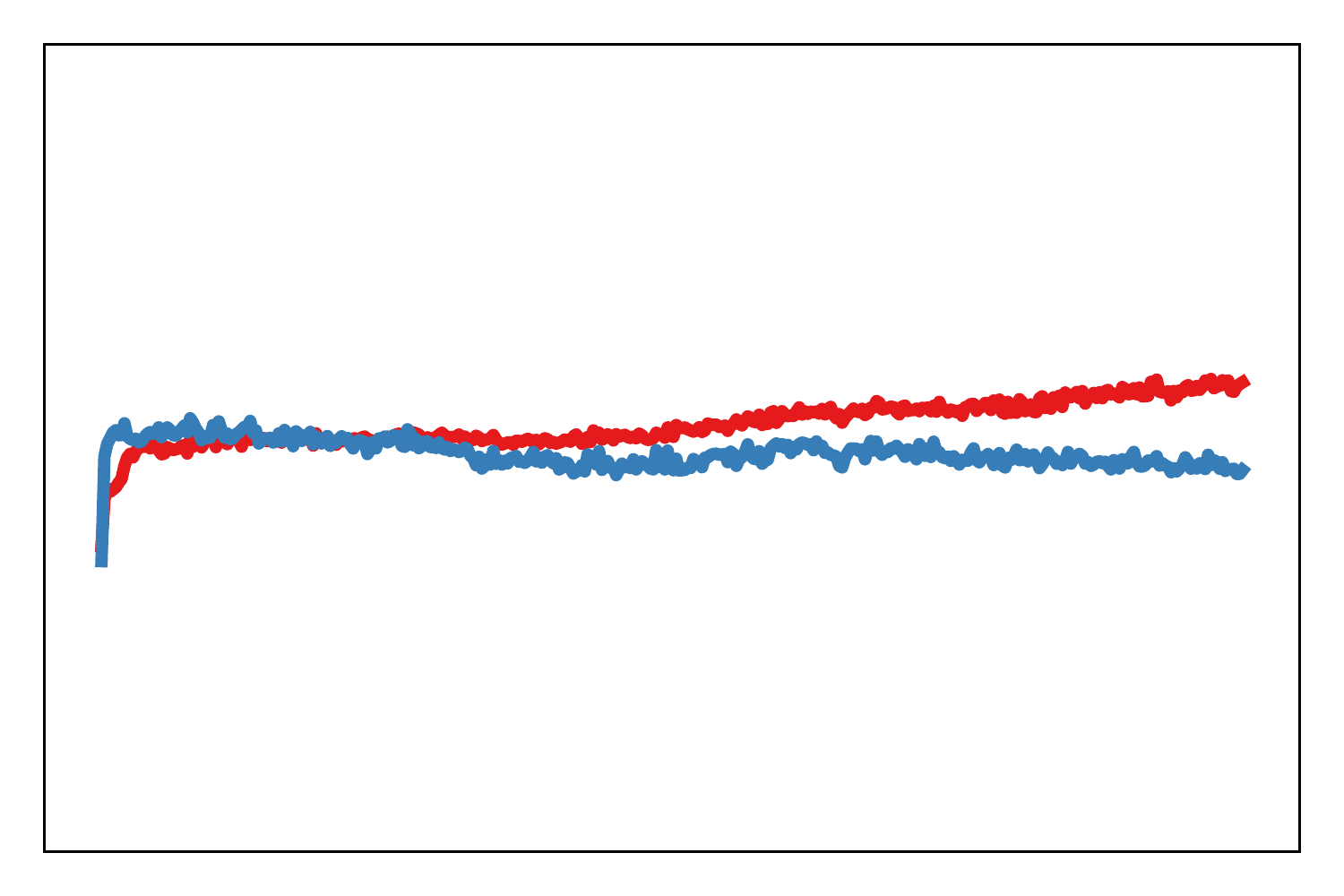}
   &
   \includegraphics[width=.090\textwidth]{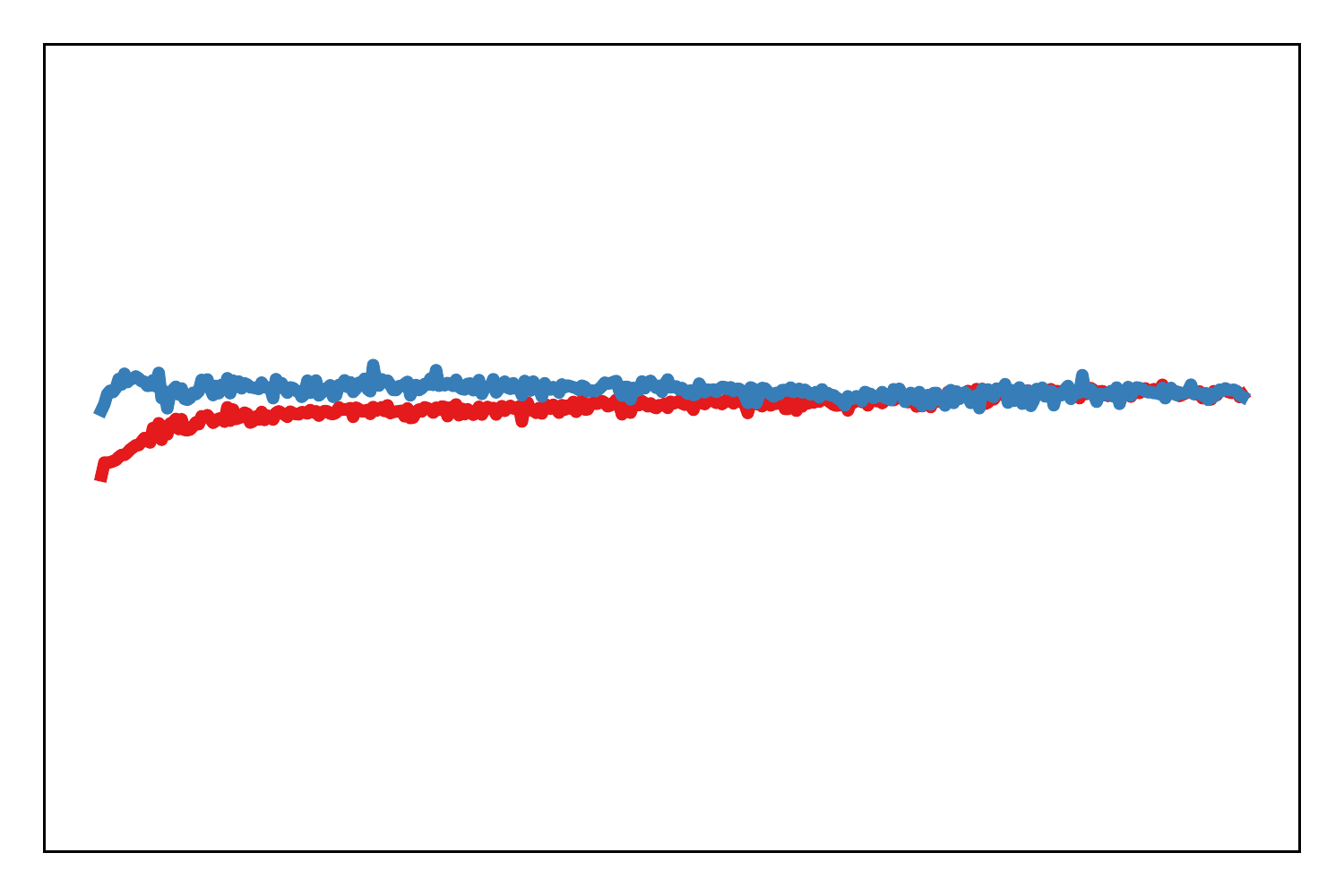}
   &
   \includegraphics[width=.090\textwidth]{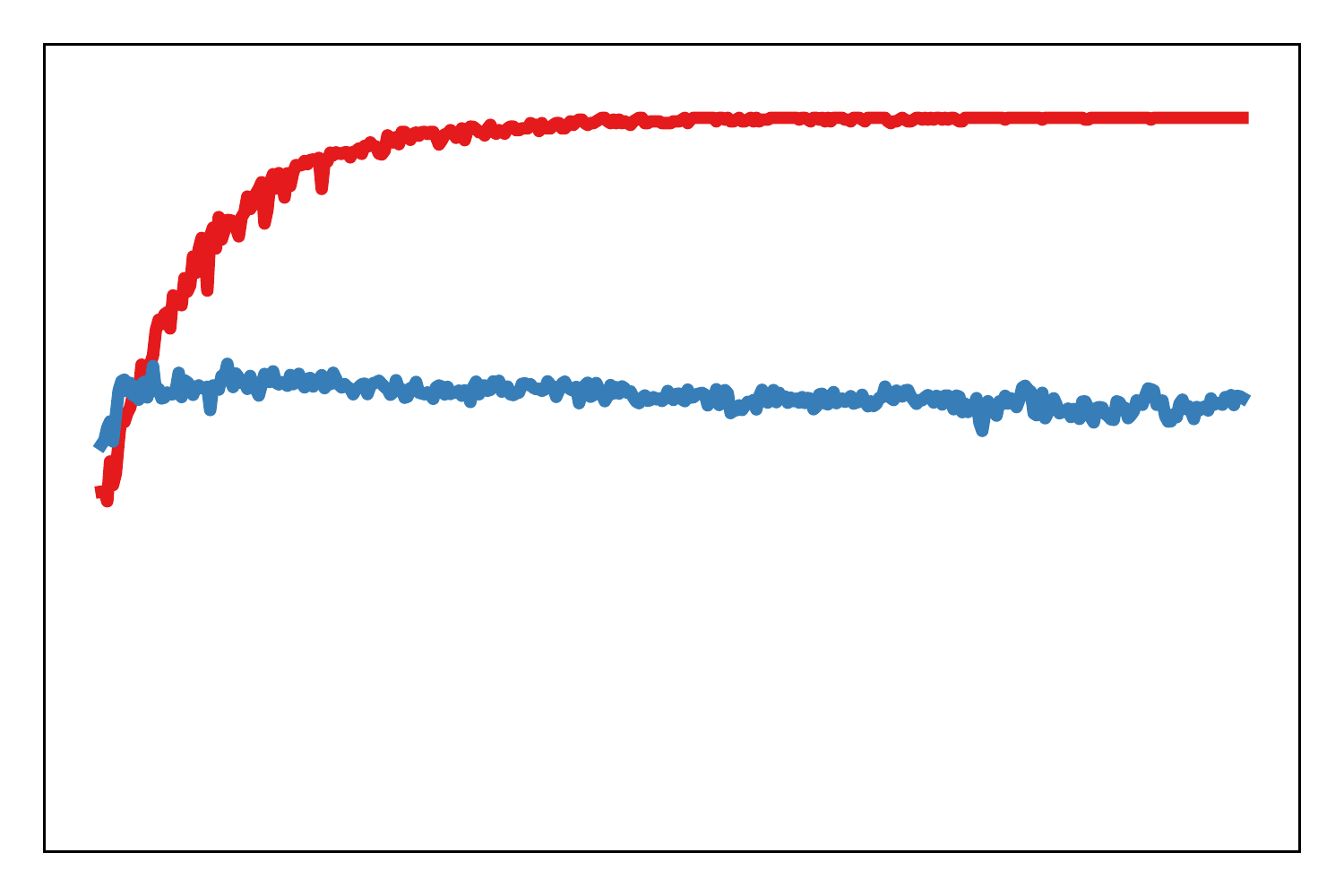}
   &
   \includegraphics[width=.090\textwidth]{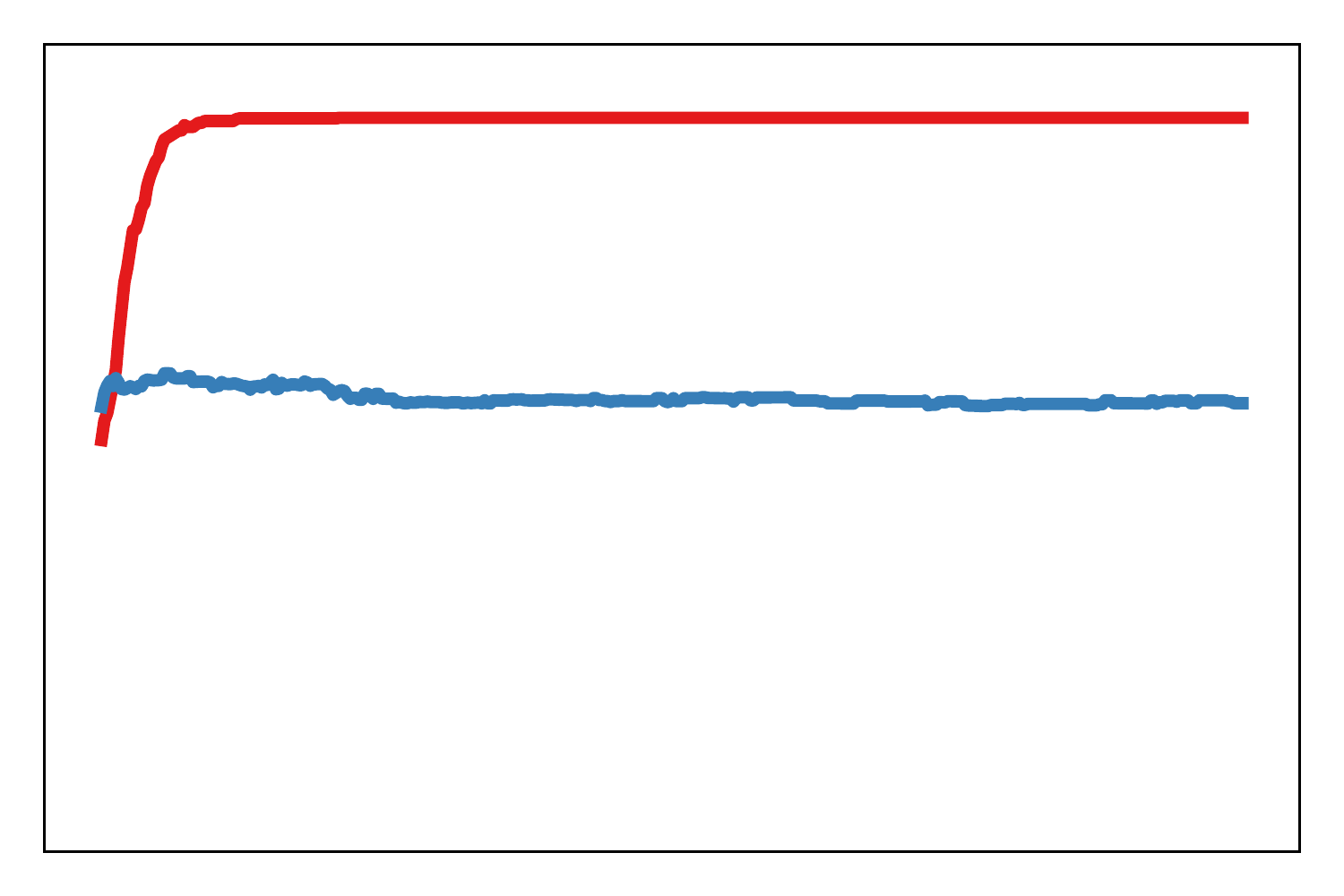}
   &
   \includegraphics[width=.090\textwidth]{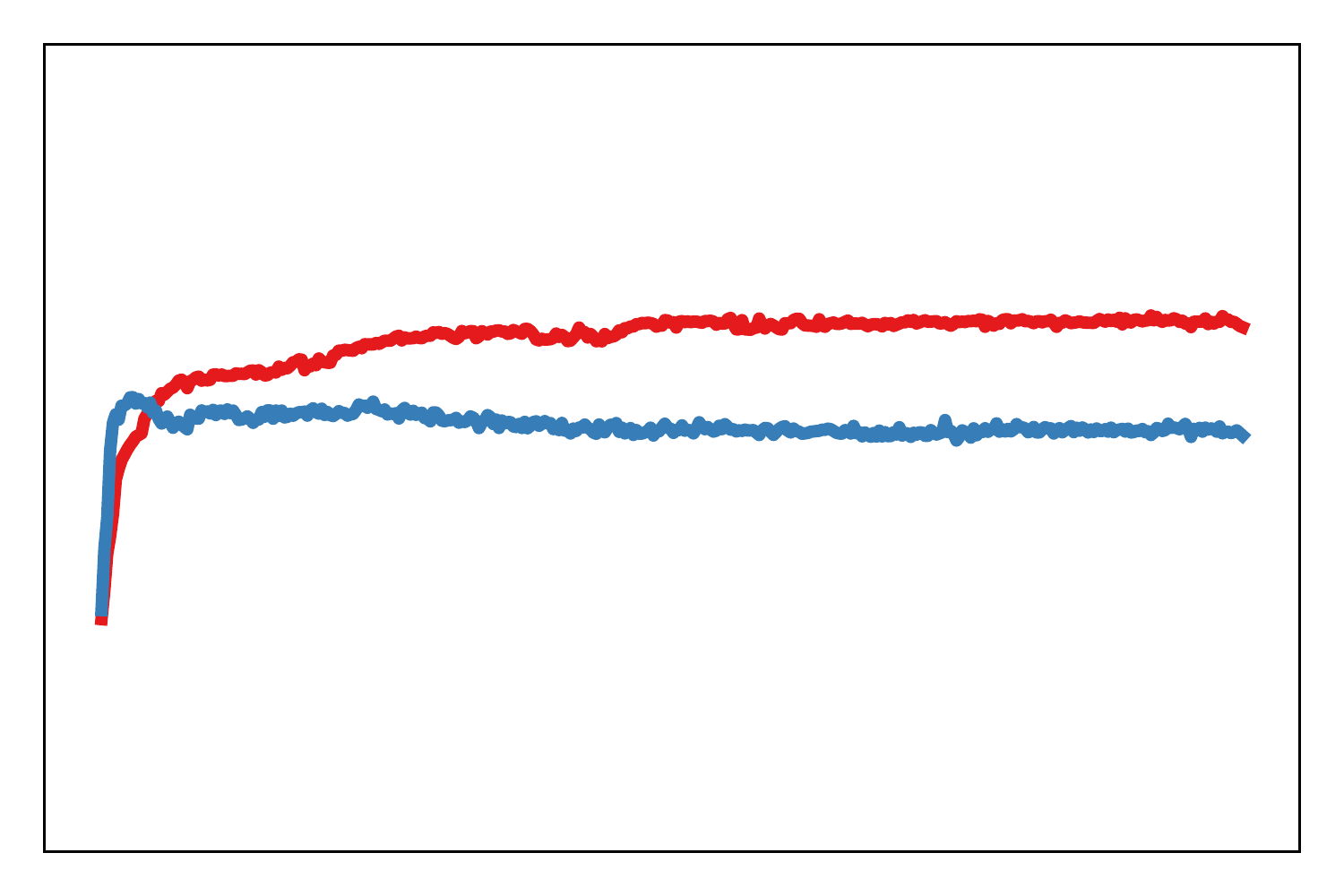}
   \\
   \raisebox{.4cm}{\tiny $50\%$} & 
   \includegraphics[width=.090\textwidth]{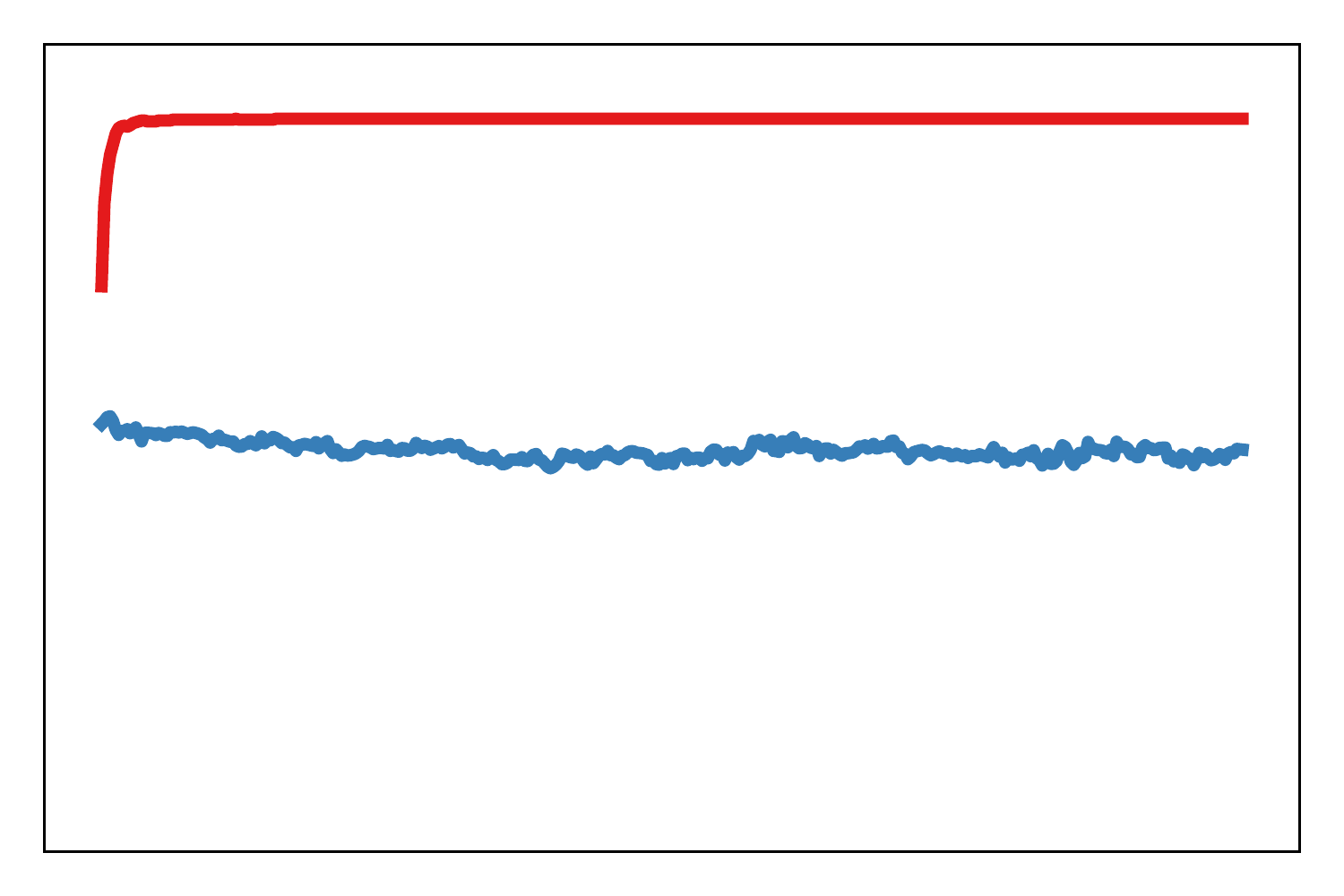}
   &
   \includegraphics[width=.090\textwidth]{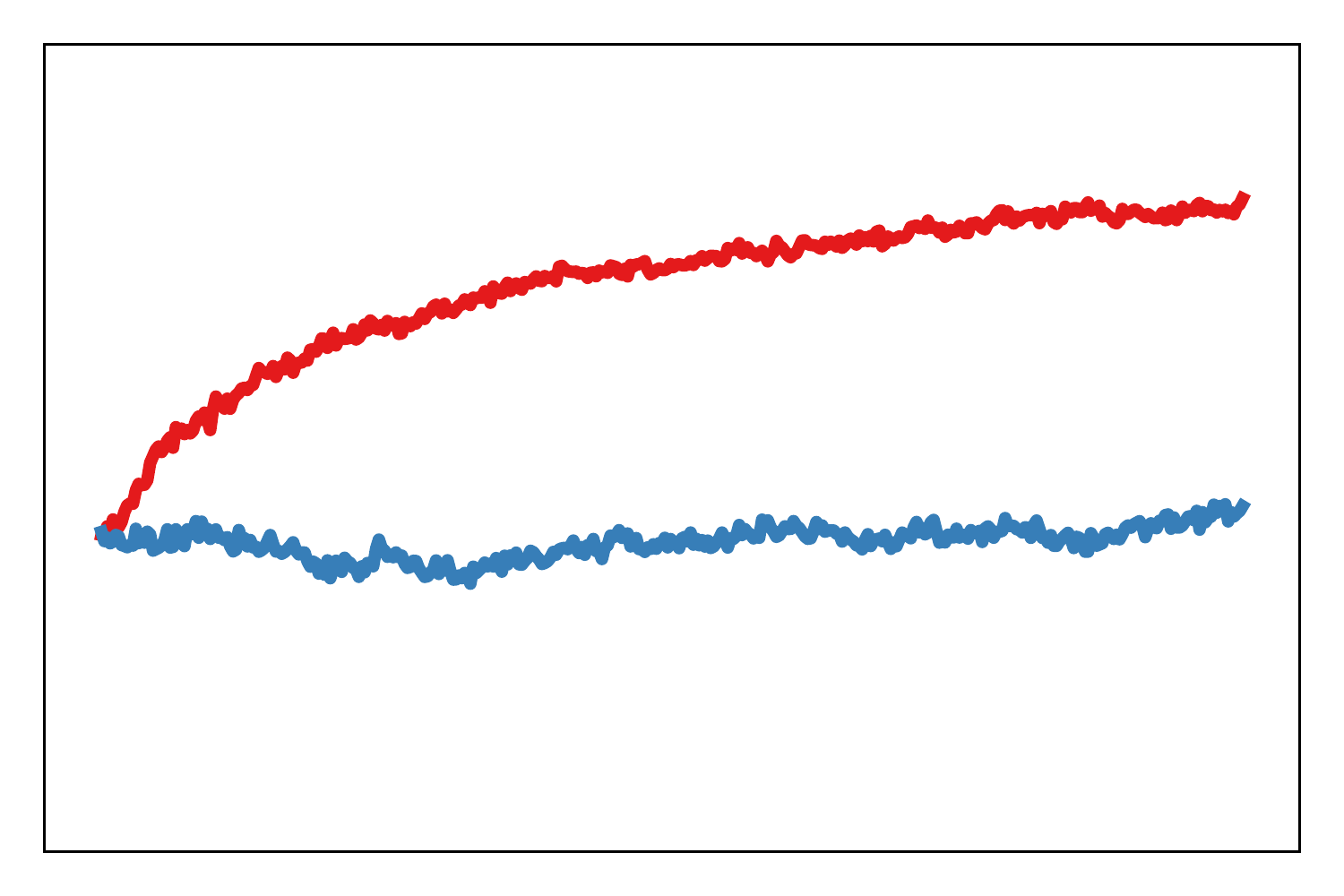}
   &
   \includegraphics[width=.090\textwidth]{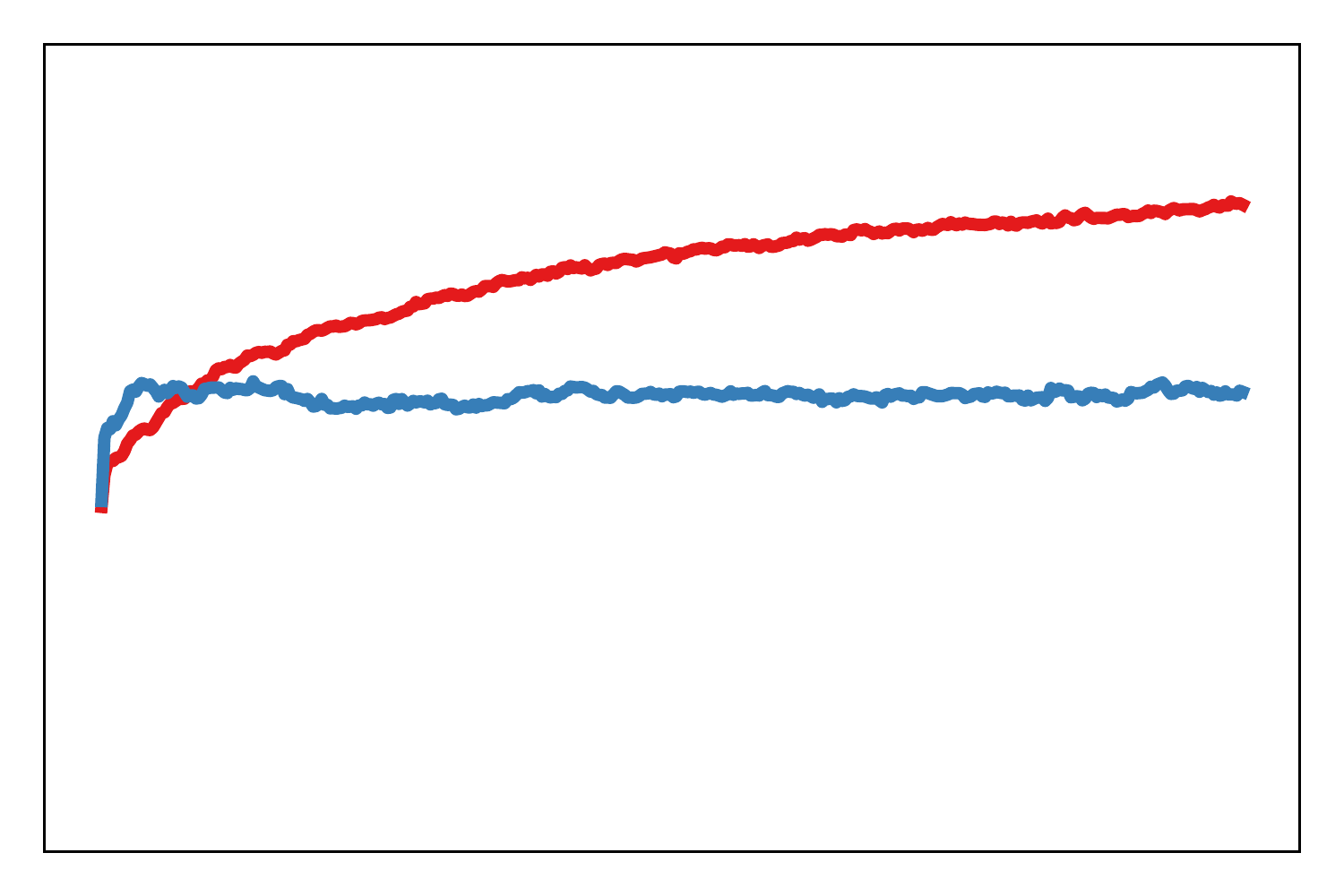}
   &
   \includegraphics[width=.090\textwidth]{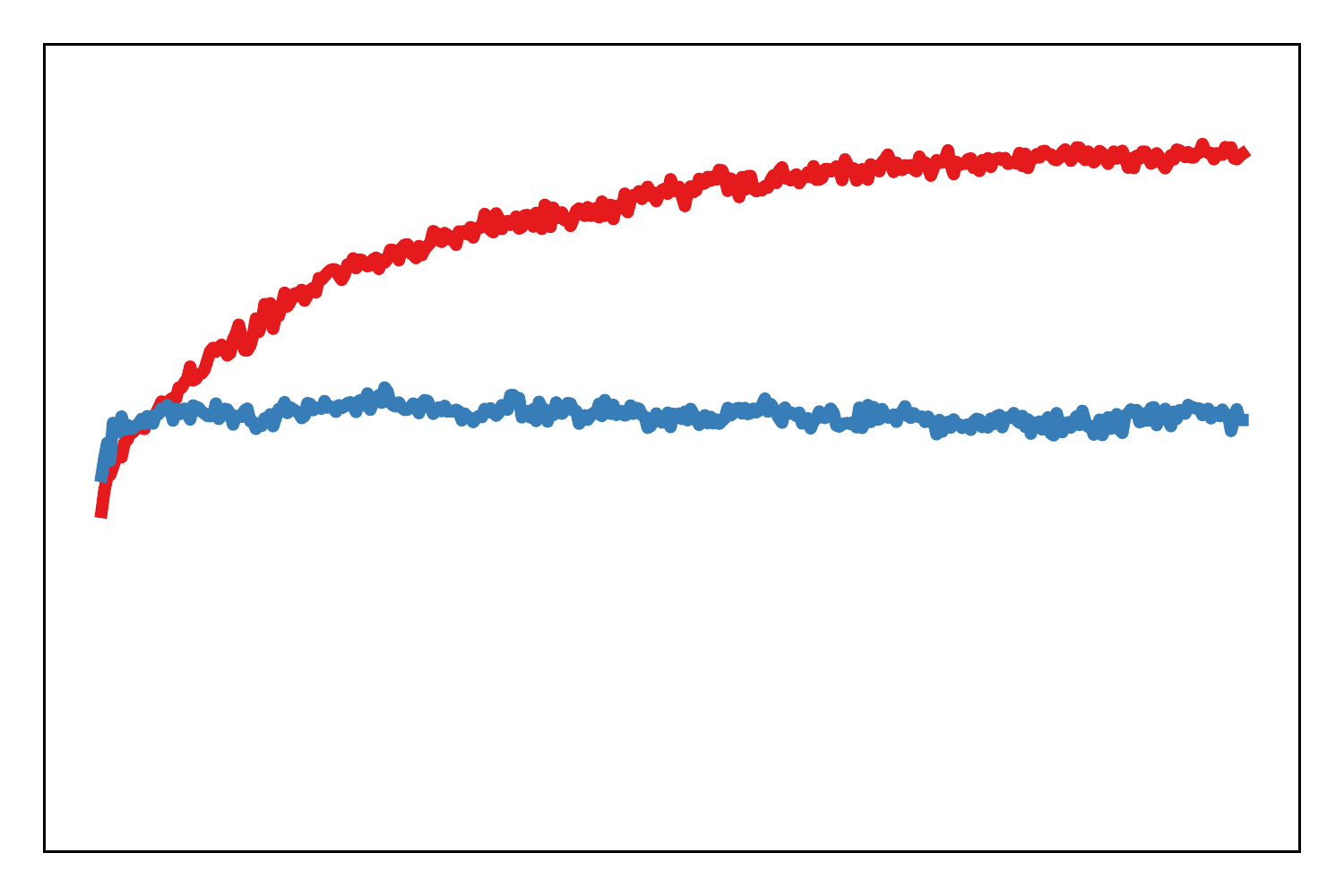}
   &
   \includegraphics[width=.090\textwidth]{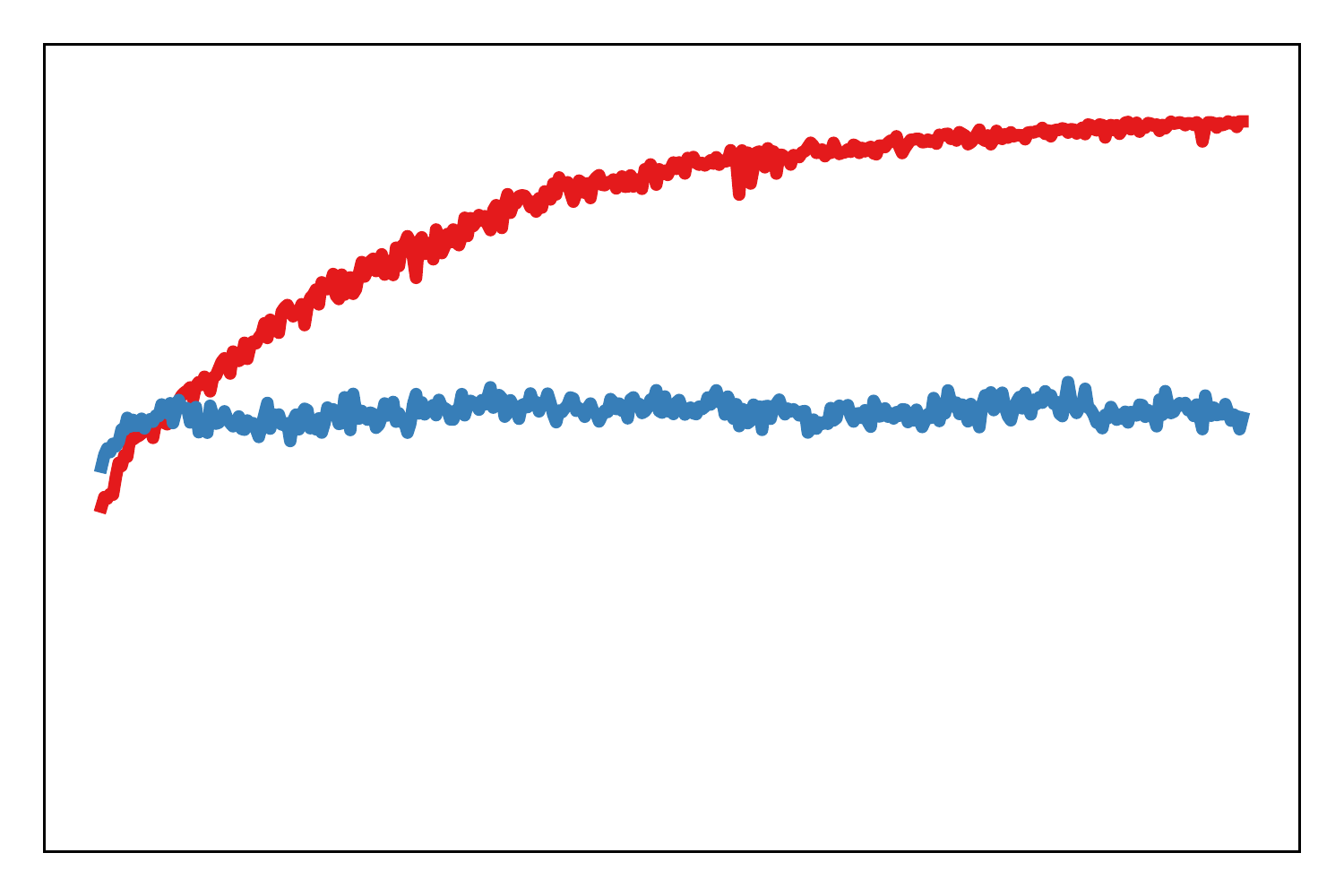}
   &
   \includegraphics[width=.090\textwidth]{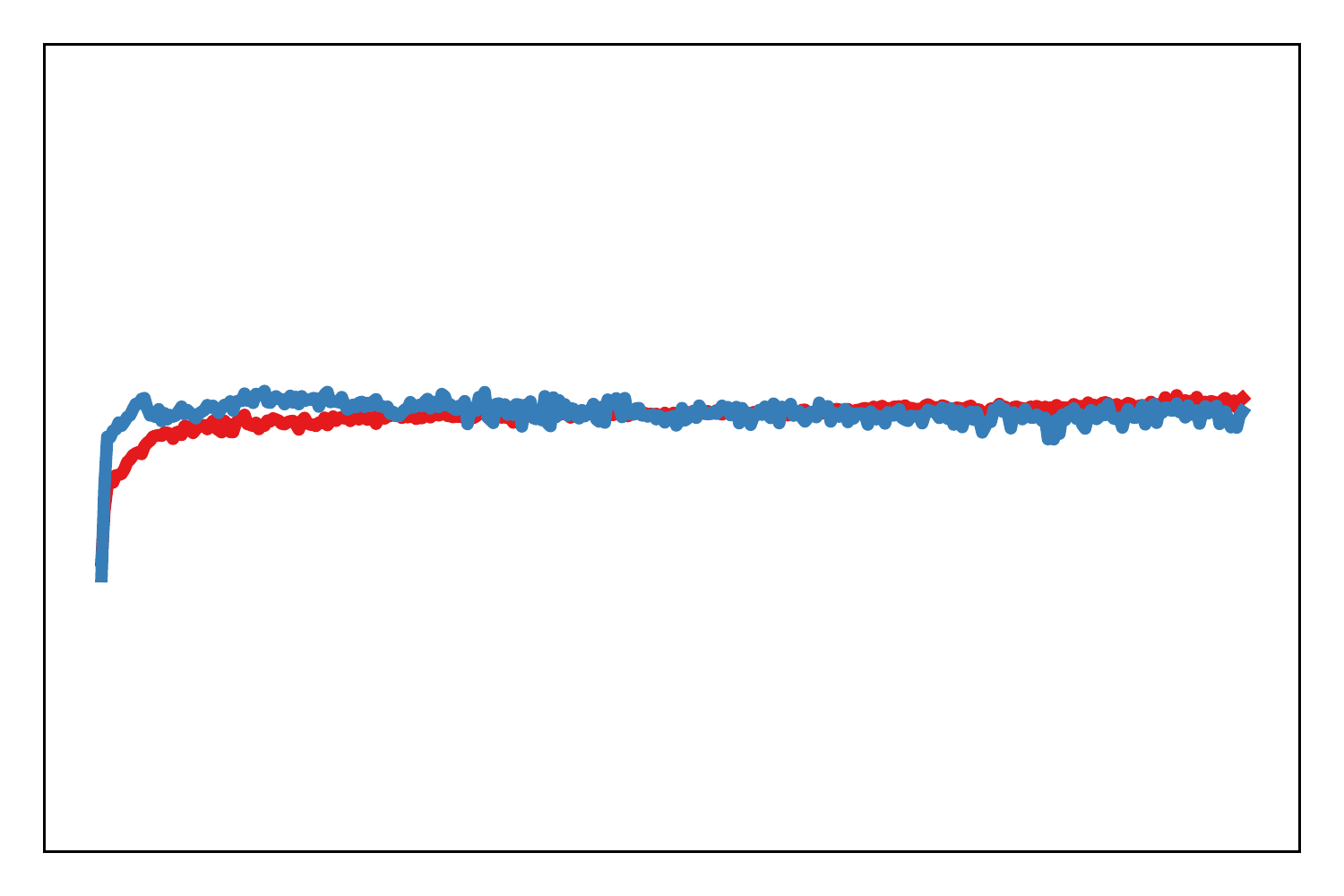}
   &
   \includegraphics[width=.090\textwidth]{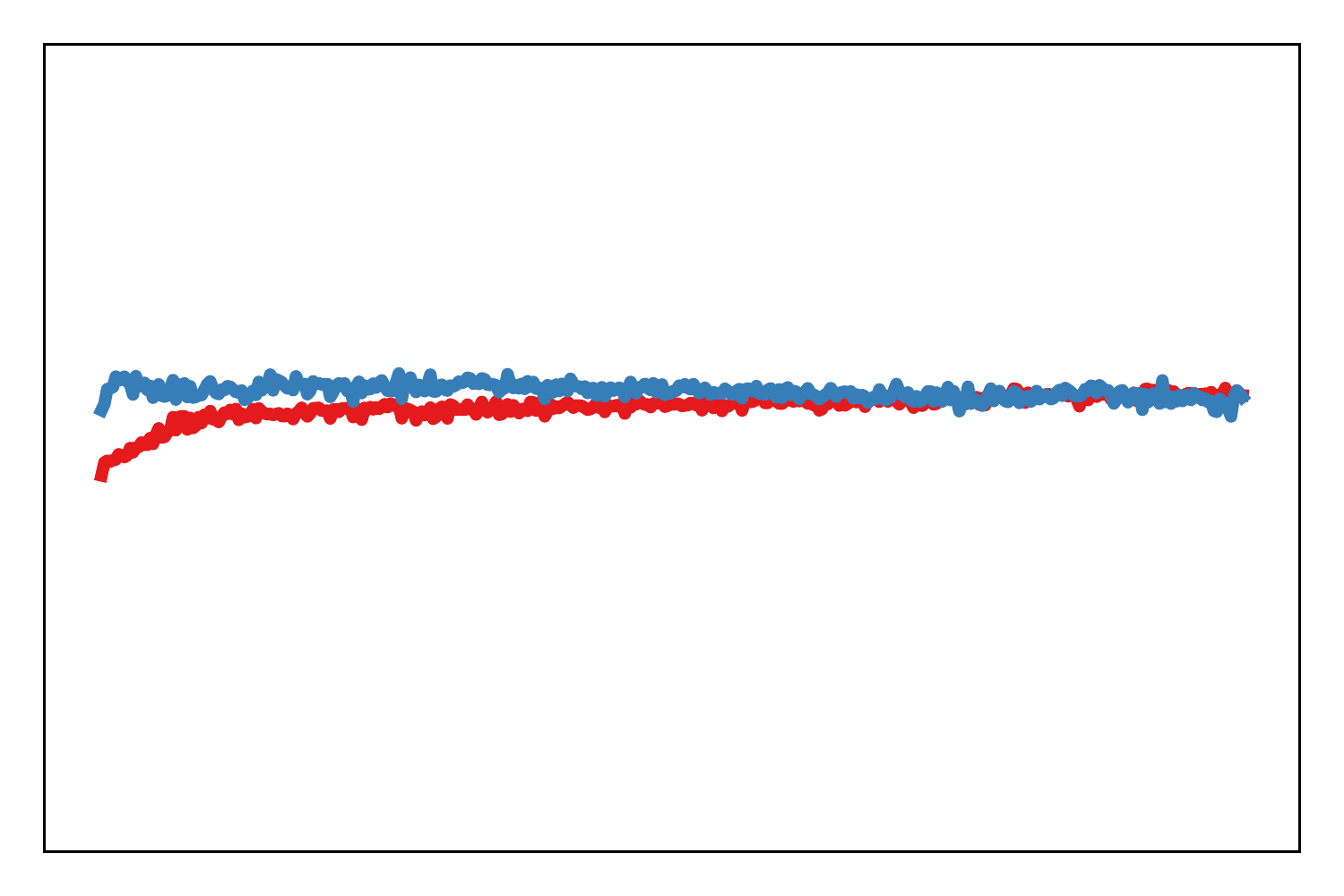}
   &
   \includegraphics[width=.090\textwidth]{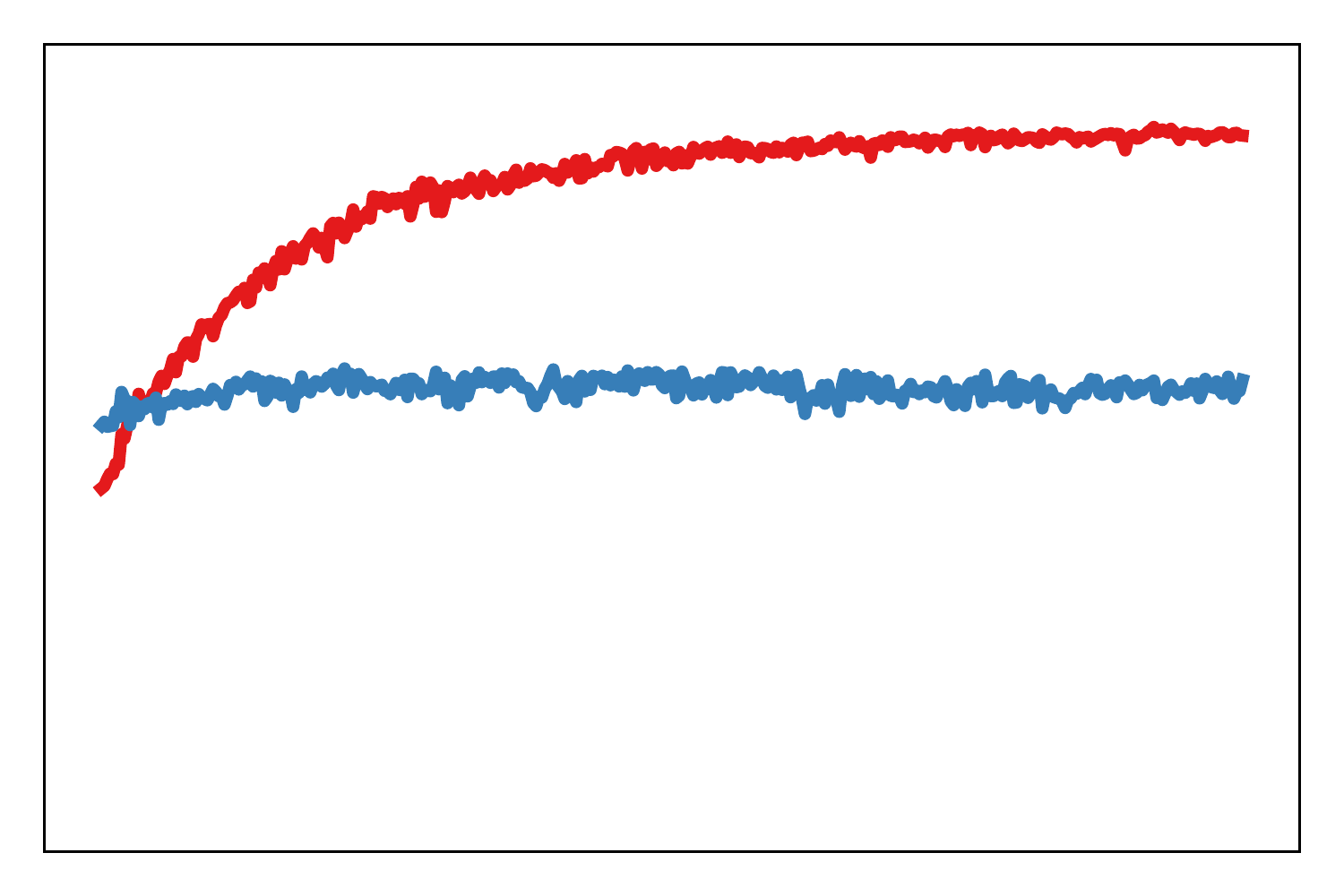}
   &
   \includegraphics[width=.090\textwidth]{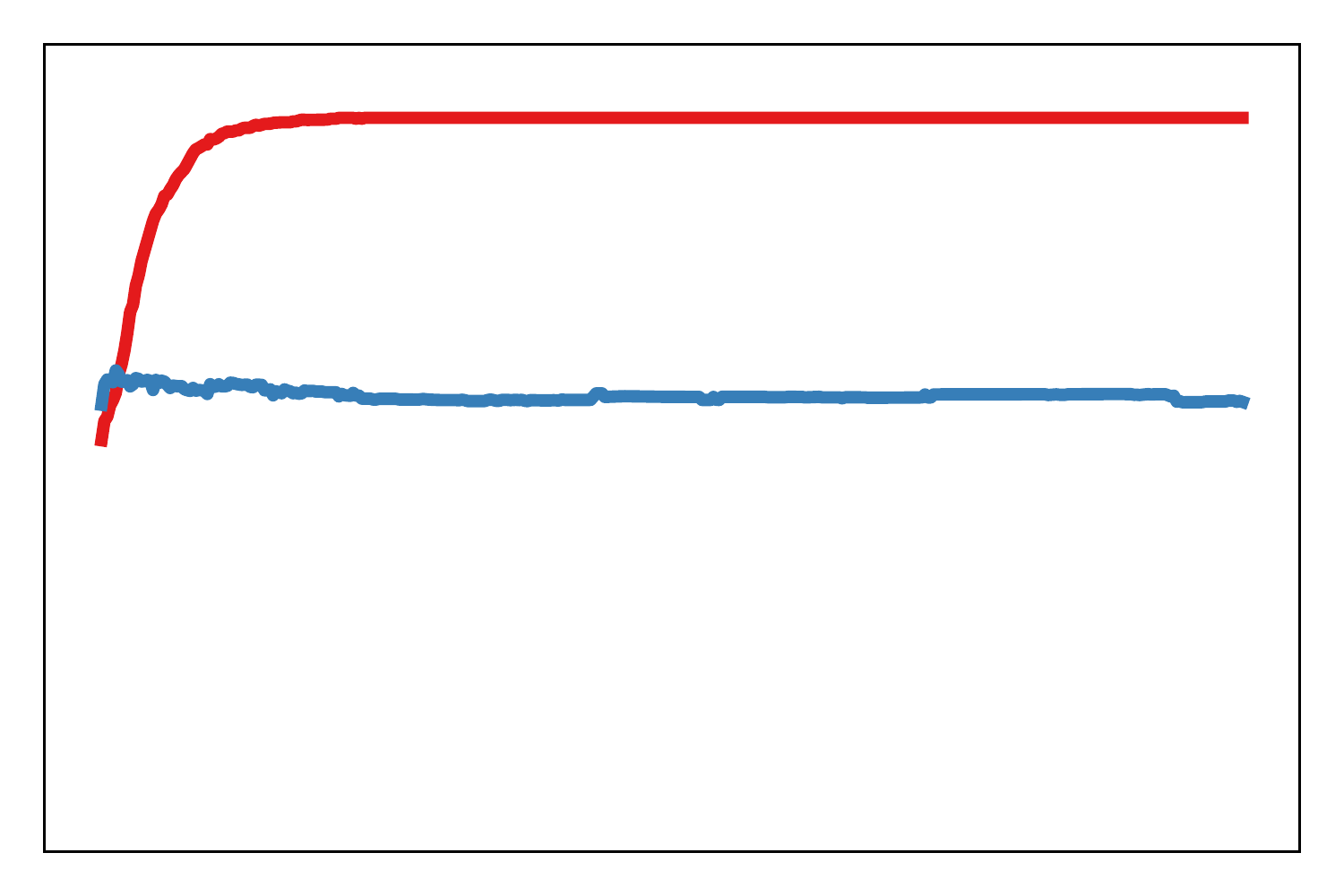}
   &
   \includegraphics[width=.090\textwidth]{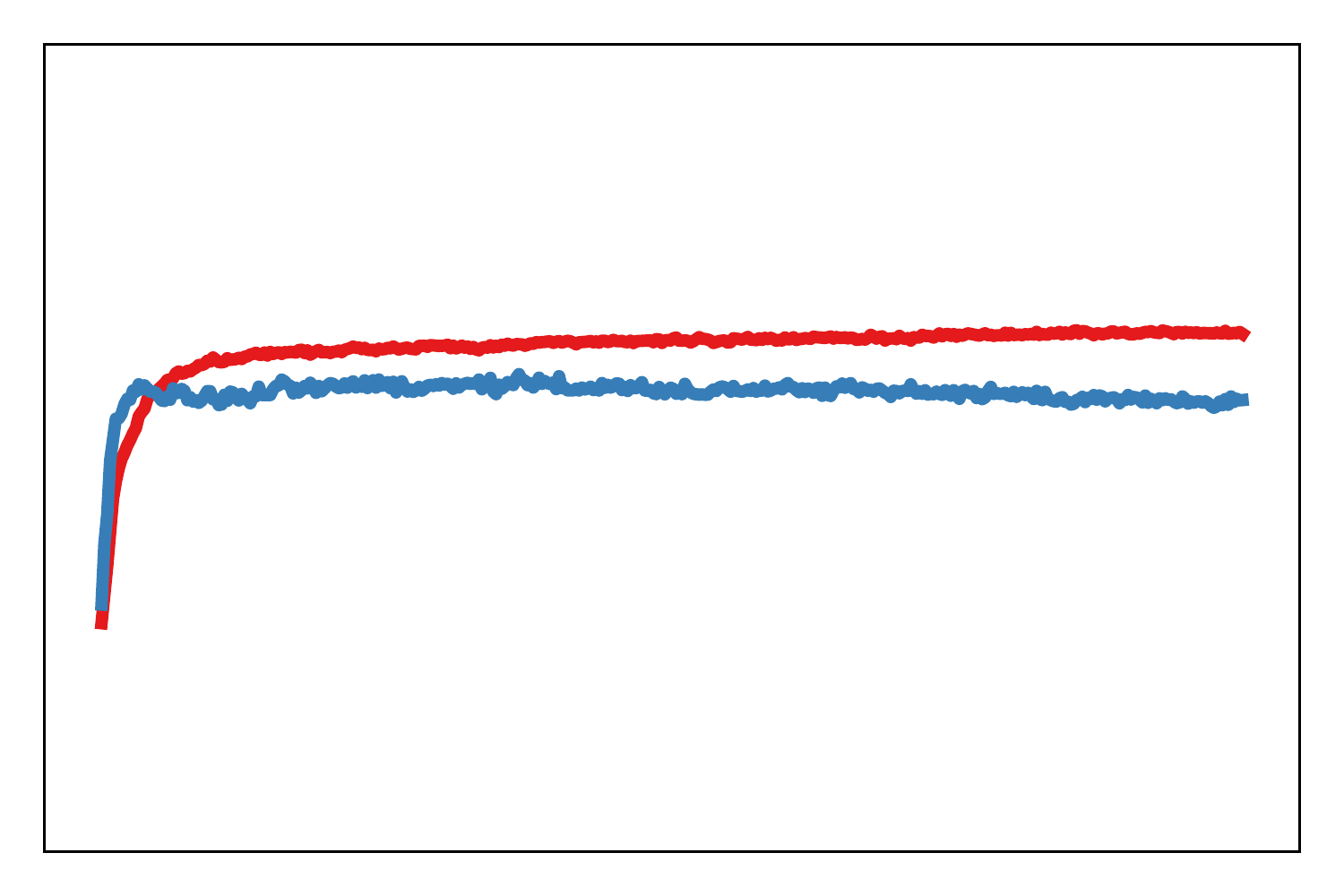}
   \\
   \raisebox{.4cm}{\tiny $75\%$} & 
   \includegraphics[width=.090\textwidth]{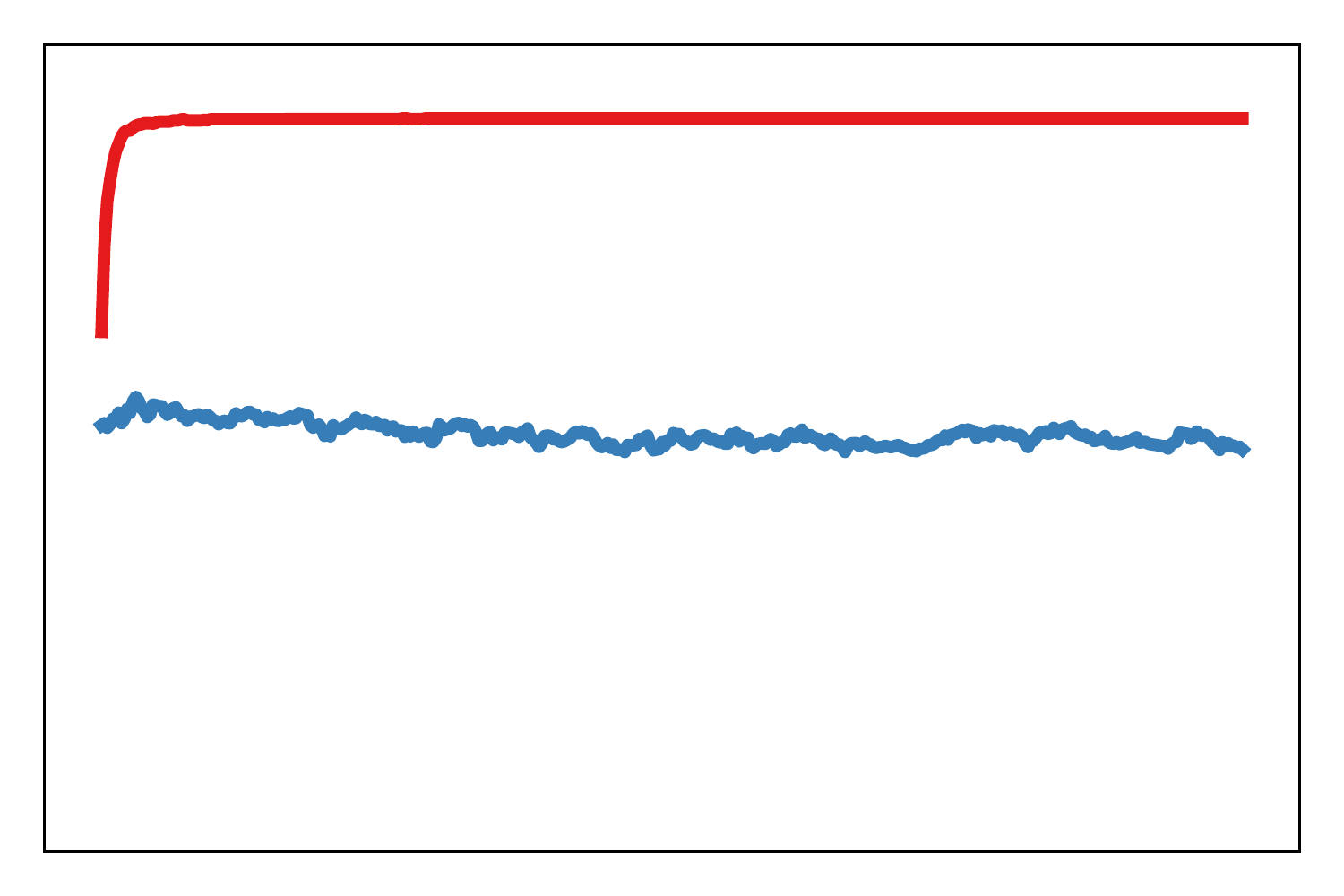}
   &
   \includegraphics[width=.090\textwidth]{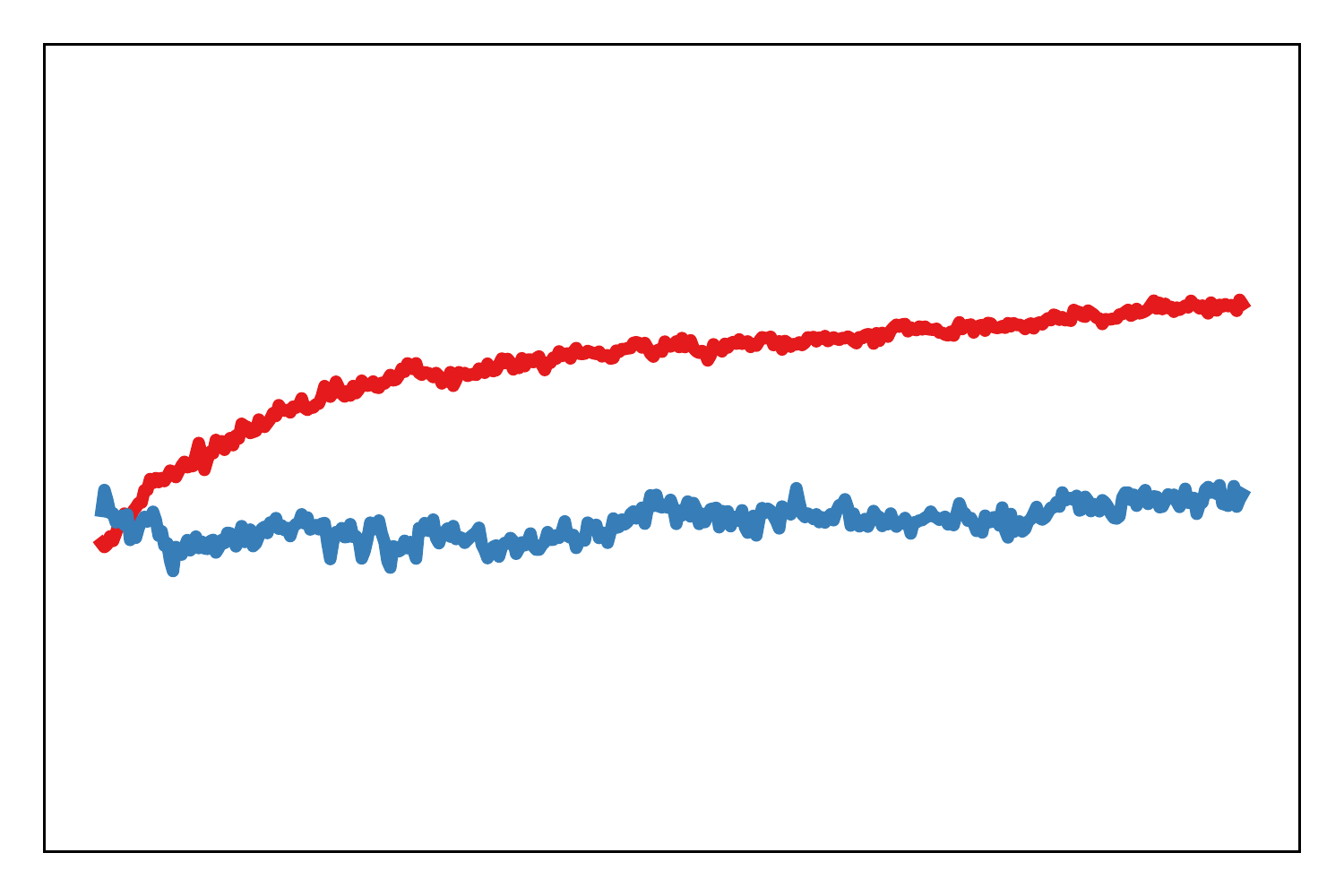}
   &
   \includegraphics[width=.090\textwidth]{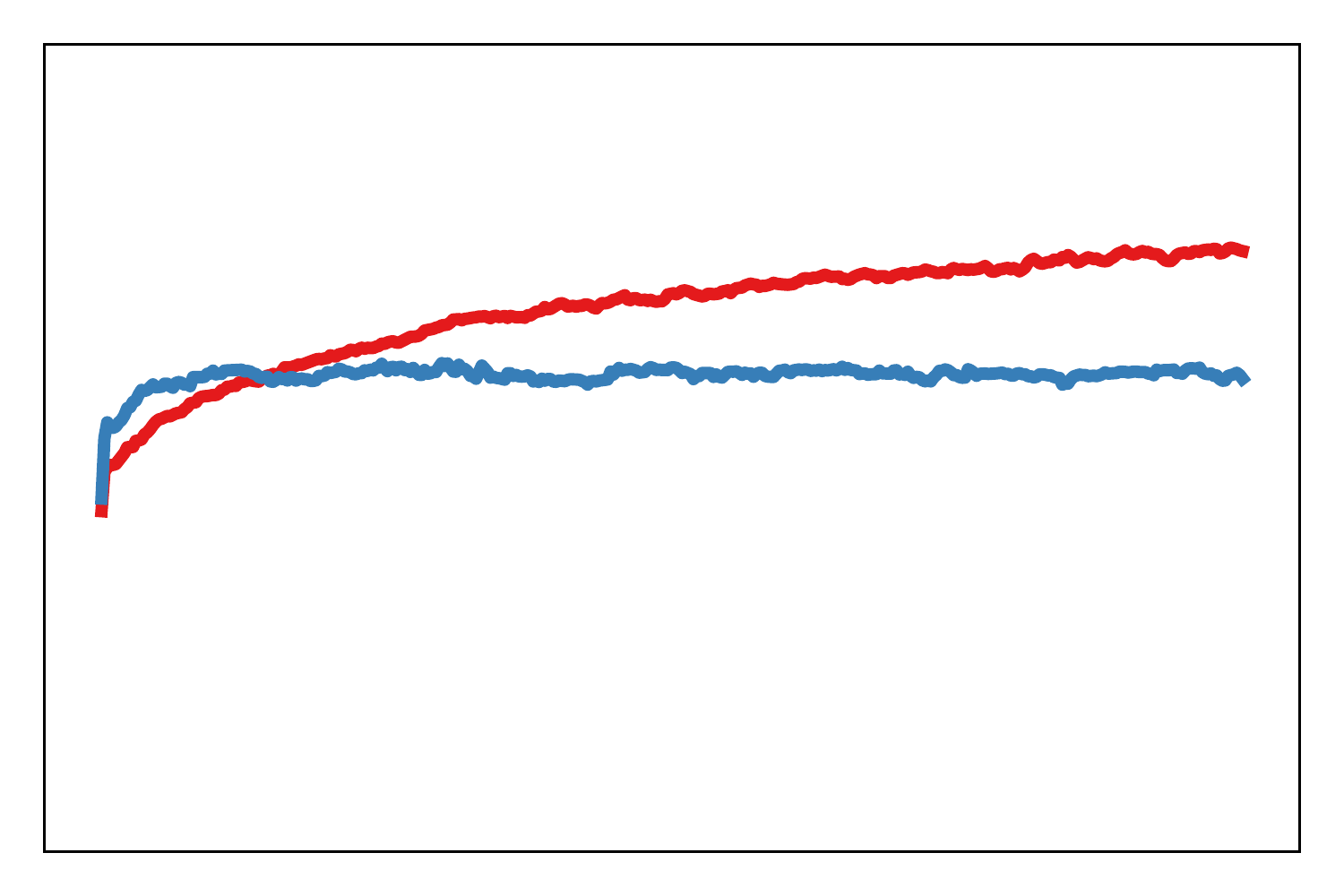}
   &
   \includegraphics[width=.090\textwidth]{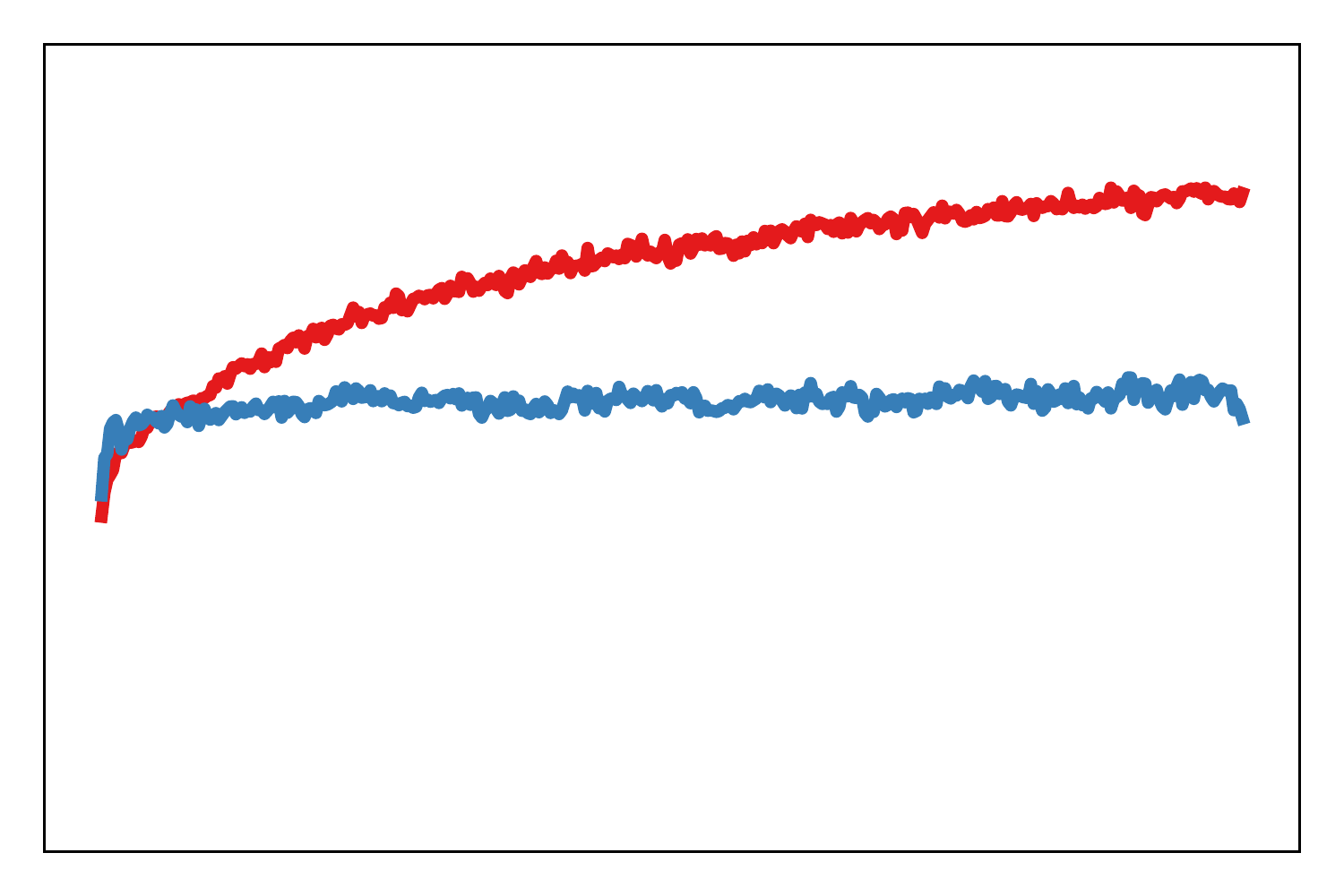}
   &
   \includegraphics[width=.090\textwidth]{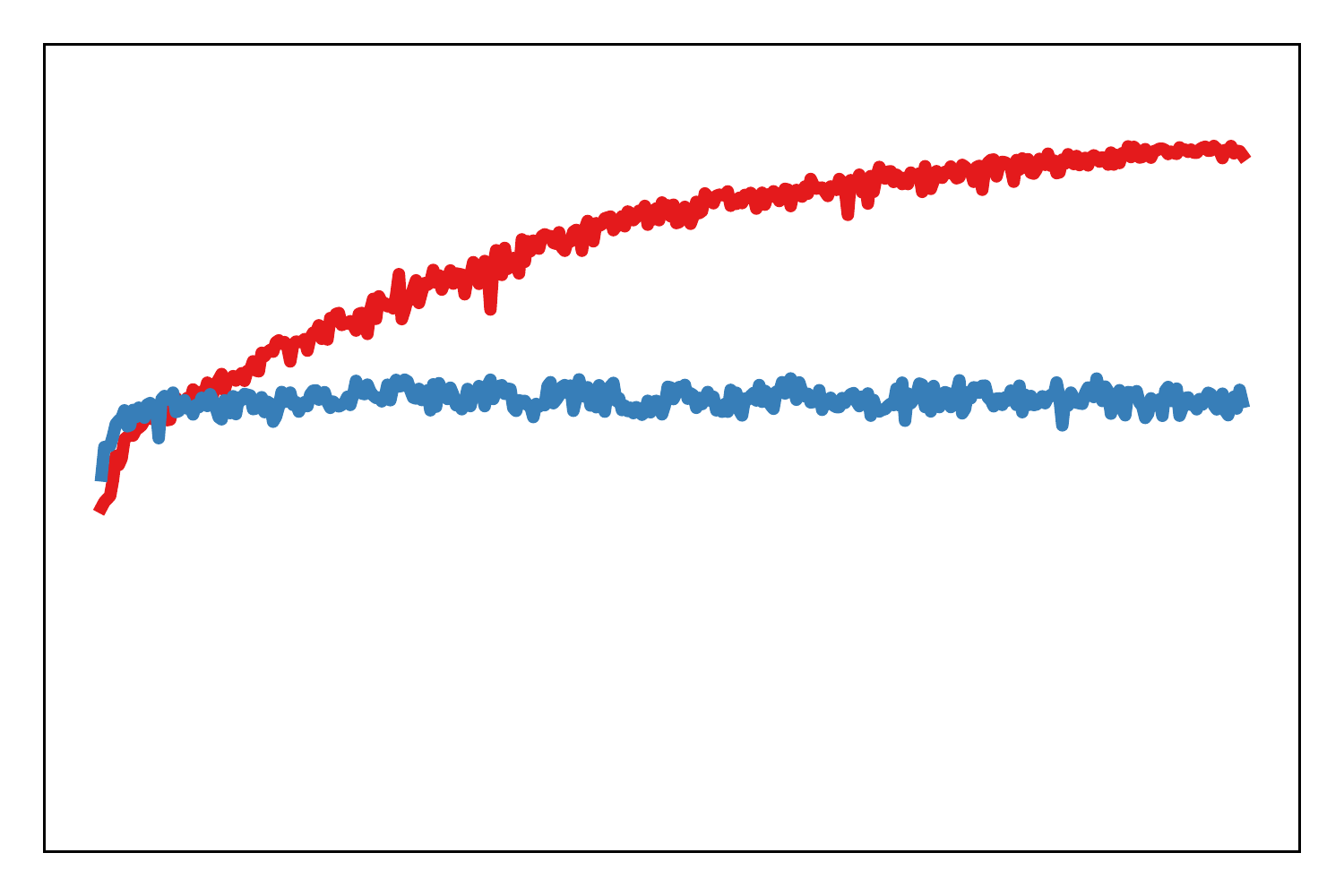}
   &
   \includegraphics[width=.090\textwidth]{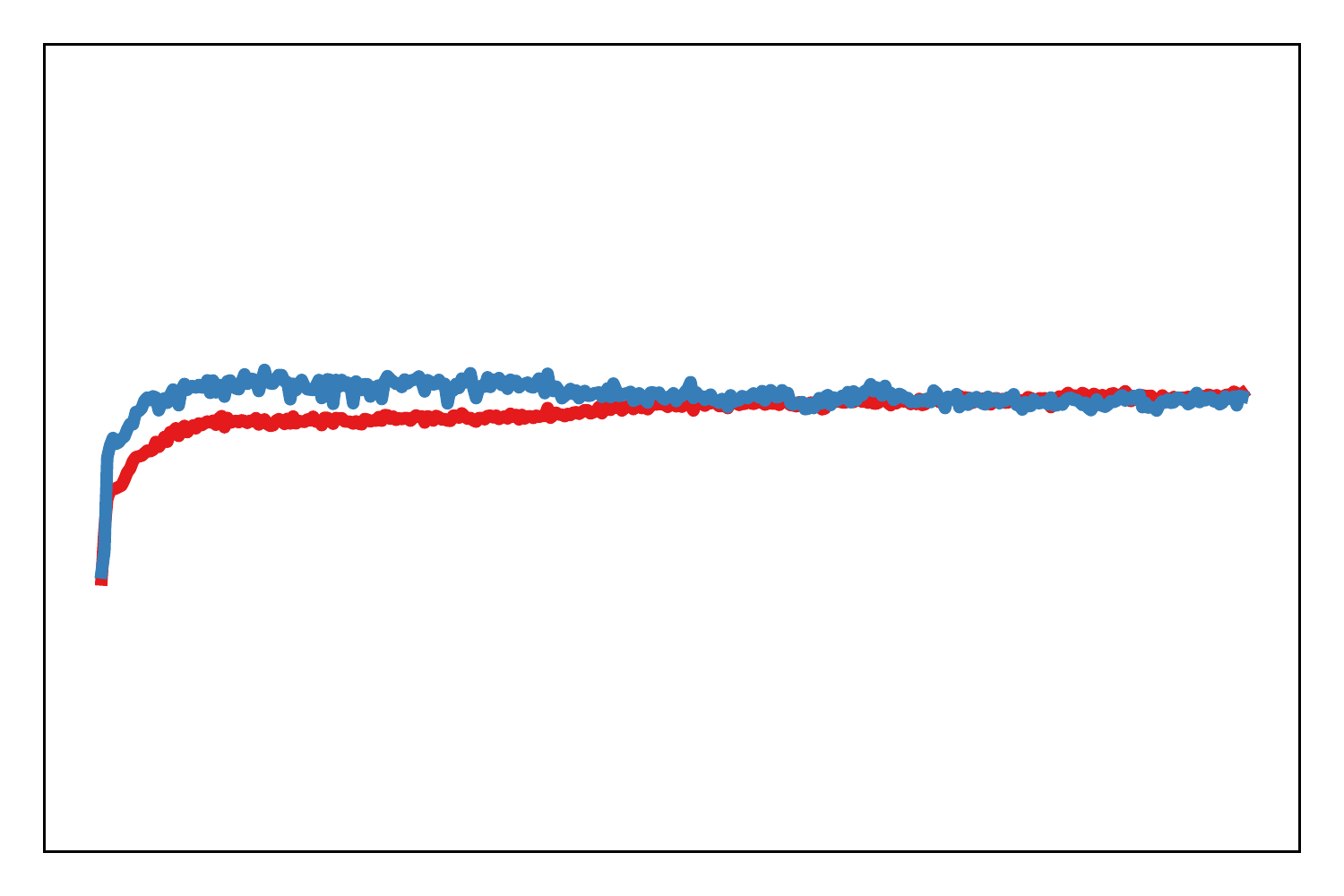}
   &
   \includegraphics[width=.090\textwidth]{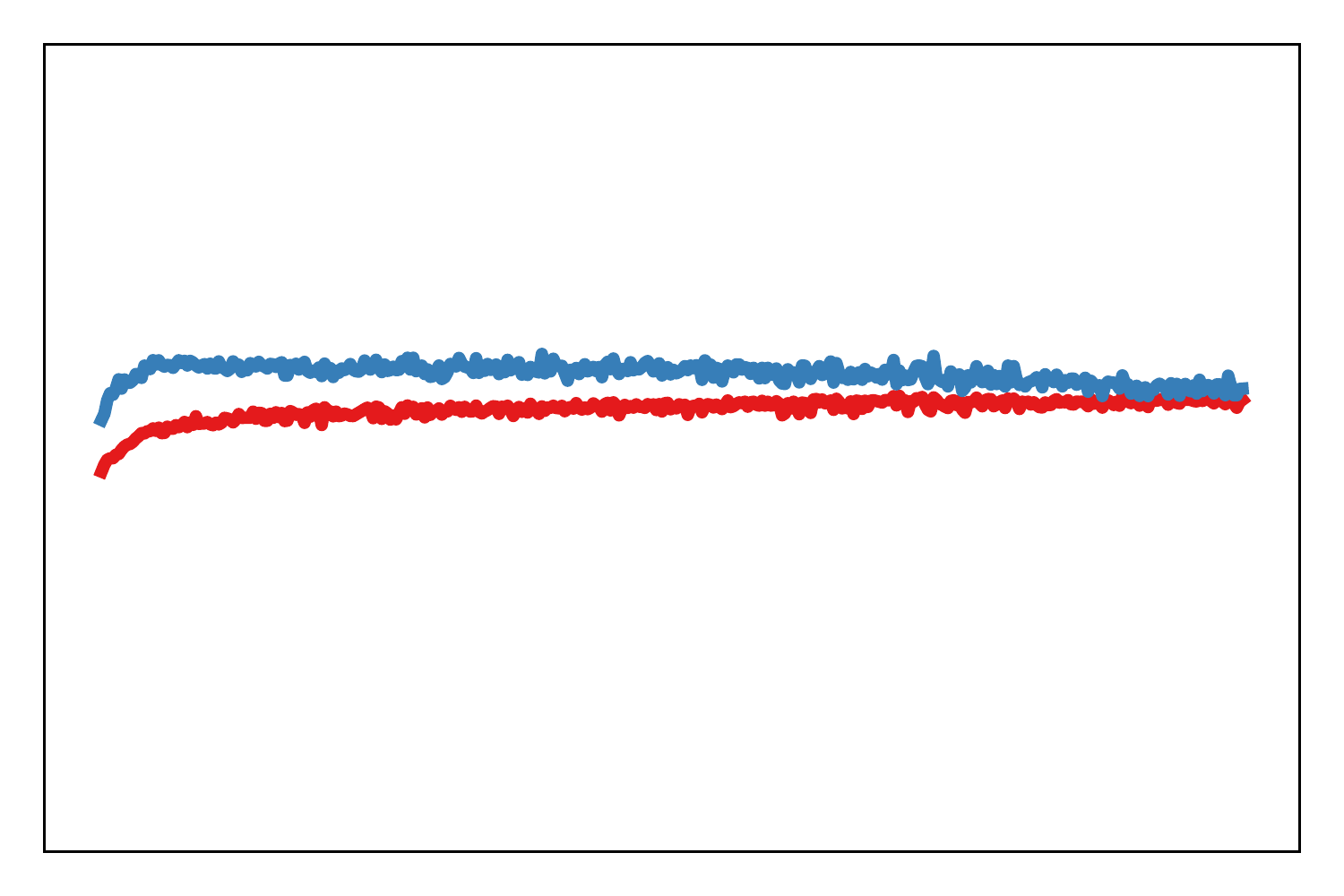}
   &
   \includegraphics[width=.090\textwidth]{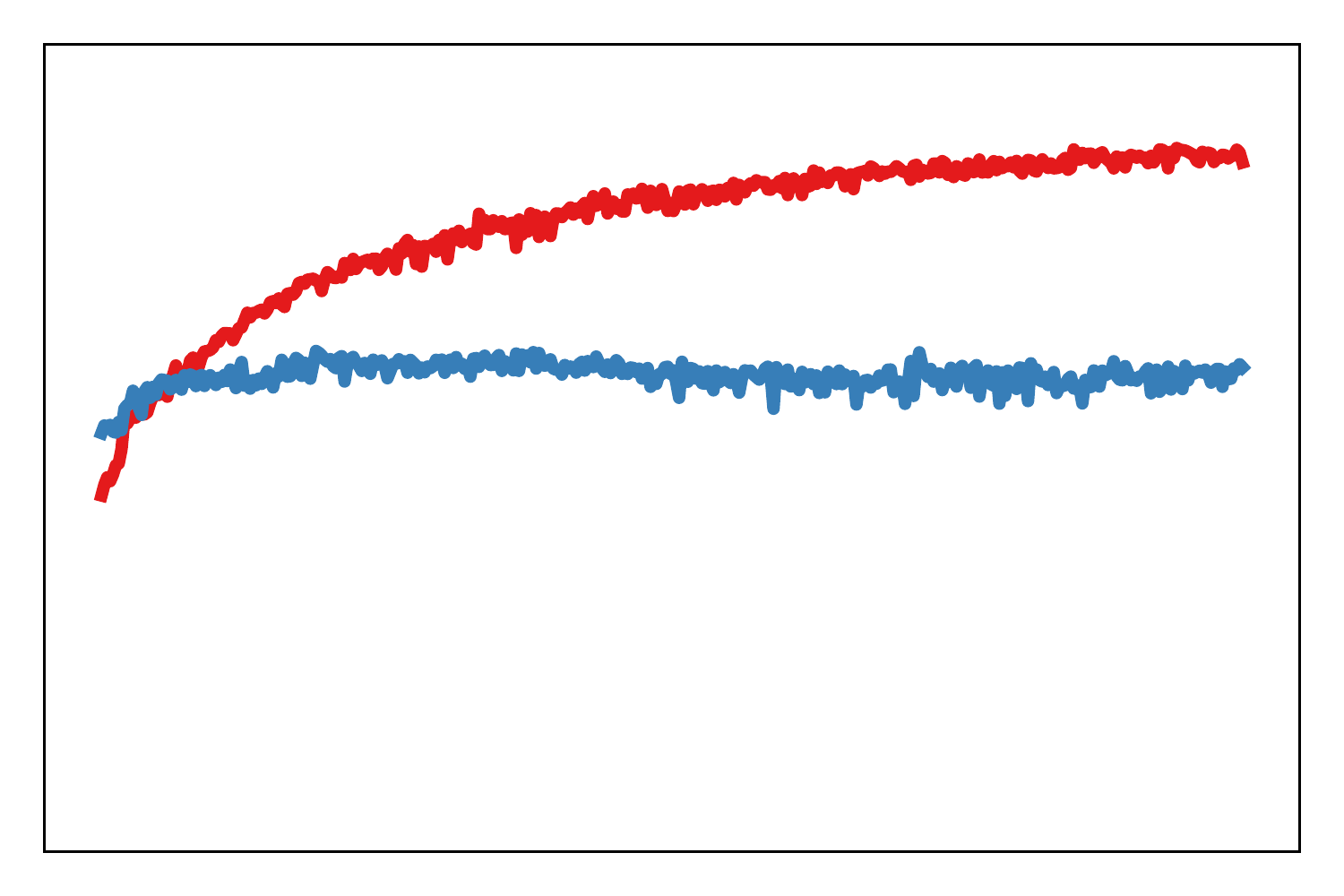}
   &
   \includegraphics[width=.090\textwidth]{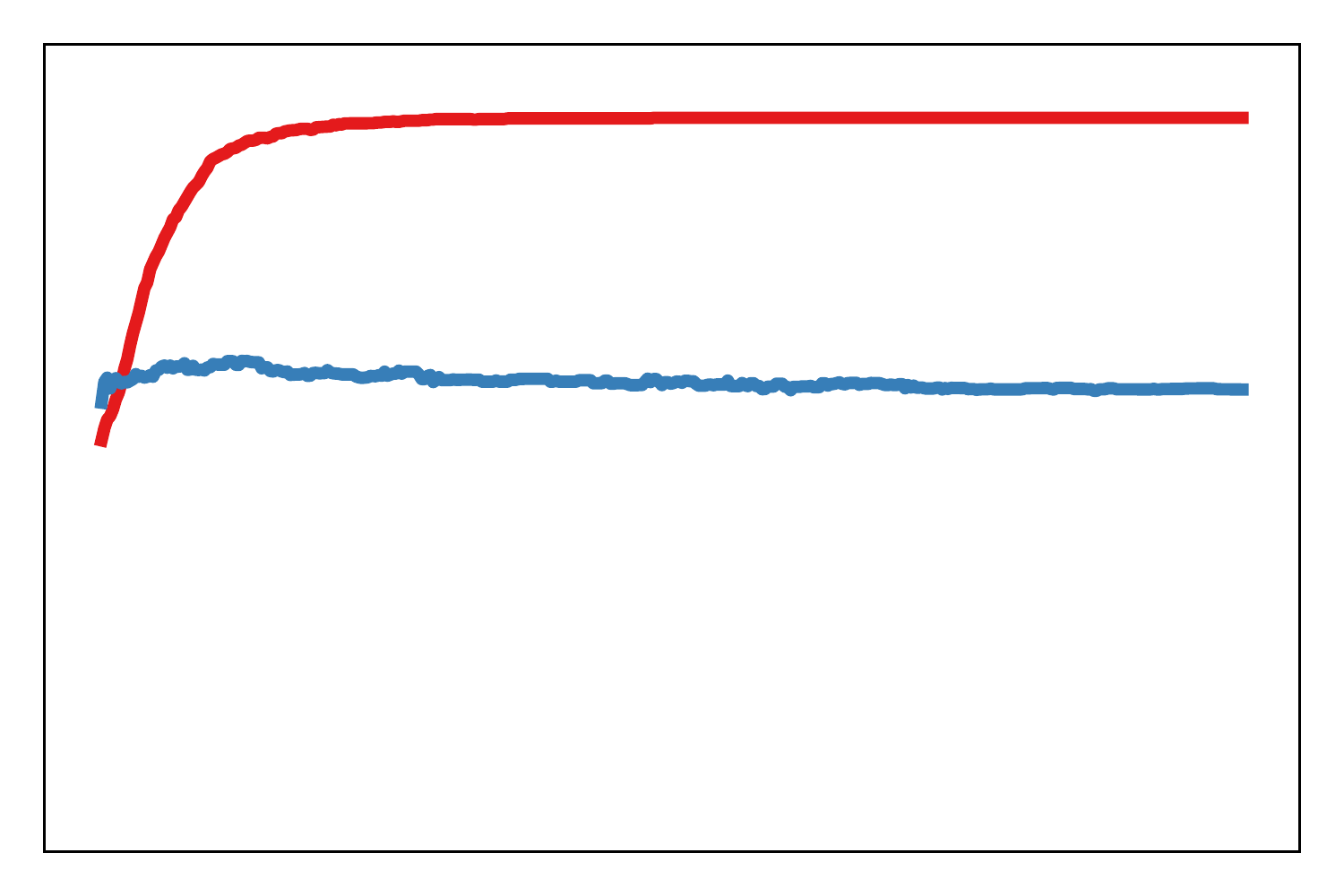}
   &
   \includegraphics[width=.090\textwidth]{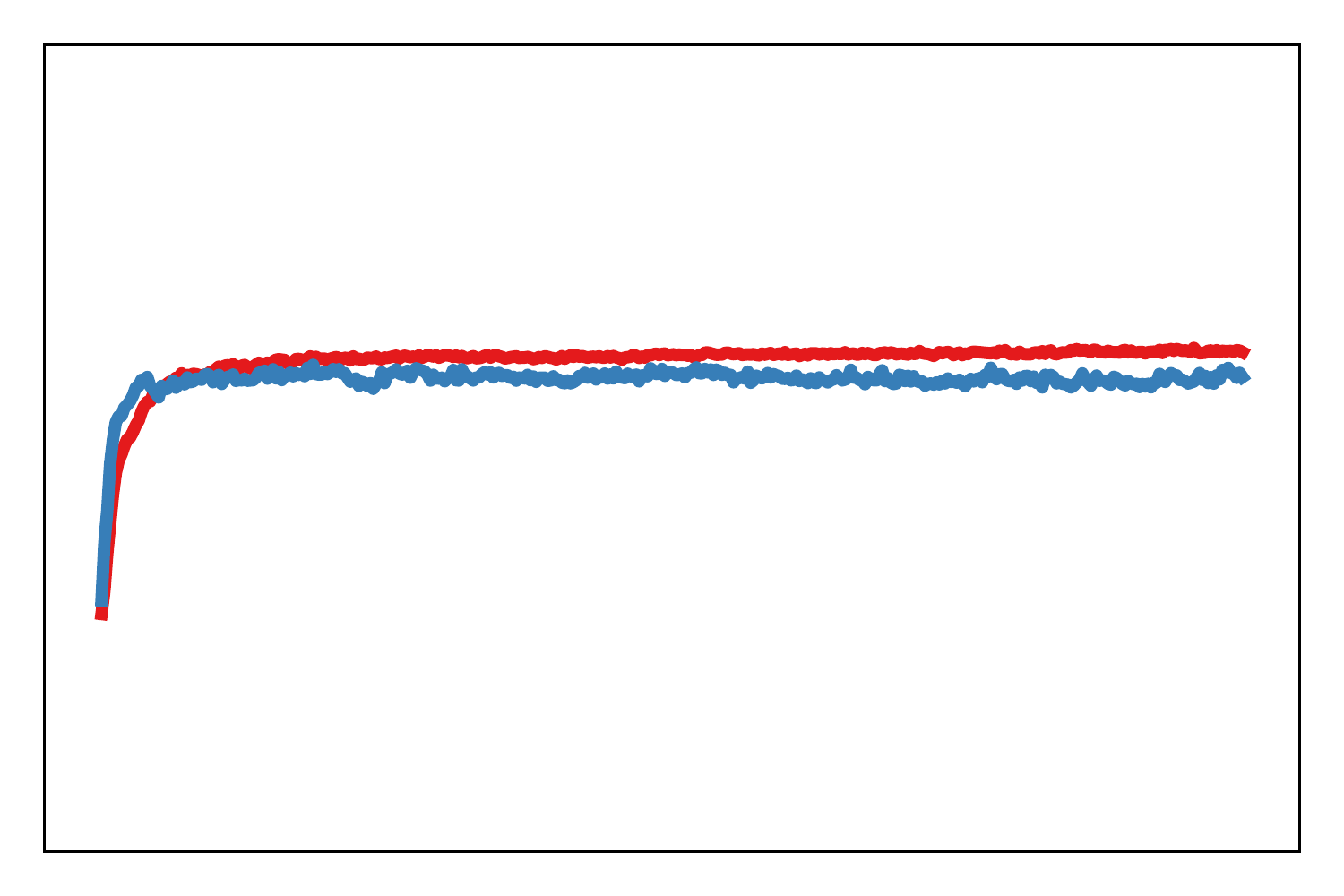}
   \\
   \raisebox{.5cm}{\tiny $100\%$} & 
   \includegraphics[width=.090\textwidth]{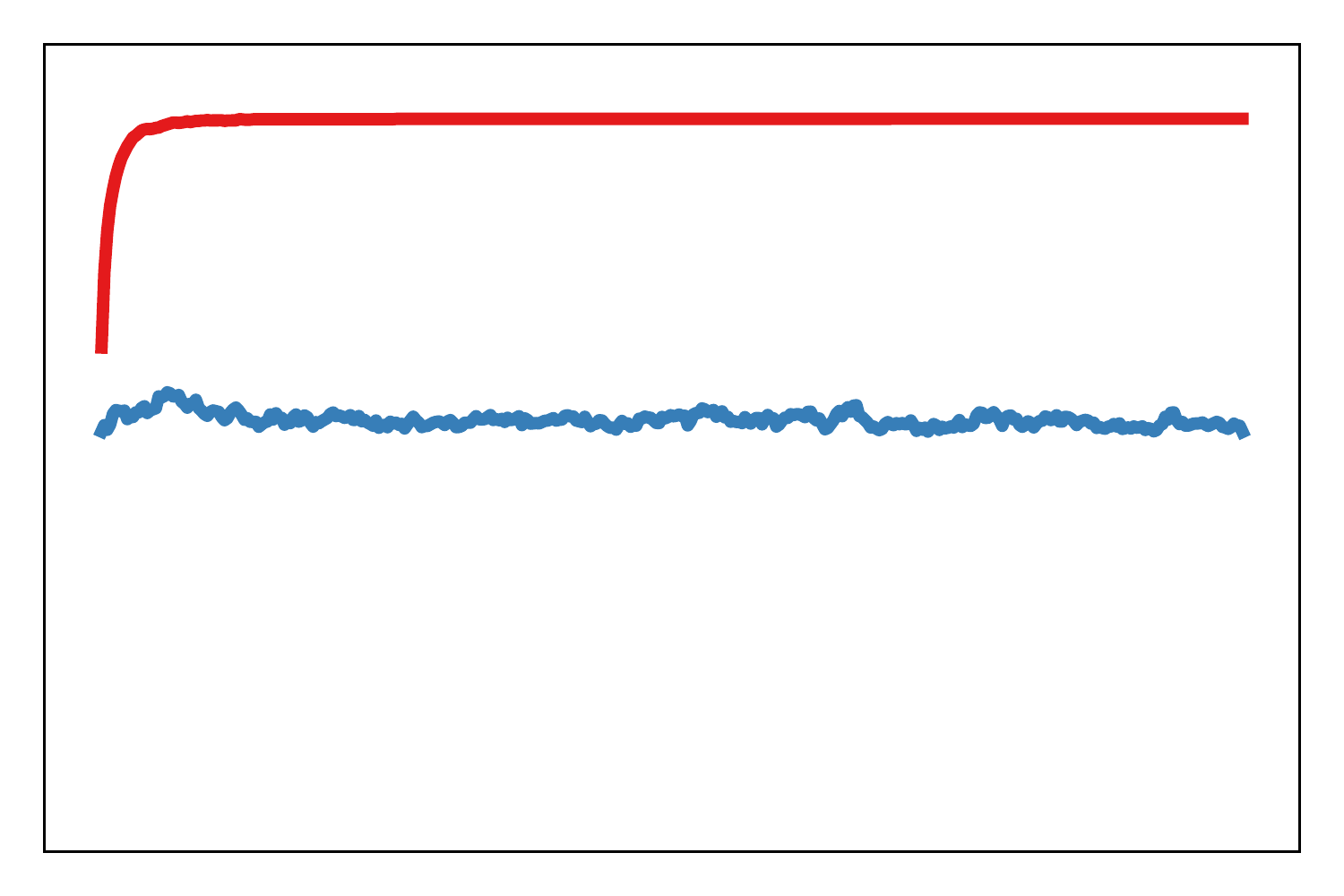}
   &
   \includegraphics[width=.090\textwidth]{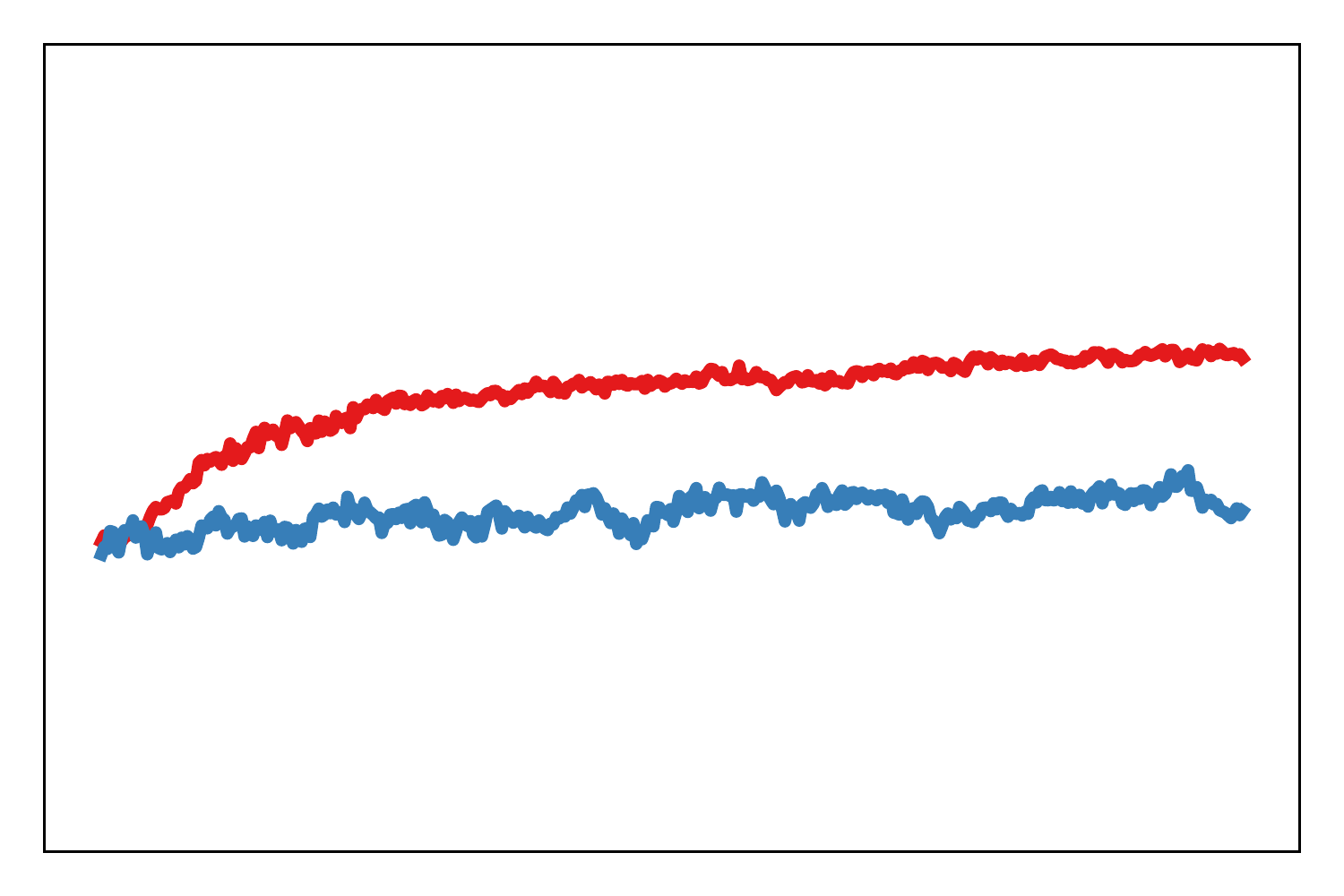}
   &
   \includegraphics[width=.090\textwidth]{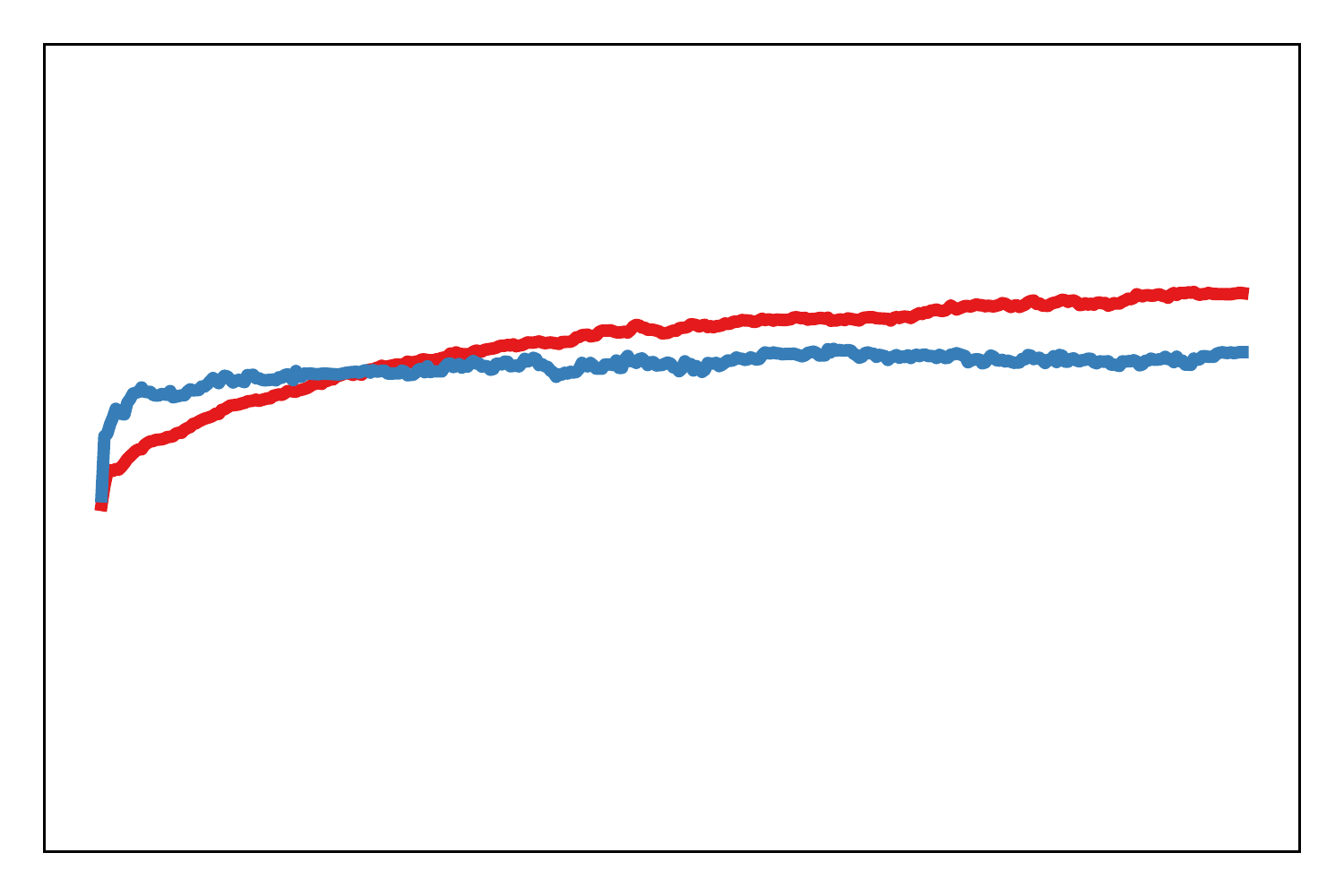}
   &
   \includegraphics[width=.090\textwidth]{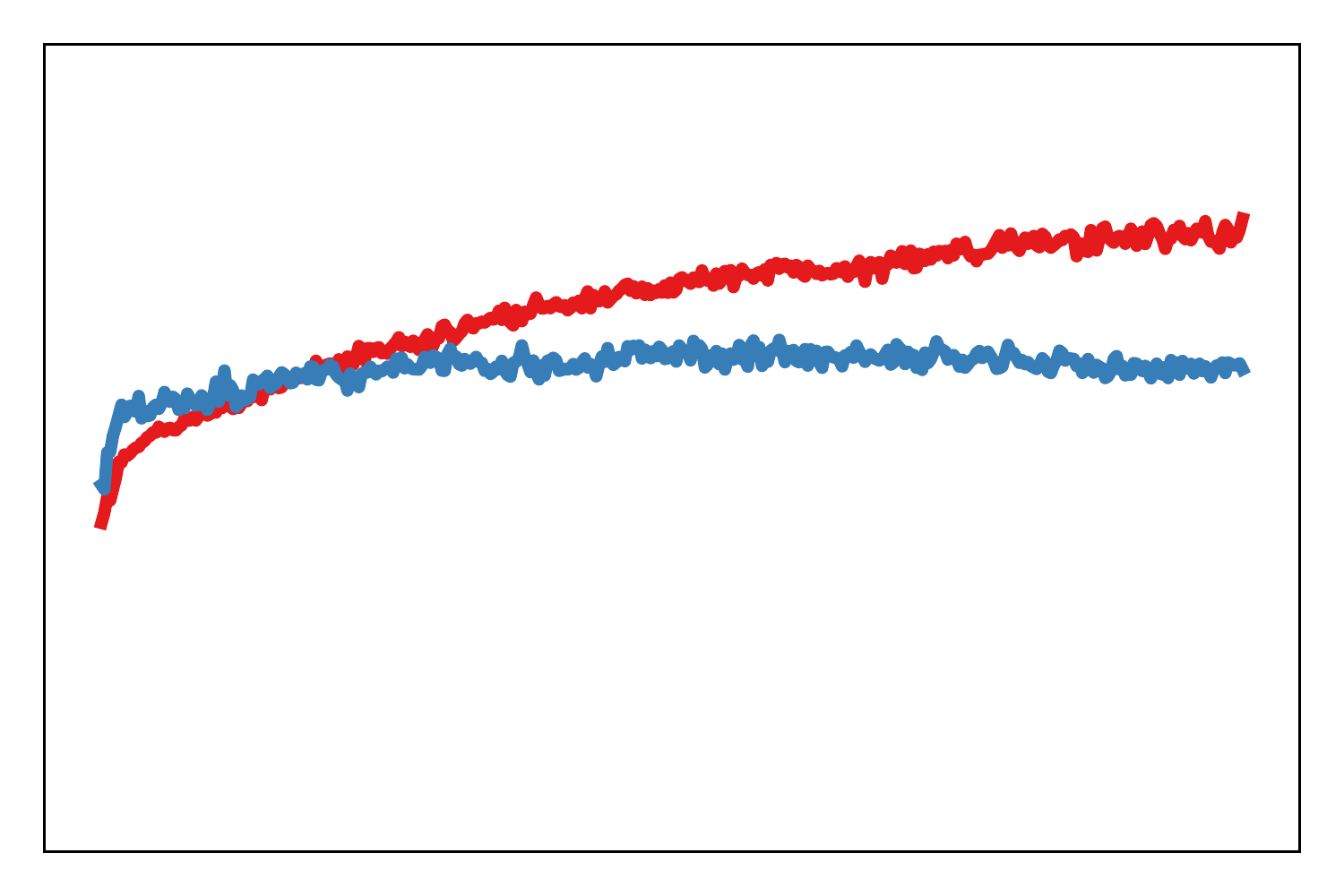}
   &
   \includegraphics[width=.090\textwidth]{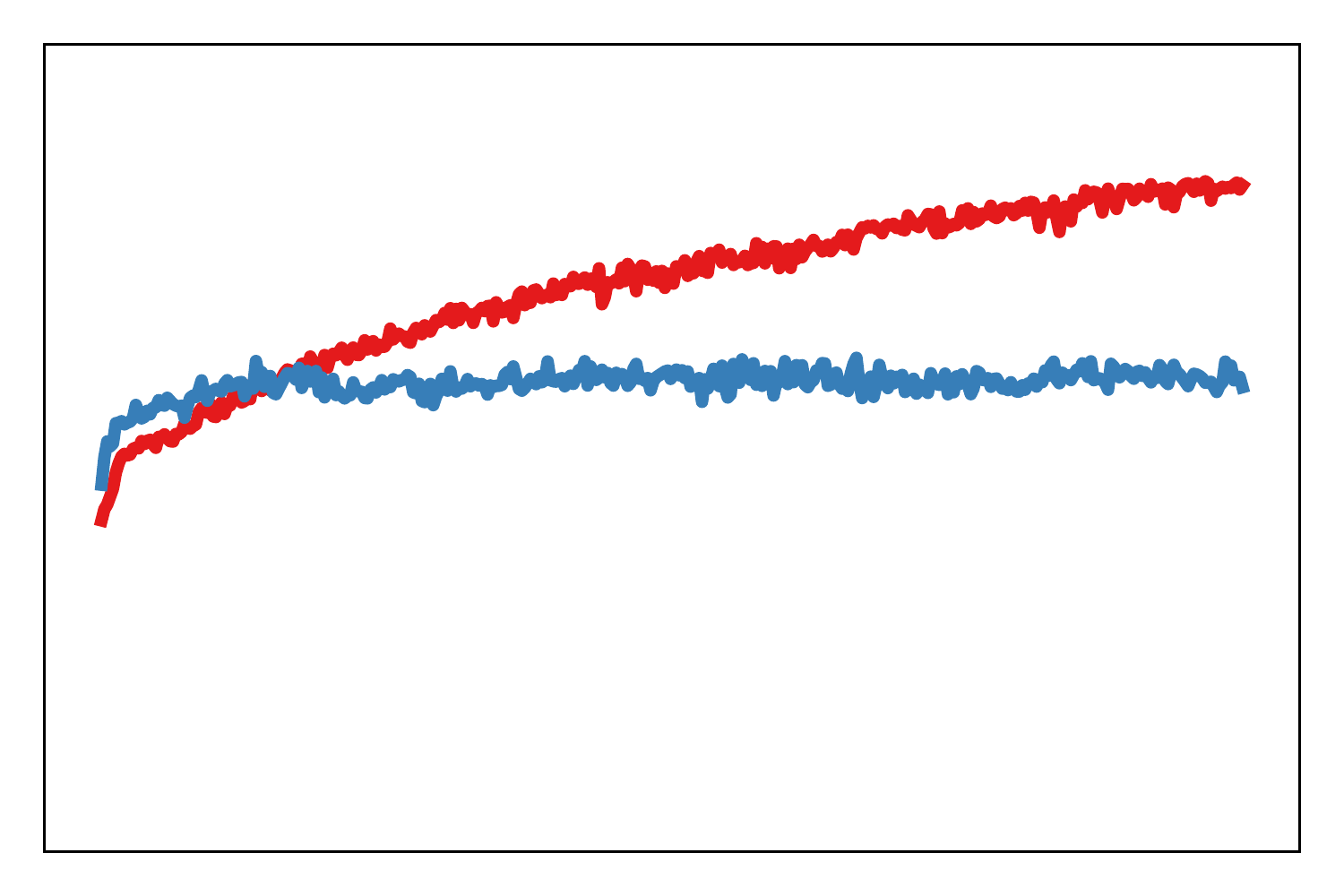}
   &
   \includegraphics[width=.090\textwidth]{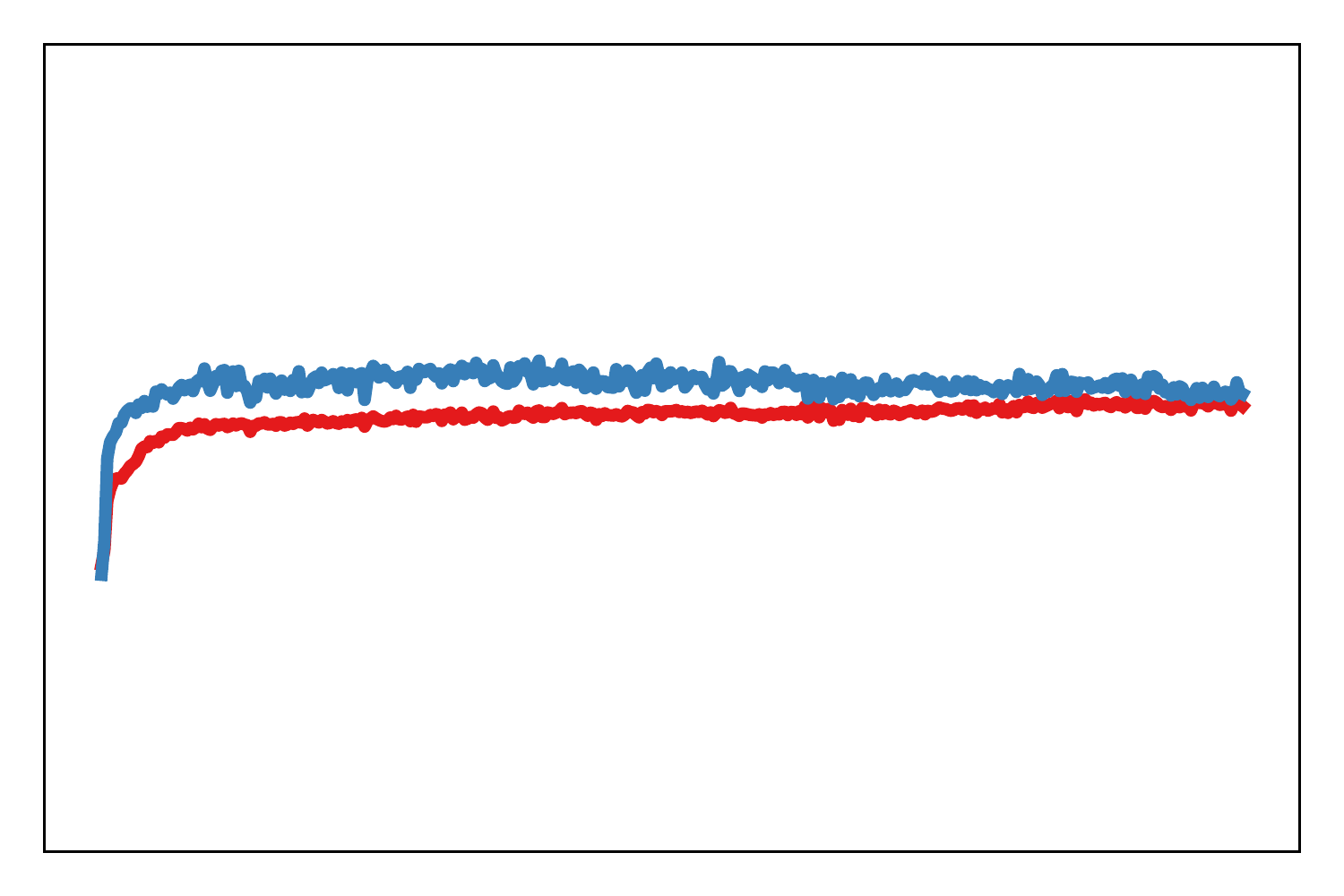}
   &
   \includegraphics[width=.090\textwidth]{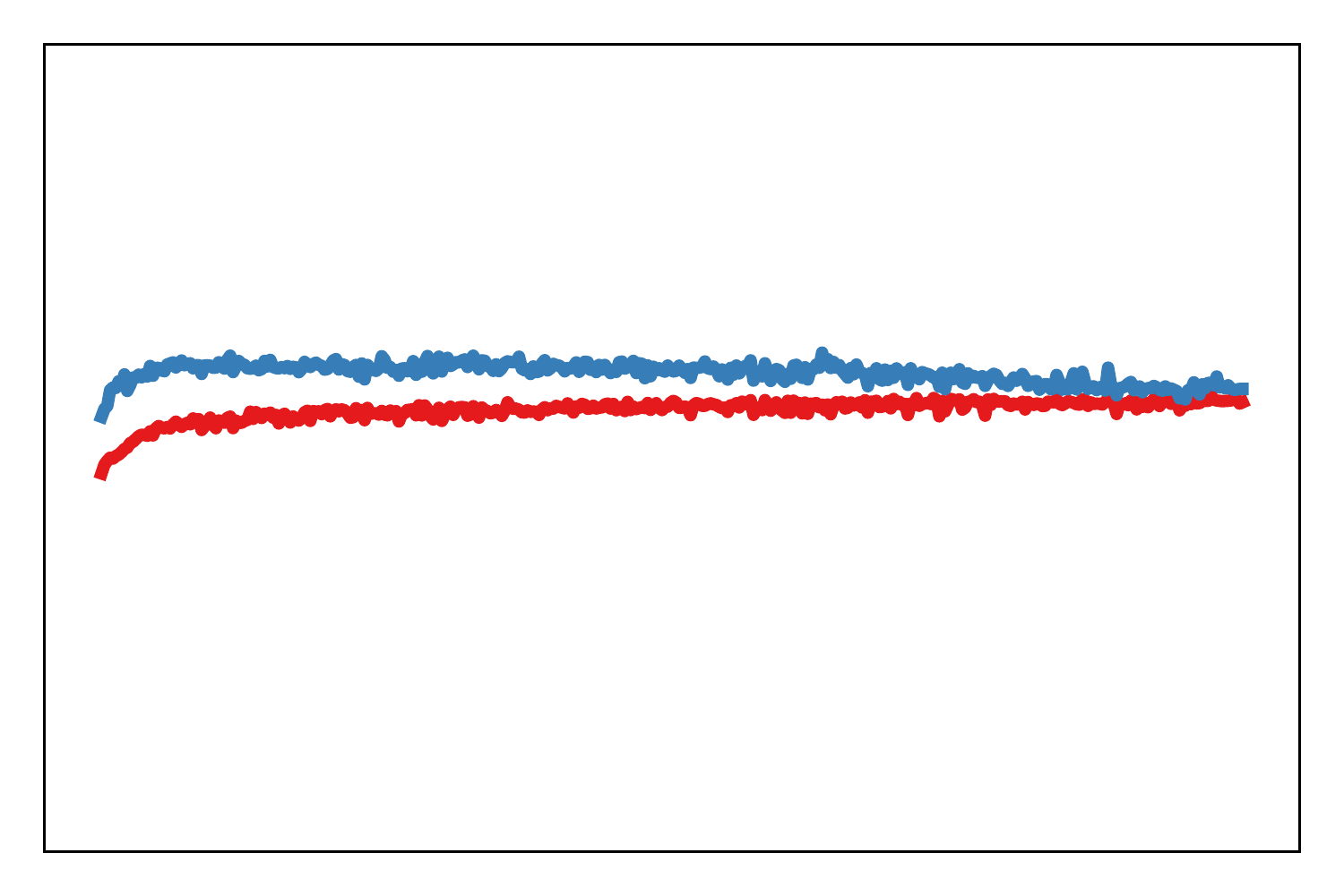}
   &
   \includegraphics[width=.090\textwidth]{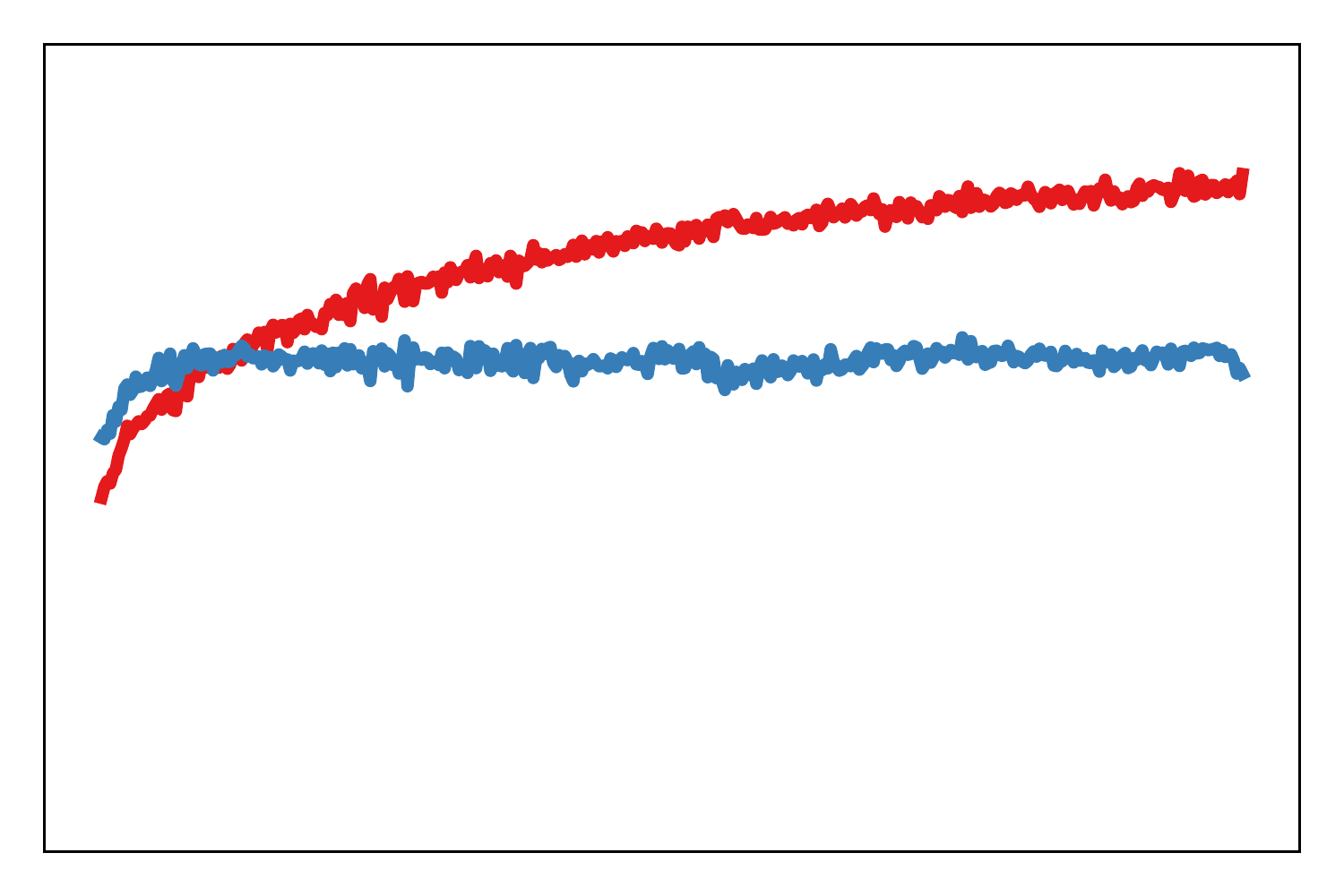}
   &
   \includegraphics[width=.090\textwidth]{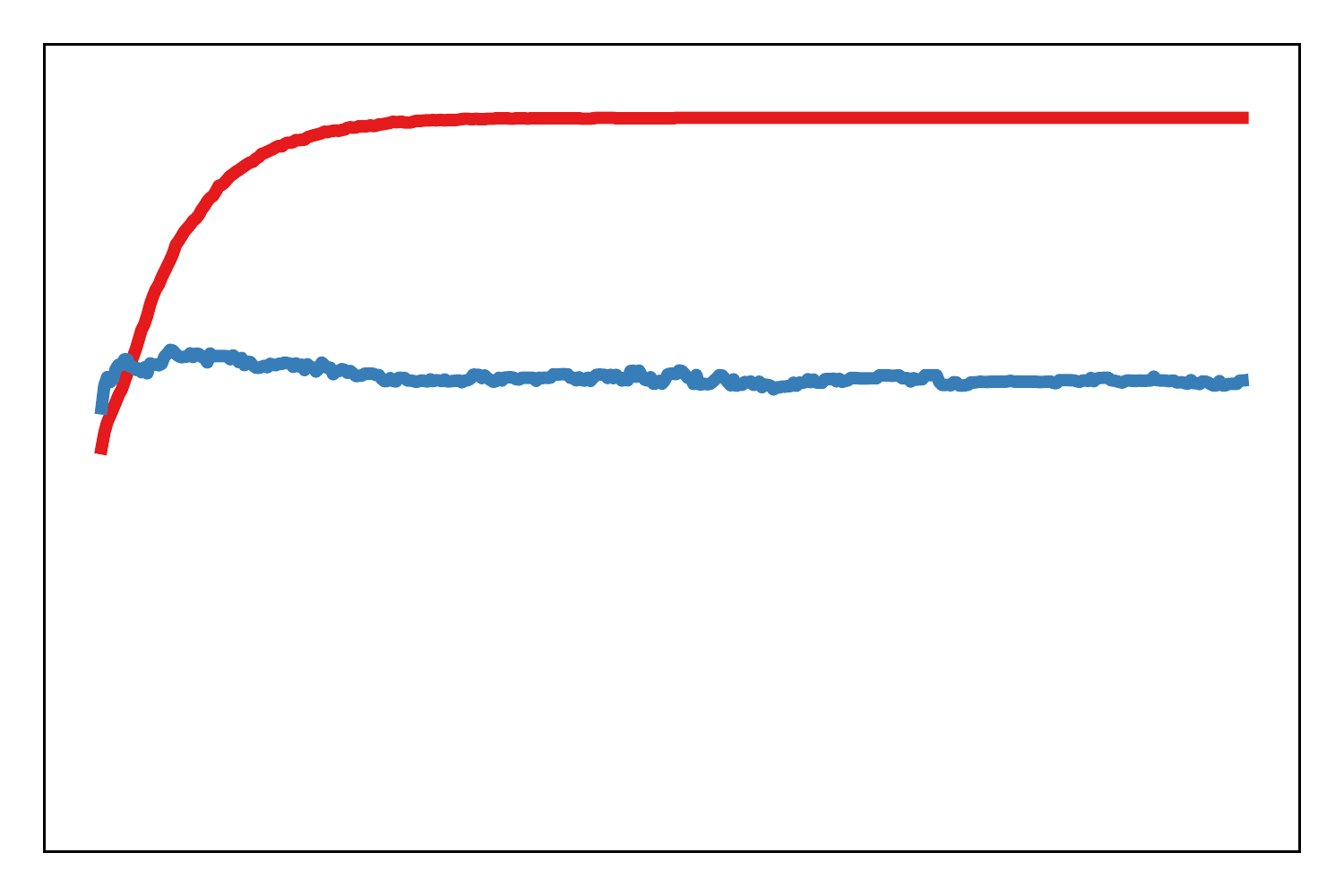}
   &
   \includegraphics[width=.090\textwidth]{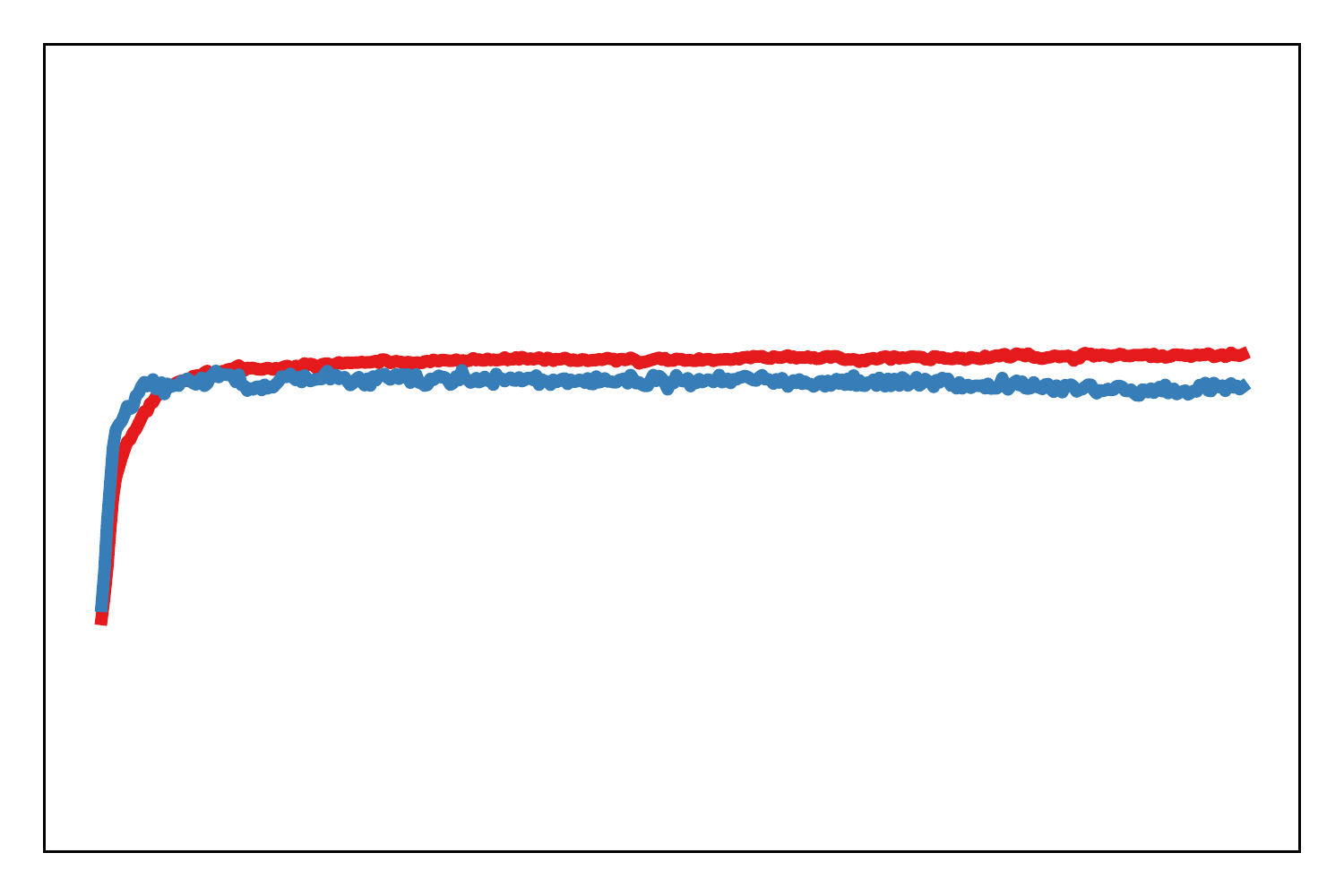} 
   \\ 
   \end{tabular}    
   \vspace*{-0.5\baselineskip}
   \caption{Training histories for various models (columns) with varying training dataset size (rows). The red and blue lines correspond to performance for training and validation datasets, respectively. Performance is measured in terms of MAP, and indicated with respect to the $y$-axes, which range from $0$ to $1$. The $x$-axes indicate the number of training iterations (epochs), and range from $0$ to $399$. The $x$- and $y$-axes are identically scaled in each of the sub-plots.}
   \label{fig:traininghistories}
\end{figure}

Figure~\ref{fig:traininghistories} illustrates the relationship, for each model, between the size of the training dataset and performance improvements over the course of training. 
We can see that most models either reach a plateau or approximately monotonically increase on the training set (shown in blue curves in Fig.~\ref{fig:traininghistories}) within the recorded training history.  There are, however, a few exceptions, namely DRMM and aNMM, which do not exhibit this desired behavior.  Another outlier is DRMM TKS, which improves at a drastically slow rate. 
It is also worth pointing out that the models DSSM and MatchPyramid overfit very quickly. This may suggest a memorization effect.

Looking at the MAP scores on the validation set (shown in blue curves in Fig.~\ref{fig:traininghistories}), we see a discrepancy from expected behavior. 
The desired behavior would be that these follow the same monotonically increasing trend as the red lines, with the gap between the two lines decreasing as the amount of training data increases.
Most of the models, however, do not behave like that. The validation lines plateau out quickly for most models, or even degrade (DRMM, aNMM).

\section{Conclusions}
\label{sec:conl}

We have briefly looked at the effects of dataset size on the neural IR task of answer selection for a number of deep architectures. The consequences of reducing the available training data logarithmically ($10\%$ versus $100\%$) are discernible, and indicate primarily a failure to generalize. This can be seen from the discrepancy between performance improvement on training data, compared to the modest improvements on validation data. 
Note that these findings are based on one particular implementation, and the inner workings of the implementation were not rigorously analyzed and verified, but were assumed to correctly enact the cited algorithms. 

These findings show that when choosing algorithms and strategies in regard to data volume, there are factors which must be considered beyond the reported benchmarks of fully trained models. The actual performance of the models during different stages of training, relative to different scales of training data, must be considered to discover any unexpected trends. 

Furthermore, performance on validation sets is clearly a very important basis for comparison, to gain an intuition about how fast models generalize from different volumes of training data, and with different numbers of training epochs. 

Future work may consist of a deeper investigation into the reproducibility of answer selection state-of-the-art results, as well as into quantifying the relationship between training dataset size and the impact of diverse neural models on generalizability.

\FloatBarrier  

\renewcommand*{\bibfont}{\scriptsize}
\bibliographystyle{abbrvnat}
\bibliography{ecir2019-datatrunc.bib}

\end{document}